\magnification=\magstep1
\font\gross=cmbx12 scaled \magstep0
\font\mittel=cmbx10 scaled \magstep0
\font\Gross=cmr12 scaled \magstep2
\font\Mittel=cmr12 scaled \magstep0
\overfullrule=0pt
\def\Bbb#1{{\bf #1}}
\baselineskip=14pt
%\nopagenumbers
\def\up{\uparrow}
\def\down{\downarrow}
\def\k{{\bf k}}

\def\p{{\bf p}}
\def\q{{\bf q}}
\def\x{{\bf x}}
\def\y{{\bf y}}
\def\I{{\rm I}}
\def\II{{\rm II}}
\def\III{{\rm III}}
\def\IV{{\rm IV}}
\def\ts{\textstyle}
\def\ds{\displaystyle}
\def\tr{\Delta}
\def\la{\langle}
\def\ra{\rangle}
\def\pro{\mathop\Pi}
\def\1cm{\hskip 1cm}
\def\1k{{\textstyle{1\over\kappa}}}
\def\db#1{{\ts{d^d\k\over (2\pi)^d}}}
\def\vp{\varphi}
\def\vep{\varepsilon}
\def\ep{\epsilon}
\def\sl{g}

\def\U{{\cal U}}
\def\V{{\cal V}}
\def\su{\mathop{\Sigma}}
\def\u{\underline}
\def\Eta#1{\u\zeta_{\phantom{.}\!#1}}
\def\Pf{{\rm Pf}}
\def\lne{\mathop{\phantom{ {{  {I\over I}  \over I}\over{I\over I}}\over 
    {{I\over I}\over{I\over I}} }\bf\vrule{\phantom{ {{I\over I}\over{I\over I}}\over 
    {{I\over I}\over{I\over I}} }}}}
\def\sq{ {{U(\k-\p)\atop\!\sim\!\sim\!\sim\!\sim\!\sim\!\sim\!\sim\!\sim\!\sim\!\sim\!}\atop 
    \phantom{I}} }
\def\s0q{ {{U({\bf 0})\atop\!\sim\!\sim\!\sim\!\sim\!\sim\!
   \sim\!\sim\!\sim\!\sim\!\sim\!}\atop 
    \phantom{I}} }

{
\nopagenumbers
\baselineskip=14pt
$$ $$
\vskip 2cm
\centerline{\Gross The Many-Electron System in the Forward,}
\smallskip
\centerline{\Gross  Exchange and BCS Approximation }
\bigskip
\bigskip
\centerline{by}
\bigskip
\bigskip
\centerline{{\Mittel Detlef Lehmann}\footnote{$\phantom{}^{*}$}{present address: 
  TU Berlin, FB Mathematik Ma 7-2, Strasse des 17. Juni 136, 
   D-10623 Berlin; e-mail: 
 lehmann@math.tu-berlin.de}}
\centerline{\Mittel University of British Columbia}
\centerline{\Mittel Department of Mathematics} 
\centerline{\Mittel Vancouver, B.C.}
\centerline{\Mittel V6T 1Z2, Canada}
\vskip 2cm 
\noindent{\bf Abstract:}  The nonrelativistic many-electron system in the forward, 
exchange and BCS approximation is considered. In this approximation, the model 
 is explicitly solvable for arbitrary space dimension $d$. 
 The partition function and 
 the correlation functions are given by 
 finite-dimensional integral representations. Renormalization effects as well as 
symmetry breaking can be seen explicitly. It is shown that the usual mean field 
approach, based on approximating the Hamiltonian by a quadratic expression, may be 
 misleading if the electron-electron interaction contains higher angular momentum 
terms and the space dimension is $d=3$. The perturbation theory of the solvable model 
is discussed. There are cases where the logarithm of the partition function has 
positive radius of convergence but the sum of all connected diagrams has radius of 
 convergence zero implying that the linked cluster theorem is not applicable in these 
 cases.

\vfill
\eject  }
%
%
%  CHAPTER I
%
%
\pageno=1
\noindent {\gross I. Introduction} 
\bigskip
In this paper we consider the nonrelativistic many-electron system in the forward, 
exchange and BCS approximation. In this approximation, which is still quartic in the
annihilation and creation operators, the model can be solved
explicitly. The partition function and the correlation functions are given by 
 finite-dimensional integral representa\-tions. We work in the quantum grand canonical 
 ensemble and start with positiv temperature $T={1\over\beta}>0$ and finite volume 
 $L^d<\infty$. 
\smallskip
The standard model of the  many-electron system in $d$ space 
  dimensions is given by the Hamiltonian 
$$H=H_0+H_{\rm int} \eqno (\I.1)$$
where 
$$H_0={\ts {1\over L^d}} \sum_{\k,\sigma} e_\k \,a_{\k\sigma}^+ 
    a_{\k\sigma}\eqno (\I.2)$$
and 
$$H_{\rm int}={\ts {1\over L^{3d}} }
   \sum_{\sigma,\tau\in \{\up,\down\}} \sum_{\k,\p,\q} \ts 
   U(\k-\p)\, a_{\k\sigma}^+ a_{\p\sigma} a_{\q-\k\tau}^+  a_{\q-\p\tau} 
    \eqno (\I.3)$$
We assume $U$ to be short range, that is $U$ is $L^1$ in coordinate space.     
The energy momentum relation $e_\k$ is given by 
$$ e_\k=\ts{\k^2\over 2m} -\mu \eqno (\I.4)$$
or may be substituted by a more general expression which satisfies $e_\k=e_{-\k}$. The 
parameter $\mu>0$, the chemical potential, is present since we are working in the grand 
canonical ensemble and is determined by the density of the system. Since we 
 are in finite volume, the spatial 
 momenta $\k$ range over some subset of $\left({2\pi\over L}\Bbb Z\right)^d$. 
The physics of the nonrelativistic many-electron system is determined by momenta 
 close to the Fermi surface 
$$ F=\bigl\{\,\k\in \ts\left({2\pi\over L}\Bbb Z\right)^d\>\bigr|\> e_\k=0\,\bigr\}
      \eqno (\I.5)$$
so we impose a fixed ultraviolet cuttoff and choose 
$$\k\in M_\omega=\bigl\{\,\k\in \ts\left({2\pi\over L}\Bbb Z\right)^d\>\bigr|\> |e_\k|
  \le \omega\,\bigr\} \eqno (\I.6)$$
In the context of conventional superconductivity the cuttoff $\omega$ is referred to 
 as the Debye frequency. 
\par
The normalizations choosen in the definition of the creation and annihilation
operators are such that the anticommutation relations read 
$$\{a_{\k\sigma},a_{\k'\sigma'}^+\}=L^d\delta_{\k,\k'}\delta_{\sigma,\sigma'}
     \eqno (\I.7)$$ 
In particular, if $\Omega_F=\!\!\ds\pro_{\k\sigma\atop e_\k<0}\!\!\!\!
    a_{\k\sigma}^+\,{\bf 1}$ 
  is the zero
temperature ground state of the noninteracting system, then 
 $(\Omega_F ,a_{\k\sigma}^+ 
    a_{\k\sigma}\Omega_F)=L^d\theta(-e_\k)$ such that 
 $(\Omega_F,H_0\Omega_F)= \sum_{\k,\sigma} e_\k\,\theta(-e_\k)\approx 
  L^d\int {d^d\k\over (2\pi)^d}\,e_\k\,\theta(-e_\k)$ 
   is indeed an extensive quantity as it should be. Here 
$\theta(\vep)$ is the step function being one for $\vep>0$ and zero otherwise.
\smallskip
We are interested in the grand canonical partition function 
$$Z=Z(\beta,L,U)=Tr\, e^{-\beta H} \eqno (\I.8)$$ 
 which may be normalized by $Tr\,e^{-\beta H_0}$ such that $Z(U=0)=1$, 
 and in particular in the two point function 
$$\la a_{\k\sigma}^+ a_{\k'\sigma'}\ra_{\beta,L}={ Tr\, e^{-\beta H} 
    a_{\k\sigma}^+ a_{\k'\sigma'}  \over Tr\, e^{-\beta H} } \eqno (\I.9)$$
which gives the momentum distribution of the system. 
Recall that in the free $(U=0)$ system, the ideal Fermi gas, 
$$\lim_{\beta\to\infty}  \la a_{\k\sigma}^+ a_{\k'\sigma'}\ra_{\beta,L}=
  L^d\delta_{\k,\k'}\delta_{\sigma,\sigma'} \theta(-e_\k)  \eqno (\I.10)$$ 
\par
Since we are in the quantum grand canonical ensemble, the traces in (I.8,9) are to be 
taken over the Fock space ${\cal F}=\oplus_{n=0}^\infty {\cal F}_n$ where 
$${\cal F}_n=\left\{ F_n\in L^2\left(
   \bigl[ [0,L]^d\times\{\uparrow,\downarrow\}\bigr]^n\right)\>
   \Bigr|\>F_n(\x_{\pi1}\sigma_{\pi1},\cdots,\x_{\pi n}\sigma_{\pi n})=
    {\rm sgn}\pi\>F_n(\x_1\sigma_1,\cdots,\x_n\sigma_n)\right\}$$
The fact that the physical system has a fixed number of particles $N$ is 
expressed by requireing that the expectation value $\la {\bf N}\ra_{\beta,L}$ of the 
number operator ${\bf N}={1\over L^d}\sum_{\k\sigma}a_{\k\sigma}^+ 
    a_{\k\sigma}$ (which is extensive, see above) is equal to the number of 
particles, $\la {\bf N}\ra_{\beta,L}=N$, which determines $\mu$ as a function 
 of the density ${N\over L^d}$. 
\bigskip
As usual, the quartic part $H_{\rm int}$ (I.3) of the Hamiltonian $H$ (I.1) may 
 be represented by the following four legged diagram: 
$$ \lne_{\phantom{I}\atop\ds \p,\sigma}^{\ds\k,\sigma
      \atop \phantom{I}}\!\!\!\! \!\! \!\! 
   \sq \!\! \!\! \!\! \!\!\!\! \!\! \!\!   \lne_{\phantom{I}\atop
    \ds \q-\p,\tau}^{\ds \q-\k,\tau \atop \phantom{I}}
   \leqno H_{\rm int}: $$
Because of conservation of momentum, there are three independent momenta here labelled 
with $\k,\p$ and $\q$. Then one can consider the following three limiting cases 
with only two independ momenta:
\goodbreak
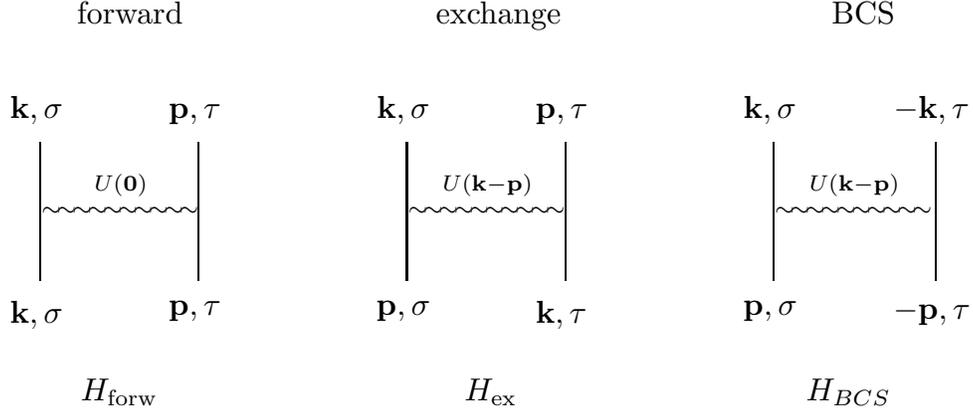
$${\rm forward}\hskip 2.8cm {\rm exchange} \hskip 3cm {\rm BCS}$$\nobreak
$$\lne_{\phantom{I}\atop\ds \k,\sigma}^{\ds\k,\sigma
      \atop \phantom{I}}\!\!\!\!\!\! \!\! 
   \s0q \!\! \!\! \!
  \!\!\!\!\!\lne_{\phantom{I}\atop
    \ds \p,\tau}^{\ds \p,\tau \atop \phantom{I}}  \hskip 1.4cm
  \lne_{\phantom{I}\atop\ds \p,\sigma}^{\ds\k,\sigma
      \atop \phantom{I}}\!\!\!\!\! \!\! \!
   \sq \!\! \!\! \!
 \!\!\!\!\!\lne_{\phantom{I}\atop
    \ds \k,\tau}^{\ds \p,\tau \atop \phantom{I}}  \hskip 1.4cm
  \lne_{\phantom{I}\atop\ds \p,\sigma}^{\ds\k,\sigma
      \atop \phantom{I}}\!\!\!\! \!\! \!\! 
   \sq \!\! \!\! \!
  \!\!\!\!\!\lne_{\phantom{I}\atop
    \ds -\p,\tau}^{\ds -\k,\tau \atop \phantom{I}}  $$
\nobreak
$$ H_{\rm forw}\hskip 3.4cm H_{\rm ex} \hskip 3.2cm H_{BCS} $$
\goodbreak
That is, one may consider the approximation 
$$ H_{\rm int}\approx H_{\rm forw}+H_{\rm ex}+H_{\rm BCS} \eqno (\I.11)$$
where 
$$\eqalignno{ H_{\rm forw}&={\ts {1\over L^{3d}} }
   \sum_{\sigma,\tau\in \{\up,\down\}} \sum_{\k,\p,\q} 
   U(\k-\p)\,\delta_{\k,\p}\, a_{\k\sigma}^+ a_{\p\sigma} a_{\q-\k\tau}^+
     a_{\q-\p\tau}  \cr
  &={\ts {1\over L^{3d}} }
   \sum_{\sigma,\tau\in \{\up,\down\}} \sum_{\k,\p} \ts 
   U({\bf 0})\, a_{\k\sigma}^+ a_{\k\sigma} a_{\p\tau}^+  a_{\p\tau} 
    & (\I.12)  \cr
 & \cr
 H_{\rm ex}&={\ts {1\over L^{3d}} }
   \sum_{\sigma,\tau\in \{\up,\down\}} \sum_{\k,\p,\q}  
   U(\k-\p)\,\delta_{\k,\q-\p}\, a_{\k\sigma}^+ a_{\p\sigma}  a_{\q-\k\tau}^+ 
    a_{\q-\p\tau} \cr
 & =
  {\ts {1\over L^{3d}} }
   \sum_{\sigma,\tau\in \{\up,\down\}} \sum_{\k,\p}  
   U(\k-\p)\, a_{\k\sigma}^+ a_{\p\sigma}  a_{\p\tau}^+ a_{\k\tau}&(\I.13) \cr
 & \cr
 H_{\rm BCS}&={\ts {1\over L^{3d}} }
   \sum_{\sigma,\tau\in \{\up,\down\}} \sum_{\k,\p,\q} 
   U(\k-\p)\,\delta_{\q,{\bf 0}}\, a_{\k\sigma}^+ a_{\p\sigma}  a_{\q-\k\tau}^+ 
        a_{\q-\p\tau} \cr
 & ={\ts {1\over L^{3d}} }
   \sum_{\sigma,\tau\in \{\up,\down\}} \sum_{\k,\p} \ts 
   U(\k-\p)\, a_{\k\sigma}^+ a_{\p\sigma}  a_{-\k\tau}^+ a_{-\p\tau} &(\I.14) \cr}$$
\par
Let us shortly make a comment on the volume factors. In (I.12-14), we only 
introduced some Kroenecker delta's but we did not cancel a volume factor $L^d$. 
That this is the right thing to do, that is, that the left hand side as well as the 
right hand side of (I.11) is indeed proportional to the volume may be seen in the 
easiest way for the forward term. On a fixed $n$ particle space ${\cal F}_n$ 
 the interacting part $H_{\rm int}$ is a multiplication operator given by 
$${H_{\rm int}}_{|_{{\cal F}_n}}
  ={\ts {1\over2}}\sum_{i,j=1\atop i\ne j}^n U(\x_i-\x_j) \eqno (\I.15)$$
Let $\delta_\y(\x)=\delta(\x-\y)$. Then 
 $\varphi(\x_1,\cdots,\x_n)=\delta_{\y_1}\wedge\cdots\wedge
  \delta_{\y_n}(\x_1,\cdots,\x_n)$ 
 is an eigenfunction of $H_{\rm int}$ with eigenvalue 
$$E={\ts {1\over2}}\sum_{i,j=1\atop i\ne j}^n U(\y_i-\y_j)=
 {\ts {1\over2 L^d}}\sum_{i,j=1\atop i\ne j}^n \sum_\q 
  e^{i(\y_i-\y_j)\q}\>U(\q)\eqno (\I.16) $$ 
which is, for $U\in L^1$, proportional to $n$ or to the volume $L^d$ for constant 
 density. One finds that $\varphi$ is also an eigenvector of the forward term, 
  $H_{\rm forw}\varphi=E_{\rm forw}\varphi$  
where $E_{\rm forw}$ is obtained from (I.16) by putting $\q=0$ without cancelling 
a volume factor, 
$$E_{\rm forw}={\ts {1\over2 L^d}}\sum_{i,j=1\atop i\ne j}^n U(\q=0)\eqno (\I.17) $$ 
which is also proportional to the volume.
\bigskip
We now come to the exact definition of the model which is solved in this paper. 
To do so, we need the functional integral representation of the perturbation series 
for the partition function. It is summarized
  in the following 
theorem which is fairly standard. One may look in [FKT1] for a nice and clean proof. 
\bigskip
\noindent{\bf Theorem I.1:} {\it Let $H=H_0+H_{\rm int}$ be the Hamiltonian (I.1-3), 
let 
$$Z=Z(\beta,L)=Tr\,e^{-\beta(H_0+H_{\rm int})}/Tr\, e^{-\beta H_0} \eqno (\I.18)$$
be the normalized grand canonical partition function and let 
$$\la a_{\k\sigma}^+ a_{\k\sigma}\ra_{\beta,L}={ Tr\, e^{-\beta H} 
    a_{\k\sigma}^+ a_{\k\sigma}  \over Tr\, e^{-\beta H} } \eqno (\I.19)$$
{\bf a)} The partition function $Z$ has the following perturbation series 
$$\eqalignno{ Z=\sum_{n=0}^\infty{\ts {(-{1\over \beta L^d})^n\over n!}}
   \sum_{\sigma_1\cdots
  \sigma_{2n}}\sum_{k_1\cdots k_{2n}\atop p_1\cdots p_{2n}}\prod_{i=1}^n&\left\{ 
  \delta_{k_{2i-1}+k_{2i},p_{2i-1}+p_{2i}}U(\k_{2i}-\p_{2i})\right\}\>\times \cr
 &\phantom{mm}
   \det\left[ \delta_{\sigma_i,\sigma_j}\delta_{k_i,p_j}\,C(k_i)\right]_{1\le i,j\le 2n}
   & (\I.20)   \cr}$$ 
Here $k=(k_0,\k)\in {\pi\over\beta}(2\Bbb Z+1)\times 
  \bigl\{\,\k\in \ts\left({2\pi\over L}\Bbb Z\right)^d\>\bigr|\> |e_\k|
  \le \omega\,\bigr\}$ and the covariance $C$ is given by 
$$C(k_0,\k)={1\over ik_0-e_\k} \eqno (\I.21)$$
If $\|U(\x)\|_{L^1}\le c\, (\beta L^d)^{-1}$ or 
 $\|U(\x)\|_{L^\infty}\le c\, (\beta L^d)^{-2}$, $c$ a constant, 
   then (I.20) converges. 
\smallskip\noindent
{\bf b)} Let 
$$d\mu_C(\psi,\bar\psi)=\pro_{k,\sigma}{\ts  {\beta L^d\over ik_0-e_\k}}\; 
   e^{-{1\over \beta L^d} \sum_{k,\sigma}(ik_0-e_\k) \bar\psi_{k,\sigma} 
   \psi_{k,\sigma} } \pro_{k,\sigma} d\psi_{k,\sigma} d\bar\psi_{k,\sigma} \eqno 
  (\I.22)$$
be the Grassmann Gaussian measure with covariance $C$. Then the perturbation series 
 (I.20) can be rewritten as 
$$Z=\int e^{-\V_{\rm int}(\psi,\bar\psi)} d\mu_C(\psi,\bar\psi) \eqno (\I.23)$$
where 
$$\V_{\rm int}(\psi,\bar\psi)={\ts {1\over (\beta L^d)^3}}
  \sum_{\sigma_1\sigma_2}\sum_{k_1,k_2\atop 
  p_1,p_2}\delta_{k_1+k_2,p_1+p_2}
    U(\k_2-\p_2)\bar\psi_{k_1\sigma_1}\bar\psi_{k_2\sigma_2}\psi_{p_1\sigma_1}
  \psi_{p_2\sigma_2}  \eqno (\I.24)$$
{\bf c)} The momentum distribution (I.19) at temperature $T={1\over\beta}$ is given by 
$$ \la a_{\k\sigma}^+ a_{\k\sigma}\ra_{\beta,L}=\lim_{\ep\to 0\atop \ep<0} 
   {\ts {1\over\beta}}\sum_{k_0\in{\pi\over\beta}(2\Bbb Z+1)} 
   \la \bar\psi_{k_0\k\sigma}
   \psi_{k_0\k\sigma}\ra_{\beta,L,\ep} \eqno (\I.25)$$
where 
$$\la \bar\psi_{k\sigma}
   \psi_{k\sigma}\ra_{\beta,L,\ep}={\ts {1\over Z_\ep}}\int \bar\psi_{k\sigma}
   \psi_{k\sigma}\> e^{-\V_{\rm int}(\psi,\bar\psi)} 
  d\mu_{C_\ep}(\psi,\bar\psi) \eqno (\I.26)$$
Here $C_\ep(k)={e^{-i\ep k_0}\over ik_0-e_\k}$ and $Z_\ep
  =\int e^{-\V_{\rm int}} d\mu_{C_\ep}$.  }
\bigskip
\noindent{\bf Remarks:} {\bf (i)}  A bound from which convergence of the
perturbation series follows for sufficiently small $U$ is easiest obtained 
in coordinate space. There the perturbation series reads 
$$\eqalignno{ Z&=\int e^{-\int d\xi_1 d\xi_2(\bar\psi\psi)(\xi_1)U(\xi_1-\xi_2)
  (\bar\psi\psi)(\xi_2)} d\mu_C(\psi,\bar\psi) \cr
 &=\sum_{n=0}^\infty {\ts {1\over n!}} \int d\xi_1\cdots d\xi_{2n}\prod_{i=1}^n 
   U(\xi_{2i-1}-\xi_{2i})\> \det\left[ C(\xi_i,\xi_j)\right]_{1\le i,j\le 2n} 
  &(\I.27) \cr}$$
where $\xi=(x_0,\x,\sigma),\;\int d\xi=\sum_{\sigma\in\{\up,\down\}}\int_{[0,\beta]}
  dx_0 \int_{[0,L]^d}d^d\x$, $U(\xi)=U(x)=\delta(x_0)U(\x)$, $C(\xi,\xi')=
  \delta_{\sigma,\sigma'} C(x-x')$ where 
$$C(x_0,\x)={\ts {1\over L^d}}\sum_{\k\in M_\omega} e^{i\k\x} e^{-e_\k x_0}[\theta(-x_0)
  \theta_\beta(-e_\k)-\theta(x_0)\theta_\beta(e_\k)] \eqno (\I.28)$$
and $\theta_\beta$ is an approximate step function given by 
 $\theta_\beta(\vep)={1\over 1+e^{-\beta\vep}}$. First observe that, since 
 $\sup_{x_0\in[0,\beta]}\bigl|e^{-e_\k x_0}[\theta(-x_0)
  \theta_\beta(-e_\k)-\theta(x_0)\theta_\beta(e_\k)]\bigr|\le 1$ for all $\k$,  one has 
$\|C\|_\infty\le {\ts {1\over L^d}}\sum_{\k\in M_\omega} 1<\infty$ if we choose a fixed
ultraviolet cuttoff in (I.6). The determinant can be bounded using 
 $|\det[\vec a_1\cdots \vec a_r]\,|\le \pro_{i=1}^r\|\vec a_i\|_2$ where the 
 $\vec a_i$ are $r$ component vectors. That is
$$\Bigl| \det\left[ C(\xi_i,\xi_j)\right]_{1\le i,j\le 2n}\Bigr|\le 
  \pro_{i=1}^{2n} \Bigl\{ \su_{j=1}^{2n} |C(\xi_i,\xi_j)|^2\Bigr\}^{1\over2}\le 
  \pro_{i=1}^{2n} (2n\|C\|_\infty^2)^{1\over2}=(2n\|C\|^2_\infty)^n \eqno (\I.29)$$
which gives 
$$|Z|\le \sum_{n=0}^\infty {\ts \left({e\over n}\right)^n} (4\beta L^d)^n  \|U\|_{L^1}^n
   (2n\|C\|^2_\infty)^n=\sum_{n=0}^\infty 
   \bigl(8e\|C\|^2_\infty\beta L^d\bigr)^n\, \|U\|_{L^1}^n 
  \eqno (\I.30)$$
\medskip
\noindent {\bf (ii)} The $\ep$-limit shows up in (I.25) because of
$$\eqalignno{ C(x_0,\x)&\buildrel x_0\ne 0\over = {\ts {1\over L^d}}\sum_\k 
  e^{i\k\x} e^{-e_\k x_0}\left[ \theta_\beta(-e_\k)\theta(-x_0)-
  \theta_\beta(e_\k)\theta(x_0)\right] \cr
   &\phantom{n } 
     ={\ts {1\over \beta L^d}} \sum_{\k,k_0} e^{i\k\x-ik_0x_0}
   \ts {1\over ik_0-e_\k} &(\I.31) \cr}$$ 
but 
$$\eqalignno{ C(0,\x)&={\ts {1\over L^d}} \sum_\k e^{i\k\x}
    \theta_\beta(-e_\k) 
 =\lim_{\ep\to 0\atop \ep<0} {\ts {1\over \beta  L^d}}\sum_{\k,k_0} 
   e^{i\k\x-ik_0\ep} \ts {1\over ik_0-e_\k}\>, &(\I.32) \cr}$$
see the proof of Lemma A1 in the appendix for a more detailed discussion. 
\medskip\goodbreak
\noindent {\bf (iii)} Whereas the momenta in the Hamiltonian are $d$ dimensional, the 
variables in the perturbation series are $d+1$ dimensional. The additional $k_0$ 
 variables in (\I.20) are the Fourier transform of the $x_0$ variables in (I.27) 
which in turn enter the perturbation series because of 
$$ \ts {d\over d\lambda}_{|_0} e^{-\beta(H_0+\lambda H_{\rm int})}=\int_0^\beta 
   e^{-x_0 H_0} H_{\rm int}\, e^{-(\beta-x_0)H_0} dx_0\>,  \eqno (\I.33)$$
that is, because $H_{\rm int}$ does not commute with $H_0$. 
\bigskip
As remarked under (iii), the momenta in 
$$H_{\rm int}={\ts {1\over L^{3d}} }
   \sum_{\sigma,\tau\in \{\up,\down\}} \sum_{\k,\p,\q} \ts 
   U(\k-\p)\, a_{\k\sigma}^+ a_{\p\sigma} a_{\q-\k\tau}^+ a_{\q-\p\tau} 
    \eqno (\I.3)$$
are $d$ dimensional, but the variables in 
$$\V_{\rm int}={\ts {1\over (\beta L^{d})^3} }
   \sum_{\sigma,\tau\in \{\up,\down\}} \sum_{k,p,q} \ts 
   U(\k-\p)\, \bar\psi_{k\sigma}\psi_{p\sigma}   \bar\psi_{q-k\tau} \psi_{q-p\tau} 
    \eqno (\I.34)$$
are $d+1$ dimensional. On the Hamiltonian level, the forward, exchange and BCS
approximation (I.11) does not lead to an explicitly solvable model, but on the 
functional integral level it does. This is the main result of this paper. 
\bigskip
\noindent{\bf Theorem I.2:}  {\it  Let 
$$\{\V_{\rm forw}+\V_{\rm ex}+\V_{\rm BCS}\}(\psi,\bar\psi)=
  {\ts {1\over(\beta L^{d})^3} }
   \sum_{\sigma\tau} \sum_{kpq} \ts 
   U(\k-\p)\,[\delta_{k,p}+\delta_{k,q-p}+\delta_{q,0}] 
  \,  \bar\psi_{k\sigma}\psi_{p\sigma}  \bar\psi_{q-k\tau}  \psi_{q-p\tau} 
    \eqno (\I.35)$$
Then the approximation 
$$\V_{\rm int}\approx \V_{\rm forw}+\V_{\rm ex}+\V_{\rm BCS} \eqno (\I.36)$$
in (I.23,26) makes the model explicitly solvable. The partition function, 
the two point functions and the generating functional for the connected amputated 
Greens functions are given by finite-dimensional integral representations. The
dimension of these representations is proportional to $J$ if the electron
electron interaction is given by 
$$U(\k-\p)=\cases{ \ds {1\over2} \sum_{\ell=-j}^j \lambda_{|\ell|}
     e^{i\ell\varphi_\k}e^{-i\ell\varphi_\p}
    +{\lambda_0\over2} & if $d=2$ \cr
 \ds \sum_{\ell=0}^j \sum_{m=-\ell}^\ell \lambda_\ell
   \bar Y_{\ell m}\left({\ts{\k'}}\right)
     Y_{\ell m}\left({\ts{\p'}}\right) & if $d=3$ \cr} \;\;=:\;
   \sum_{l=0}^J\lambda_l\> y_l(\k)\>\bar y_l(\p)\eqno(\I.37) $$
Here $\k=|\k|(\cos\vp_\k,\sin\vp_\k)$ for $d=2$ and $\k'=\k/|\k|$ for $d=3$. }
\bigskip 
The general solution is written down in Theorem II.1. In the following, we summarize
the results for $\V_{\rm int}\approx \V_{\rm BCS}$  and for an electron-electron
interaction given by (I.37). This case is treated in section III.1.  
\smallskip
In that case the two point function (I.26) becomes (we suppress the $\ep$):
$$\eqalignno{   \la \bar\psi_{p\up} 
    \psi_{p'\up} \ra_{\beta,L}&
 =  \beta  L^d\delta_{p,p'} { \int^{\phantom{I}} F_{\p}^{\up\up}(\phi)\>
   e^{-\beta L^d V(\phi)}\ds \pro_{\sigma\tau} 
  \pro_{l=0}^J du_{\sigma\tau}^l dv_{\sigma\tau}^l \over 
   \int e^{-\beta L^d V(\phi)} \ds \pro_{\sigma\tau} 
  \pro_{l=0}^J du_{\sigma\tau}^l dv_{\sigma\tau}^l }&(\I.38)  \cr}$$
Here $\phi^l_{\sigma\tau}=u^l_{\sigma\tau}+i v^l_{\sigma\tau}$, $\sigma\tau\in 
  \{\up\up,\down\down,\up\down\}$,  
$$F_{\p}^{\up\up}(\phi)=\ts { (ip_0+e_\p) [ p_0^2+e_\p^2+ \Phi_{-\p\up\down} 
     \bar\Phi_{-\p\up\down}+\Phi_{\p\down\down}\bar\Phi_{\p\down\down}] 
  \over (p_0^2+e_\p^2+\Omega_\p^+)(p_0^2+e_\p^2+\Omega_\p^-) }\eqno  (\I.39) $$
$$\Phi_{\k\up\down}=\sum_{l=0}^J\lambda_l^{1\over2}\phi^l_{\up\down} 
    \, y_l(\k')\>,\;\;\;\;
  \bar\Phi_{\k\up\down} 
   =\sum_{l=0}^J \lambda_l^{1\over2} \bar\phi^l_{\up\down} 
    \, \bar y_l(\k')\>,\eqno (\I.40)$$
$$\Phi_{\k\sigma\sigma}=\sum_{l=0}^J \lambda_l^{1\over2}
    \phi_{\sigma\sigma}^l\,{\ts [y_l(\k')-y_l(-\k')] } \>,\;\;\;
  \bar\Phi_{\k\sigma\sigma}=\sum_{l=0}^J \lambda_l^{1\over2} 
     \bar\phi_{\sigma\sigma}^l\,  {\ts [\bar y_l(\k')-\bar y_l(-\k')]}\eqno (\I.41) $$
and $\Omega^{\pm}_\k$ are the solutions of the quadratic equation 
$$\eqalignno{  \Omega^2-&\Bigl( \bar\Phi_{\k\up\down} 
  \Phi_{\k\up\down}+\bar\Phi_{-\k\up\down} \Phi_{-\k\up\down}
  +\Phi_{\k\up\up}\bar\Phi_{\k\up\up} 
  +\Phi_{\k\down\down}\bar\Phi_{\k\down\down}\Bigr)\Omega   
  +\Phi_{\k\up\up}\Phi_{\k\down\down}
    \bar\Phi_{\k\up\down}\bar\Phi_{-\k\up\down}  \cr  
 &+\bar\Phi_{\k\up\up}\bar\Phi_{\k\down\down}
    \Phi_{\k\up\down}\Phi_{-\k\up\down} 
   +\Phi_{\k\up\up}\Phi_{\k\down\down}
    \bar\Phi_{\k\up\up}\bar\Phi_{\k\down\down} 
  +\bar\Phi_{\k\up\down}  \Phi_{\k\up\down}
   \bar\Phi_{-\k\up\down}  \Phi_{-\k\up\down} =0 &(\I.42)  \cr}$$
Observe that $\bar\Phi_{\k\sigma\tau}$ is not necessarily the complex conjugate 
 of $\Phi_{\k\sigma\tau}$, depending on the signs of the coupling constants 
 $\lambda_l$.       
The effective potential $V_\beta$ is given by 
$$V_\beta(\phi )= 
   \sum_{l=0}^J\left(  |\phi_{\up\down}^l|^2+ |\phi_{\up\up}^l|^2+
    |\phi_{\down\down}^l|^2\right)   -\sum_{\epsilon\in\{+,-\}} 
    \int_{M_\omega}   \ts 
   {d^d\k\over (2\pi)^d}\>{1\over\beta} 
    \log\left[ {\cosh({\beta\over2}\sqrt{ e_\k^2+\Omega^\epsilon_\k }) 
    \over \cosh {\beta\over 2} e_\k} \right]\; \eqno (\I.43) $$
\smallskip
For a pure even interaction, that is, if $\lambda_\ell=0$ for all odd angular 
 momentum $\ell$, one has, if $y_l(-\k)=(-1)^ly_l(\k)$, 
   $\Phi_{\k\up\up}=\Phi_{\k\down\down}=0$ and 
 the expectation value simplifies to 
$$\eqalignno{   \la \bar\psi_{p,\sigma} 
    \psi_{p',\sigma} \ra_{\beta,L}
 &=  \beta  L^d\delta_{p,p'} \> 
  { \int {-ip_0-e_\p \over p_0^2+e_\p^2+\Phi_{\p\up\down}\bar\Phi_{\p\up\down}} 
    e^{ -\beta L^d V_\beta(\phi_{\up\down}) }
      \ds\pro_{l=0}^J du^l_{\up\down} dv^l_{\up\down}
   \over \int e^{ -\beta L^d V_\beta(\phi_{\up\down} ) } 
    \ds\pro_{l=0}^J du^l_{\up\down} dv^l_{\up\down} }\> &(\I.44)  \cr}$$
with an effective potential 
$$V_\beta(\phi_{\up\down})= \sum_{l=0}^J |\phi_{\up\down}^l|^2-
   \int_M \ts 
   {d^d\k\over (2\pi)^d}\>{1\over\beta} 
    \log\left[ {\cosh({\beta\over2}
     \sqrt{ e_\k^2+\Phi_{\k\up\down}\bar\Phi_{\k\up\down} }) 
    \over \cosh {\beta\over 2} e_\k} \right]^2 \; \eqno (\I.45) $$
Recall that the free two point function is given by $\la \bar\psi_{p,\sigma} 
    \psi_{p',\sigma} \ra_{\beta,L}=\beta L^d\delta_{p,p'} 1/(ip_0-e_\p)=
  -\beta L^d\delta_{p,p'} (ip_0+e_\p)/(p_0^2+e_\p^2)$. 
In particular, for a delta function interaction one obtains the two dimensional 
 integral representation 
$$\eqalignno{   \la \bar\psi_{p,\sigma} 
    \psi_{p',\sigma} \ra_{\beta,L}&=-\beta L^d \delta_{p,p'}\> 
  { \int { ip_0+e_\p
    \over p_0^2+e_\p^2+\lambda(u^2+v^2)} \> e^{-\beta L^d 
   V_{\beta}(u,v)} dudv \over \int e^{-\beta L^d V_{\beta}(u,v)} dudv }
    & (\I.46) \cr}$$
where
$$V_{\beta}(u,v)=u^2+v^2-
     \int_{M_\omega}   \ts {d^d\k\over (2\pi)^d}{1\over\beta}  \log\left[ 
   {\cosh({\beta\over2}\sqrt{ e_\k^2+\lambda(u^2+v^2)})\over 
    \cosh {\beta\over2} e_\k}\right]^2 \eqno (\I.47)$$
A positive $\lambda$ corresponds to an attractive interaction. 
\medskip\goodbreak
Since the only place where the volume $L^d$ shows up in the 
 integral  representations (I.44,46) is the prefactor in the exponential 
 $e^{-\beta L^d V(\phi)}$, the computation of the infinite volume 
 limit comes down to the determination of the global minimum of the real part 
 of the effective potential $V$ as a function of the $\phi^l_{\sigma\tau}$'s. 
\vskip 2cm
$$ figure \;1$$
\vskip 2cm
%\vfill\eject
For repulsive coupling $\lambda<0$ $V_\beta$ (I.47) has a global minimum at $u=v=0$ 
which results in 
$$\lim_{L\to\infty} {e^{-\beta L^d 
   V_{\beta}(u,v)} dudv \over \int e^{-\beta L^d V_{\beta}(u,v)} dudv
   }=\delta(u)\delta(v) \eqno (\I.48)$$
and 
$$\lim_{L\to\infty} \la \bar\psi_{p,\sigma} 
    \psi_{p',\sigma} \ra_{\beta,L}=-\beta\delta_{p_0,p_0'}\delta(\p-\p')\>
  \ts {ip_0+e_\p\over p_0^2+e_\p^2} \eqno (\I.49)$$
For attractive coupling $\lambda>0$ and sufficiently small $T={1\over\beta}$
 the effective potential  (I.47) has the form of a mexican hat. This results in 
$$\lim_{L\to\infty} \la \bar\psi_{p,\sigma} 
    \psi_{p',\sigma} \ra_{\beta,L}=-\beta\delta_{p_0,p_0'}\delta(\p-\p')\>
  \ts {ip_0+e_\p\over p_0^2+e_\p^2+\Delta^2} \eqno (\I.50)$$
where $\Delta^2=\lambda \rho_0^2\sim e^{-{1\over\lambda}}$ and $\rho_0^2=u_0^2+
 v_0^2$ is the value where $V_\beta$ takes its global minimum which lies on a circle
in the $u,v$ plane.  
 \bigskip
The expectation values $\la \psi\psi\ra$ and $\la\bar\psi\bar\psi\ra$ 
 can also be computed. 
 To make them nonzero,  
  we introduce a small external field $r=|r| e^{i\alpha}$. 
That is, we substitute $\V_{\rm BCS}$ by 
  (we do not consider here $\la a_\up a_\up \ra$ expectations)
$$\V_{{\rm BCS},r}=\V_{\rm BCS}+{\ts {1\over \beta L^d}} \sum_k [ 
   r \,\psi_{k\up} \psi_{-k\down}-\bar r\, \bar\psi_{k\up} \bar\psi_{-k\down}] 
   \eqno (\I.51)$$
One obtains again a finite-dimensional integral representation. 
For a delta function interaction, that is for $J=0$ in (I.37),  it 
 reduces to a two dimensional integral: 
$$\eqalignno{   \la \bar\psi_{p,\up} 
    \bar\psi_{-p',\down} \ra_{\beta,L}&=\beta L^d \delta_{p,p'}\> 
  { \int {-i\sqrt\lambda \> e^{i\alpha} 
   (u+iv) \over p_0^2+e_\p^2+\lambda(u^2+v^2)} \> e^{-\beta L^d 
   V_{\beta,r}(u,v)} dudv \over \int e^{-\beta L^d V_{\beta,r}(u,v)} dudv }
    & (\I.52) \cr}$$
where 
$$V_{\beta,r}(u,v)=u^2+{\ts \left( v+{|r|\over \sqrt \lambda}\right)^2 
   -{2\over\beta} } \int_M \ts {d^d\k\over (2\pi)^d} \log\left[ 
   {\cosh({\beta\over2}\sqrt{ e_\k^2+\lambda(u^2+v^2)})\over 
    \cosh {\beta\over2} e_\k}\right] \eqno (\I.53)$$
\par
Consider an attractive $\lambda>0$ and sufficiently small $T={1\over\beta}$ such that 
the global minimum of the effective potential moves away from zero.  
For $r=0$, the global minimum of $V_{\beta,r}$ is degenerated and lies on a 
 circle in the $u,v$ plane. In particular, $V_{\beta,0}$  is an even function 
 of $u$ and $v$ and 
 $\la \bar\psi_{p\up} \psi_{-p\down}\ra$ vanishes by symmetry. 
\par
\vskip 2cm
$$ figure\; 2$$
\vskip 2cm
For $r\ne 0$,   $V_{\beta,r}$  has a unique 
global minimum at $(u,v)=(0,v_0)$ where $v_0$ is given by the negative solution
of 
$$ v_0\left\{ \lambda \int_M \ts {d^d\k\over (2\pi)^d} 
   {\tanh({\beta\over2}\sqrt{e_\k^2+\lambda v_0^2})\over 2\sqrt{ e_\k^2+\lambda 
   v_0^2}}-1 \right\}=2|r|  \eqno (\I.54)$$
which, in the limit $r\to 0$,  becomes the BCS equation 
 for $|\tr|^2=\lambda v_0^2$. Thus 
$$\lim_{|r|\to 0}\lim_{L\to\infty} {\ts {e^{-\beta L^d V_{\beta,r}(u,v)} \over 
   \int e^{-\beta L^d V_{\beta,r}(u,v)} dudv }  }=
   \lim_{|r|\to 0}\lim_{L\to\infty} \ts {e^{-\beta L^d
      [V_{\beta,r}(u,v)-V_{\beta,r}(0,v_0)] } \over 
   \int e^{-\beta L^d [V_{\beta,r}(u,v)-V_{\beta,r}(0,v_0)]} dudv }
  =\delta(u)\delta\left(v+{|\tr|\over \sqrt\lambda}\right)  \eqno (\I.55) $$
and $\la \bar\psi_{p\up} \bar\psi_{-p\down}\ra$ becomes nonzero. 
\par
For repulsive $\lambda<0$, $V_{\beta,r}$ is complex and the real part of 
 $V_{\beta,r}$ has a unique global minimum at $(u,v)=(0,0)$ which results in 
 $\lim_{|r|\to 0}\lim_{L\to\infty}\ts {e^{-\beta L^d V_{\beta,r}(u,v)}\over \int 
   e^{-\beta L^d V_{\beta,r}(u,v)}dudv }=\delta(u)\delta(v)$ 
and $\lim_{|r|\to 0}\lim_{L\to\infty}\la \bar\psi_{p\up}\bar\psi_{-p\down}\ra=0$. 
\bigskip
These results of course are also obtained if one applies the usual mean field
formalism which is based on approximating the quartic Hamiltonian by a 
 quadratic expression. However, the situation is different if the electron-electron 
interaction contains higher angular momentum terms and the space dimension is 3. In 
that case the standard mean field formalism [AB,BW] predicts an angle dependent 
gap but, if $e_\k$ has $SO(3)$ symmetry, this is not the case. 
\par
Suppose that  $\lambda_\ell>0$ is attractive and 
$$U(\k-\p)=\lambda_\ell\sum_{m=-\ell}^\ell \bar Y_{\ell m}(\k') Y_{\ell m}(\p') 
   \eqno (\I.56)$$
Then the Anderson Balian Werthammer mean field formalism gives 
$$\lim_{L\to\infty} {\ts{1\over L^d}} 
  \la a_{\k\sigma}^+ a_{\k\sigma} \ra=\ts {1\over2}\left( 1-e_\k 
   \Bigl[ {\tanh({\beta\over2}\sqrt{ e_\k^2+\Delta_\k^*\Delta_\k})\over 
      \sqrt{ e_\k^2+\Delta_\k^*\Delta_\k}} \Bigr]_{\sigma\sigma}\right)
   \eqno (\I.57 )$$
 where the $2\times 2$ matrix $\Delta_\k$, $\Delta_\k^T=-\Delta_{-\k}$, is 
 a solution of the gap equation 
$$\Delta_\p=\int_{M_\omega} \ts{d^d\k\over (2\pi)^d}\, U(\p'-\k')\, \Delta_\k 
   {\tanh({\beta\over2}\sqrt{ e_\k^2+\Delta_\k^*\Delta_\k})\over 
     2 \sqrt{ e_\k^2+\Delta_\k^*\Delta_\k}}  \eqno (\I.58)$$
In 3 dimensions, it has been proven  [FKT2]  
 that for all $\ell \ge 2$ (IV.58) does not have unitary isotropic 
  ($\Delta_\k^*\Delta_\k=const\,Id$) solutions. That is, the gap in (I.57) is 
angle dependent. However in Theorem III.3 it is shown that, for even $\ell$, 
$$\eqalignno{ \lim_{L\to\infty} {\ts{1\over\beta L^d}}
    \la \bar\psi_{k\sigma} \psi_{k\sigma}  \ra_{\beta,L}&= 
    \int_{S^2}\ts {ik_0+e_\k\over k_0^2+e_\k^2+
   \lambda_\ell \rho_0^2 | \su_m  \alpha_m^0 Y_{\ell m}(\x)|^2} 
   \, {d\Omega (\x)\over 4\pi} &(\I.59)  \cr}$$
which gives, using (I.25), 
$$\lim_{L\to\infty}{\ts {1\over L^d}} \la a_{\k\sigma}^+ a_{\k\sigma} \ra=\int_{S^2}
   \ts {1\over2}\left( 1-e_\k 
      {\tanh({\beta\over2}\sqrt{ e_\k^2+|\Delta(\x)|^2})\over 
      \sqrt{ e_\k^2+|\Delta(\x)|^2}} \right)\, {d\Omega (\x)\over 4\pi}
   \eqno (\I.60 )$$
if $\Delta(\x)=\lambda_\ell^{1\over2} \rho_0 
   \su_m  \alpha_m^0 Y_{\ell m}(\x)$ and $\rho_0$ and $\alpha^0$ are values at the
global minimum. 
The point is that for $SO(3)$ symmetric $e_\k$ also the effective potential has 
 $SO(3)$ symmetry which means that also the global minimum has $SO(3)$ symmetry. 
Since in the infinite volume limit the integration variables are forced to be 
 at the global minimum, the integral over the sphere in (I.59,60) is the averaging 
 over all global minima. 
\bigskip
There is some physics literature [B,BZT,BR,G,Ha,W] which investigates 
  the relation 
between the reduced but still quartic and not solvable BCS Hamiltonian 
$$H_{\rm BCS}=H_0+{\ts {1\over L^{3d}} }
   \sum_{\sigma,\tau\in \{\up,\down\}} \sum_{\k,\p} \ts 
   U(\k-\p)\, a_{\k\sigma}^+ a_{-\k\tau}^+ a_{\p\sigma} a_{-\p\tau}
  \eqno (\I.61) $$
and the quadratic, explicitly diagonalizable mean field 
 Hamiltonian 
$$\eqalignno{ H_{\rm MF}& =H_0+ {\ts {1\over L^{3d}} }
   \sum_{\sigma,\tau\in \{\up,\down\}} \sum_{\k,\p} \ts 
   U(\k-\p)\,\Bigl(  a_{\k\sigma}^+   
   a_{-\k\tau}^+ \la a_{\p\sigma} a_{-\p\tau}\ra +
 \la a_{\k\sigma}^+   
   a_{-\k\tau}^+ \ra a_{\p\sigma} a_{-\p\tau} \cr
 &\phantom{mmH_0+ {\ts {1\over L^{3d}} }
   \sum_{\sigma,\tau\in \{\up,\down\}} \sum_{\k,\p} \ts 
   U(\k-\p)\,\Bigl( }  - 
  \la a_{\k\sigma}^+  a_{-\k\tau}^+ \ra  
   \la a_{\p\sigma} a_{-\p\tau}\ra \Bigr) & (\I.62a )  \cr} $$
where the numbers $\la a_{\p\sigma} a_{-\p\tau}\ra$ have to be
determined by the condition 
$$\la a_{\p\sigma} a_{-\p\tau}\ra={ Tr\> e^{-\beta H_{\rm MF}}
   a_{\p\sigma} a_{-\p\tau} \over Tr\> e^{-\beta H_{\rm MF}} } 
   \eqno (\I.62b )$$
which is equivalent to the gap equation (I.58) if one defines 
 $\Delta_{\p\sigma\tau}={1\over L^d} 
   \sum_{\k}U(\p-\k){1\over L^d}\la a_{\k\sigma} a_{-\k\tau} \ra$. 
One has, if $H':=H_{\rm BCS}-H_{\rm MF}$, 
$$H'={\ts {1\over L^{3d}} }
   \sum_{\sigma,\tau\in \{\up,\down\}} \sum_{\k,\p} \ts 
   U(\k-\p)\Bigl( a_{\k\sigma}^+ a_{-\k\tau}^+ - \la a_{\k\sigma}^+   
   a_{-\k\tau}^+ \ra \Bigr) \Bigl( 
    a_{\p\sigma} a_{-\p\tau}- \la a_{\p\sigma} a_{-\p\tau}\ra \Bigr)
  \eqno (\I.63) $$
It is claimed that, in the infinite volume limit, the correlation 
 functions of both models should coincide. More precisely, it is
claimed that 
$$\lim_{L\to \infty} \ts {1\over L^d}
    \log {Tr\, e^{-\beta H_{\rm BCS}} 
  \over Tr\, e^{-\beta H_{\rm MF}} } \eqno (\I.64 )$$
vanishes. To this end it is argued that each order   of perturbation 
 theory (with respect to 
 $H'$) of $Tr\, e^{-\beta (H_{\rm MF}+H')} 
    / Tr\, e^{-\beta H_{\rm MF}} $ is finite as the volume goes to
infinity. The Haag paper argues that spatial averages of field operators like 
 ${1\over L^d}\int_{[0,L]^d} d^d\x a_\up(\x) a_\down(\x)$ may be substituted by 
 numbers in the infinite volume limit, since commutators with them 
 have an extra one over volume factor, but there is 
 no rigorous controll of the error. 
 At least for a more complicated electron-electron 
 interaction (I.56), this reasoning cannot be correct 
 in view of (I.57-60). 
\par
Namely, consider the $\la a_{\k\sigma}^+ a_{\k\sigma}\ra_{\beta,L}$ 
expectation. In terms of Grassmann variables it is given by (I.25,26). 
Assume first a delta function interaction $U(\k-\p)=\lambda$. 
For the full model (I.3), by making a Hubbard Stratonovich
transformation, (I.26) can be rewritten as 
$$\la \bar\psi_{p\sigma} \psi_{p\sigma}\ra=
  {\int F(\phi)  \>\ds e^{-V(\phi)} \pro_{q_0\q} d\phi_{q_0\q} 
  d\bar\phi_{q_0\q}  \over
  \int e^{-V(\phi)}\ds \pro_{q_0\q} d\phi_{q_0\q} 
  d\bar\phi_{q_0\q} } \eqno(\I.65)$$
where the integrand is given by 
$$F(\phi)={ {\partial\over \partial s_{p\sigma}}_{|s=0} \det\left[ 
 \matrix{ A+S_\up & i( {\lambda\over \beta L^d})^{1\over2}\,\phi\cr
  i( {\lambda\over \beta L^d})^{1\over2}\,\bar\phi & 
   \bar A+S_\down \cr}\right] \over 
  \det\left[ \matrix{ 
  A & i( {\lambda\over \beta L^d})^{1\over2}\,\phi\cr
  i( {\lambda\over \beta L^d})^{1\over2}\,\bar\phi & 
   \bar A \cr} \right]  } \eqno(\I.66) $$ 
and the effective potential reads 
$$V(\phi)=\sum_{q_0\q}|\phi_{q_0\q}|^2-\log{ 
  \det\left[ \matrix{ 
  A & i( {\lambda\over \beta L^d})^{1\over2}\,\phi\cr
  i( {\lambda\over \beta L^d})^{1\over2}\,\bar\phi & 
   \bar A \cr} \right] \over 
  \det\left[ \matrix{ 
  A & 0 \cr
  0 & \bar A \cr} \right] }\eqno(\I.67)$$
Here $A$ and $S_\sigma$ 
 are diagonal matrices with entries $A_k=ik_0-e_\k$ and $s_{k\sigma}$ 
respectively 
 and $\phi$ is a short notation for the matrix $(\phi_{k-p})_{k,p}$ 
where $k=(k_0,\k)\in {\pi\over\beta}(2\Bbb Z+1)\times M_\omega$. 
  Of course, for example 
 $\det\left[ {
  A \; 0\atop  
  0 \; \bar A } \right]=\pro_k (k_0^2+e_\k^2)$ does not make sense, 
    but the quotients in (I.66,67) are well defined. 
\par
The correlation functions of the reduced, but still quartic 
BCS Hamiltonian (I.61) are obtained from (I.65-67) by first assuming that 
 the global minimum $\phi_q^{\rm min}$ of $V(\phi)$ is 
 proportional to $\delta_{\q,{\bf 0}}\phi_{q_0}$ and second by 
suppressing the quantum fluctuations around $\phi_{\q q_0}^{\rm min}=
 0$ for all $\q\ne 0$. That is, for the model (I.61) one finds 
$$\la \bar\psi_{p\sigma} \psi_{p\sigma}\ra=
  {\int F_{\p}(\phi)  \>\ds e^{-L^d V(\phi)} \pro_{q_0} d\phi_{q_0} 
  d\bar\phi_{q_0}  \over
  \int e^{-L^d V(\phi)}\ds \pro_{q_0} d\phi_{q_0} 
  d\bar\phi_{q_0} }  \eqno (\I.68)$$
where the integrand is given by 
$$F_{\p}(\phi)={ {\partial\over \partial s_{p_0\sigma}}_{|s=0} 
   \det\left[ 
 \matrix{ A_\p+S_{\p\up} & i( {\lambda\over \beta})^{1\over2}\,\phi\cr
  i( {\lambda\over \beta})^{1\over2}\,\bar\phi & 
   \bar A_\p +S_{\p\down} \cr}\right] \over 
  \det\left[ \matrix{ 
  A_\p & i( {\lambda\over \beta})^{1\over2}\,\phi\cr
  i( {\lambda\over \beta})^{1\over2}\,\bar\phi & 
   \bar A_\p \cr} \right]  }  \eqno (\I.69)$$ 
and the effective potential reads 
$$V(\phi)=\sum_{q_0}|\phi_{q_0}|^2-{\ts {1\over L^d}}\sum_{\k} \log{ 
  \det\left[ \matrix{ 
  A_\k & i( {\lambda\over \beta})^{1\over2}\,\phi\cr
  i( {\lambda\over \beta })^{1\over2}\,\bar\phi & 
   \bar A_\k \cr} \right] \over 
  \det\left[ \matrix{ 
  A_\k & 0 \cr
  0 & \bar A_\k \cr} \right] } \eqno (\I.70)$$
The volume factor $L^{-{d\over 2}}$ in the determinant in (I.66,67)  
  has been transformed
away by a substitution of variables such that it shows up in the exponent 
 in front of the effective potential in  
 (I.68).  
The matrices in (I.69,70) are labelled only by the $k_0,p_0$ variables, 
that is, $A_\k$ is the diagonal matrix with entries 
 $A_{\k,k_0}=ik_0-e_\k$ and $\phi$ is a short notation for the matrix 
 $(\phi_{k_0-p_0})_{k_0,p_0}$. 
\par
Contrary to the full model, 
 the volume dependence of the model (I.61) or (I.68-70) is such that in the infinite 
 volume limit the integration variables $\phi_{q_0}$ in (I.68) are 
forced to be at the global minimum of (I.70). 
\par
The model discussed in this paper (that is, only the BCS part) 
  is obtained from (I.68-70) by assuming that 
 the global minimum of (I.70) is proportional to $\delta_{q_0,0}$. In that case, 
the only integration variable which is left in (I.68) is the $q_0=0$ mode $\phi=
 \phi_0$ and the expressions (I.68-70) reduce to the integral 
 representation (I.46,47). 
\smallskip
Now assume that the elctron electron interaction is given by (I.56) and suppose 
for simplicity that $\ell$ is even which suppresses $\up\up\up\up$ and 
 $\down\down\down\down$ contributions in (I.3) and (I.61). In that case the 
 model (I.61) gives 
$$\la \bar\psi_{p\sigma} \psi_{p\sigma}\ra=
  {\int F_{\p}(\phi)  \>\ds e^{-L^d V(\phi)}\pro_{m=-\ell}^\ell \pro_{q_0} 
  d\phi_{q_0}^m 
  d\bar\phi_{q_0}^m  \over
  \int e^{-L^d V(\phi)}\ds \pro_{m=-\ell}^\ell \pro_{q_0} d\phi_{q_0}^m 
  d\bar\phi_{q_0}^m } \eqno (\I.71)$$
where the integrand is given by 
$$F_{\p}(\phi)={ {\partial\over \partial s_{p_0\sigma}}_{|s=0} 
   \det\left[ 
 \matrix{ A_\p+S_{\p\up} & i( {\lambda\over \beta})^{1\over2}\,\Phi_\p\cr
  i( {\lambda\over \beta})^{1\over2}\,\bar\Phi_\p & 
   \bar A_\p +S_{\p\down} \cr}\right] \over 
  \det\left[ \matrix{ 
  A_\p & i( {\lambda\over \beta})^{1\over2}\,\Phi_\p \cr
  i( {\lambda\over \beta})^{1\over2}\,\bar\Phi_\p & 
   \bar A_\p \cr} \right]  } \eqno (\I.72) $$ 
and the effective potential reads 
$$V(\phi)=\sum_{m=-\ell}^\ell \sum_{q_0}|\phi_{q_0}^m|^2
     -{\ts {1\over L^d}}\sum_{\k} \log{ 
  \det\left[ \matrix{ 
  A_\k & i( {\lambda\over \beta})^{1\over2}\,\Phi_\k\cr
  i( {\lambda\over \beta })^{1\over2}\,\bar\Phi_\k & 
   \bar A_\k \cr} \right] \over 
  \det\left[ \matrix{ 
  A_\k & 0 \cr
  0 & \bar A_\k \cr} \right] } \eqno (\I.73)$$
Here $\Phi_\k$ denotes the matrix  with 
 entries $\Phi_{\k,p_0p_0'}=\sum_{m=-\ell}^\ell \phi_{p_0-p_0'}^m Y_{\ell m}(\k)$, 
 labelled by the $p_0,p_0'$ variables.  
The model discussed in this paper is obtained from (I.71-73) by 
  assuming that the 
global minimum of (I.73) has only nonzero $\phi_{q_0=0}^m$ modes. In that 
case (I.71-73) reduce to the integral representation (I.44,45) which further 
  can be reduced, 
 by Theorem III.3,  to (I.59).  However, the argument used in Theorem III.3 that the 
 $\la \bar\psi_{p\sigma} \psi_{p\sigma}\ra$ expectation has to be $SO(3)$ 
 invariant if $e_\k$ is $SO(3)$ invariant still applies to (I.71), since (I.73) is 
  invariant 
under simultanious transformation of the $\phi_{q_0}^m$'s to 
 $\sum_{m'} U(R)_{mm'}\phi_{q_0}^{m'}$ where $U(R)$ is the unitary
representation of $SO(3)$ given by $Y_{\ell m}(R\k)=\sum_{m'} U(R)_{mm'} 
  Y_{\ell m'}(\k)$. 
\par
That is, if the BCS equation (I.58) or (I.62b) of the quadratic mean field model
 (I.62)          
 has a solution such that $|\Delta_\k|^2$ is not $SO(3)$ invariant (and, by 
 [FKT2], this is necessarily the case for any nonzero solution for 
 $d=3$ and $\ell\ge 2$), then the $\la a_{\k\sigma}^+ a_{\k\sigma}\ra$ 
 expectation of the quadratic mean field model (I.62) does not coincide with 
 the corresponding expectation of the quartic reduced BCS Hamiltonian (I.61).  
\par
In [AB], Anderson and Brinkmann used the quadratic mean field Hamiltonian 
with an $\ell =1$ interaction to describe the properties of superfluid Helium 3. 
The basic quantities in their analysis are the 
  $\la a_{\k\sigma} a_{-\k\tau}\ra$ expectations or the matrix
$\Delta_{\k\sigma\tau}$  
which is obtained as a solution of the gap equation of the quadratic model. 
 In view of the discussion above, one may regard as the more natural approach 
 to take the quartic BCS Hamiltonian (I.61), to add in a symmetry breaking term 
 which breaks the U(1) particle symmetry as  well as the spatial SO(3) 
 symmetry and then to compute (in the approximation $\phi_{q_0}^m= 
 \delta_{q_0,0}\phi^m$) the infinite volume limit followed by the limit 
 symmetry breaking term $\to 0$. In particular, besides the usual U(1) symmetry 
 breaking one may expect SO(3) symmetry breaking in the sense that 
probably also for the $\la a^+ a\ra$ expectations the above two limits 
do not commute. That is, if $B$ denotes the SO(3) symmetry breaking term, 
 whereas $\lim_{L\to\infty}\lim_{B\to 0}\la a_{\k\sigma}^+ a_{\k\sigma}\ra$ has 
SO(3) symmetry, $\lim_{B\to 0}\lim_{L\to\infty}\la a_{\k\sigma}^+ a_{\k\sigma}\ra$ 
may have not. However, to what extend the quantities of the [AB] paper are 
related to these expectations is not clear. For example, for $\ell \ge 2$ 
 ([AB] has $\ell=1$, but still uses the quadratic mean field formalism) 
any nontrivial solution of the quadratic model (I.62) is 
 necessarily anisotropic, without any external SO(3) symmetry 
  breaking field at all.  
The quantities $\lim_{B\to 0}\lim_{L\to\infty}\la a_{\k\sigma}^+ a_{\k\sigma}\ra$ 
for the quartic model (I.61) or the model discussed in this paper of course would 
depend on the direction of $B$, if they become anisotropic. 
A more careful analysis of this question we defer to another paper. 
\bigskip
\bigskip
\noindent{\bf Acknowledgements} 
\bigskip
I am grateful to Horst Kn\"orrer and Eugene Trubowitz and to the 
 Forschungsinstitut f\"ur Mathematik at ETH Z\"urich for the hospitality and 
 the support during the summer 1996. Furthermore I would like to thank 
 Joel Feldman who made it possible for me to visit the University of British 
 Columbia in Vancouver in the academic year 1996/97. 
\bigskip
\bigskip
\bigskip
\goodbreak
\magnification=\magstep1
\font\gross=cmbx12 scaled \magstep0
\font\mittel=cmbx10 scaled \magstep0
\font\Gross=cmr12 scaled \magstep2
\font\Mittel=cmr12 scaled \magstep0
\overfullrule=0pt
\def\Bbb#1{{\bf #1}}
\def\blacksquare{\bullet}
%\baselineskip=12pt
%\nopagenumbers
\def\up{\uparrow}
\def\down{\downarrow}
\def\k{{\bf k}}

\def\p{{\bf p}}
\def\q{{\bf q}}
\def\x{{\bf x}}
\def\y{{\bf y}}
\def\I{{\rm I}}
\def\II{{\rm II}}
\def\III{{\rm II}}
\def\ts{\textstyle}
\def\ds{\displaystyle}
\def\tr{\Delta}
\def\la{\langle}
\def\ra{\rangle}
\def\pro{\mathop\Pi}
\def\1cm{\hskip 1cm}
\def\1k{{\textstyle{1\over\kappa}}}
\def\db#1{{\ts{d^d\k\over (2\pi)^d}}}
\def\vp{\varphi}
\def\sl{g}

\def\U{{\cal U}}
\def\V{{\cal V}}
\def\su{\mathop{\Sigma}}
\def\u{\underline}
\def\Eta#1{\u\zeta_{\phantom{.}\!#1}}
\def\Pf{{\rm Pf}}

%\pageno=13

\noindent {\gross II. Solution of the Model in the 
  Forward, Exchange }
\par\noindent{\gross$\phantom{II.}$ and BCS Approximation} 
\bigskip
\bigskip
Let $H=H_0+H_{\rm int}$ where
$$H_0={\ts {1\over L^d}}\sum_{\sigma\in \{\uparrow,\downarrow\}} 
    \sum_{\k\in M} e_\k \,a_{\k\sigma}^+ 
    a_{\k\sigma} \eqno(\I.2)$$
and
$$H_{\rm int}=  {\ts {1\over L^{3d}}} 
        \sum_{\sigma,\tau\in\{\uparrow,\downarrow\}} 
   \sum_{\k,\p,\q} U(\k-\p) \> a_{\k,\sigma}^+ a_{\p,\sigma} a_{{\q}-\k,\tau}^+ 
    a_{{\q}-\p,\tau} \eqno(\I.3) $$
Then 
$$Z={Tr\,e^{-\beta (H_0+H_{int})}\over Tr\,e^{-\beta H_0}}=
    \int e^{-\V_{\rm int}(\psi,\bar\psi)}  d\mu_C(\psi,\bar\psi)\eqno(\I.23)$$
where the exponent in the fermionic integral is given by ($\kappa=\beta L^d$)
$$\V_{\rm int}(\psi)={\ts {1\over \kappa^3}}\sum_{\sigma,\tau\in\{\up,\down\}}
    \sum_{k,p,q} U(\k-\p)\>\bar\psi_{k,\sigma}  \psi_{p,\sigma} 
   \bar\psi_{{q}-k,\tau}
  \psi_{{q}-p,\tau}\eqno(\I.24)$$
and $k=(k_0,\k)\in{\pi\over\beta}(2\Bbb Z+1)\times M$. 
\par
The forward, exchange and BCS approximation is obtained by restricting the 
above sum to the following terms
$${\rm forward:}\;\;\;  \delta_{k,p}\>,\hskip 1cm  {\rm exchange:}
  \;\; \;\delta_{k,q-p} \>,\hskip 1cm {\rm BCS:}\;\;\;
 \delta_{q,0} \>. \eqno(\I.35)$$
We consider the approximation 
 $\V_{\rm int}\approx\V_{\rm forw}+\V_{\rm ex}+\V_{\rm BCS}\equiv
  \U(\psi,\bar\psi)$. That is,  
$$ \U(\psi,\bar\psi)={\ts {1\over \kappa^3}}
   \sum_{\sigma,\tau\in\{\up,\down\}}
    \sum_{k,p,q} U(\k-\p)\bigl[\delta_{k,p}+\delta_{k,q-p}
  +\delta_{q,0} \bigr] \>\bar\psi_{k,\sigma} \psi_{p,\sigma} 
   \bar\psi_{{q}-k,\tau}
  \psi_{{q}-p,\tau}   \eqno(\II.1)$$
and 
$$U(\k-\p)=-\cases{ \ds {1\over2} \sum_{\ell=-j}^j \lambda_{|\ell|}
     e^{i\ell\varphi_\k}e^{-i\ell\varphi_\p}
    +{\lambda_0\over2} & if $d=2$ \cr
 \ds \sum_{\ell=0}^j \sum_{m=-\ell}^\ell \lambda_\ell
   \bar Y_{\ell m}\left({\ts{\k'}}\right)
     Y_{\ell m}\left({\ts{\p'}}\right) & if $d=3$ \cr} \;\;=:\;
   -\sum_{l=0}^J\lambda_l\> y_l(\k')\>\bar y_l(\p')\eqno(\I.37) $$
and we abbreviated 
$$\kappa=\beta L^d \eqno (\II.2)$$
This model can be solved explicitly. Before we write down the general solution 
we shortly indicate the computation for $\V_{\rm int}\approx \V_{\rm BCS}$ and 
$U(\k-\p)=-\lambda$. In that case it comes down to the usual effective 
 potential computation with `constant $\phi$'. 
 That is, using the identity ($\phi=u+iv$, $\bar\phi=u-iv\in \Bbb C$)
$$e^{2ab}=\ts {1\over 2\pi}\int_{\Bbb R^2} e^{a\phi+b\bar\phi}  e^{-{1\over2} 
    |\phi|^2} dudv  \eqno (\II.3)$$
one obtains 
$$\eqalignno{ Z&=
   \int e^{-\left({\lambda\over \kappa}\right)^{1\over2} {1\over\kappa }
   \sum_{k} \bar\psi_{k\up}\bar\psi_{-k\down}\times 
     \left({\lambda\over \kappa}\right)^{1\over2}{1\over \kappa}
   \sum_p \psi_{p\up}\psi_{-p\down} } 
   d\mu_C(\psi,\bar\psi)   \cr
 &=\int_{\Bbb R^2}
    \int e^{i\left({\lambda\over\kappa}\right)^{1\over2} 
    \phi\, {1\over\kappa}\sum_k\psi_{k\up}\psi_{-k\down}\,+\,
   i\left({\lambda\over\kappa}\right)^{1\over2} 
   \bar\phi \,{1\over\kappa}\sum_k\bar\psi_{k\up}\bar\psi_{-k\down}  } 
   d\mu_C(\psi,\bar\psi) \ts {1\over 2\pi} e^{-{1\over2} 
    |\phi|^2} dudv  \cr
 &=\int_{\Bbb R^2} \pro_k{\ts{1\over k_0^2+e_\k^2}} \ts  \det\left(
 { ik_0-e_\k\;\;\;\; i
    \lambda^{1\over2}\bar\phi \atop 
   i\lambda^{1\over2} \phi \;\;\;\; -ik_0-e_\k} \right)
   \ts {\kappa\over \pi} e^{-\kappa |\phi|^2 } dudv  \cr
  &={\ts {\kappa\over \pi}}\int_{\Bbb R^2} e^{-\kappa V_\beta(u,v)} 
      dudv    &(\II.4) \cr}$$
\par
In the following theorem the $\xi$ variables take care of the exchange contributions, 
 the $w$ variable makes the forward contribution quadratic in the fermion fields 
 and the 
 $\phi$ variables, as in (II.3,4) above, sum up BCS contributions. 
\bigskip
%
%  THEOREM II 1
%
\noindent{\bf Theorem II.1:} {\it  Let $\U(\psi,\bar\psi)$ be given by (II.1) and let 
$$\eqalignno{ Z(\beta,L,\{s_{k,\sigma}\})&=\int e^{-\U(\psi,\bar\psi)+{1\over\kappa}
    \sum_{k,\sigma} s_{k,\sigma}\bar\psi_{k,\sigma}\psi_{k,\sigma} }
    d\mu_C(\psi,\bar\psi) & (\II.5)   \cr
   \ts \la \bar\psi_{p\sigma}\psi_{p\sigma}\ra&=\ts 
 \kappa\> {\partial\over \partial s_{p\sigma}}_{|s=0} \log Z(\beta,L,\{s_{k\sigma}\})
   &(\II.6) \cr}$$
Let $w\in\Bbb R$, $\xi^l_{\sigma\tau}=a^l_{\sigma\tau}+ib^l_{\sigma\tau}$, 
 $\phi^l_{\sigma\tau}=u^l_{\sigma\tau}+iv^l_{\sigma\tau}\in \Bbb C$ for 
 $0\le l\le J$, $(\sigma\tau)\in\{\up\up,\down\down,\up\down\}$ and define the 
 fields 
$$\Xi_{\k\up\down}=\sum_{l=0}^J (2\lambda_l)^{1\over2} 
    \xi_{\up\down}^l\,y_l(\k')\>,\;\;\;
  \bar\Xi_{\k\up\down}=\sum_{l=0}^J (2\lambda_l)^{1\over2} 
     \bar\xi_{\up\down}^l\, \bar y_l(\k')  \eqno $$
$$ \Xi_{\k\sigma\sigma}=\sum_{l=0}^J \lambda_l^{1\over2} 
    \xi_{\sigma\sigma}^l\,y_l(\k')\>, 
  \;\;\;
  \bar\Xi_{\k\sigma\sigma}=\sum_{l=0}^J \lambda_l^{1\over2}
          \bar\xi_{\sigma\sigma}^l\,\bar y_l(\k')\>,  \eqno (\II.7)$$
$$\Gamma_{\k\up}=U_0^{1\over2}w
      +i\Xi_{\k\up\up}+i\bar\Xi_{\k\up\up}\>,\;\;\;
  \Gamma_{\k\down}=U_0^{1\over2}w
      +i\Xi_{\k\down\down}+i\bar\Xi_{\k\down\down}\>,  \eqno (\II.8)$$
$$\Phi_{\k\up\down}=\sum_{l=0}^J (2\lambda_l)^{1\over2} 
    \phi_{\up\down}^l\,y_l(\k')\>,\;\;\;
  \bar\Phi_{\k\up\down}=\sum_{l=0}^J (2\lambda_l)^{1\over2} 
     \bar\phi_{\up\down}^l\, \bar y_l(\k')  \eqno (\II.9)$$
$$\Phi_{\k\sigma\sigma}=\sum_{l=0}^J \lambda_l^{1\over2} \phi_{\sigma\sigma}^l\,
  {\ts [y_l(\k')-y_l(-\k')] } \>,\;\;\;
  \bar\Phi_{\k\sigma\sigma}=\sum_{l=0}^J \lambda_l^{1\over2} 
     \bar\phi_{\sigma\sigma}^l\,  {\ts [\bar y_l(\k')-\bar y_l(-\k')]}\>. \eqno
(\II.10) $$
where $U_0=U(\k={\bf 0})=\int d^d\x\,U(\x)$. 
For each $k=(k_0,\k)$, let $S_k$ be the $8\times 8$ skew symmetric matrix 
$$S_k=-i \pmatrix{0&-{1\over i}A_{k\up}&\bar\Phi_{\k\up\down}&
    0&\bar\Xi_{\k\up\down}&
   0&0&\bar\Phi_{\k\up\up} \cr 
  {1\over i}A_{k\up} &0&0&\Phi_{\k\up\down}&0&-\Xi_{\k\up\down}&
   \Phi_{\k\up\up}&0\cr
  -\bar\Phi_{\k\up\down}&0&0&-{1\over i}A_{-k\down}&0&
   -\bar\Phi_{\k\down\down}
    &\Xi_{-\k\up\down}&0 \cr
 0&-\Phi_{\k\up\down}&{1\over i}A_{-k\down}&0&-\Phi_{\k\down\down}&0
    &0&-\bar\Xi_{-\k\up\down}\cr
  -\bar\Xi_{\k\up\down}&0&0&\Phi_{\k\down\down}&0
    &{1\over i}A_{k\down}&-\Phi_{-\k\up\down}&0\cr
   0&\Xi_{\k\up\down}&\bar\Phi_{\k\down\down}&0&-{1\over i}A_{k\down}&0
     &0&-\bar\Phi_{-\k\up\down}\cr
 0&-\Phi_{\k\up\up}&-\Xi_{-\k\up\down}&0
     &\Phi_{-\k\up\down}&0&0&{1\over i}A_{-k\up}\cr
 -\bar\Phi_{\k\up\up}&0&0&\bar\Xi_{-\k\up\down}&
   0&\bar\Phi_{-\k\up\down}&-{1\over i}A_{-k\up}
  &0\cr}  $$
where  $A_{k\sigma}=a_k-s_{k\sigma}-\Gamma_{k\sigma}=
     ik_0-e_\k-s_{k\sigma}-\Gamma_{k\sigma}$ 
and let ${\rm Pf}S_k$ be the Pfaffian of $S_k$ given by Lemma II.2 below. Then:
\item{\bf a)}
$$ Z(\beta,L,\{s_{k,\sigma}\})= 
  \int \prod_{k_0>0\atop\k\in M}
    \biggl\{ {\ts \left({1\over a_k a_{-k}}\right)^2 }
    {\rm Pf} S_k \biggr\} d\nu_\kappa(w,\xi,\phi)  \eqno (\II.11)   $$
where 
$$\eqalignno{ d\nu_\kappa(w,\xi,\phi)&=  
   {\ts  \left( {\kappa\over{4\pi}}\right)^{1\over2}  }
    \,e^{-{\kappa\over4}w^2}dw\, \pro_{l=0}^J \pro_{\sigma\tau }  \Bigl\{
  \ts \left({\kappa\over\pi}\right)^2 \, e^{-\kappa(|\phi_{\sigma\tau}^l|^2
    +|\xi_{\sigma\tau}^l|^2)}
    du_{\sigma\tau}^ldv_{\sigma\tau}^l da_{\sigma\tau}^l
    db_{\sigma\tau}^l   \Bigr\} \cr
  & &(\II.12)\cr}$$
\item{\bf b)}
$$\la \bar\psi_{p\sigma}\psi_{p'\sigma}\ra=\beta L^d\delta_{p,p'} \, 
   {  \int { {\partial\over \partial s_{p\sigma}}{\rm Pf}S_p \over {\rm Pf}S_p} 
    e^{-\kappa V(w,\xi,\phi)}dw \pro d\xi d\phi \over 
    \int^{\phantom{I}} e^{-\kappa V(w,\xi,\phi)}dw \pro d\xi d\phi } \eqno (\II.13) $$
where the effective potential $V$ is given by 
$$V(w,\xi,\phi)={\ts {1\over4}} w^2+\sum_{\sigma\tau} \sum_{l=0}^J 
  ( |\phi_{\sigma\tau}^l|^2+|\xi_{\sigma\tau}^l|^2)-{\ts {1\over\beta L^d} }
     \sum_{\k\in M}\log\pro_{k_0>0}\ts {{\rm Pf}S_k\over a_k^2a_{-k}^2}
   \eqno(\II.14)  $$
and $\pro d\xi d\phi =\ds\pro_{\sigma\tau}\pro_{l=0}^J  d\xi_{\sigma\tau}^l
      d\phi_{\sigma\tau}^l$. Here $\sigma\tau\in \{\up\up,\down\down,\up\down\}$. }
\bigskip
\bigskip
\noindent{\bf Remark:} The product ${\ds\pro_{k_0>0}} { {\rm Pf}S_k\over a_k^2 
  a_{-k}^2}$ in the effective potential (II.14) 
   where ${\rm Pf}S_k={\rm Pf}S_k(a_k,a_{-k})$ has to be computed 
 according to the rule 
$$ \prod_{k_0>0}{ { {\rm Pf}S_k\over a_k^2  a_{-k}^2}}=
   \lim_{\epsilon\to 0\atop \epsilon<0} \prod_{k_0>0}
   { { {\rm Pf}S_k(e^{ik_0\epsilon}a_k,e^{-ik_0\epsilon}a_{-k})
     \over a_k^2  a_{-k}^2}}  \eqno (\II.15)$$
That there are cases where it is necessary to make the phase factors explicit 
can be seen from the discussion given in the proof of Lemma A1 in the appendix. 
%
%  PROOF 
%
\bigskip
\noindent{\bf Proof:} Since we assume $U(\p-\k)=U(\k-\p)$, one has 
$$\eqalignno{ \U(\psi,\bar\psi)
   &={\ts {1\over\kappa^3}}\sum_{\sigma,\tau}\sum_{k,p}
   U({\bf 0})\>\bar\psi_{k,\sigma}
   \bar\psi_{p,\tau}\psi_{k,\sigma} 
  \psi_{p,\tau} +{\ts {1\over \kappa^3}}\sum_{\sigma,\tau}
     \sum_{k,p}U(\k-\p)\>\bar\psi_{p,\sigma}
   \bar\psi_{k,\tau}\psi_{k,\sigma} 
  \psi_{p,\tau}  \cr
 &\phantom{=}+{\ts {1\over \kappa^3}}\sum_{\sigma,\tau}
    \sum_{k,p}U(\k-\p)\>\bar\psi_{p,\sigma}
   \bar\psi_{-p,\tau}\psi_{k,\sigma} 
  \psi_{-k,\tau}  \cr
&=-{\ts {1\over\kappa^2}{U_0\over\kappa}}
  \sum_{k,p} \Bigl(\bar\psi_{k\up}\psi_{k\up}+\bar\psi_{k\down}\psi_{k\down}\Bigr)
  \Bigl(\bar\psi_{p\up}\psi_{p\up}+\bar\psi_{p\down}\psi_{p\down}\Bigr) \cr
 &\phantom{=}+{\ts {1\over \kappa^3}}\sum_{k,p}U(\k-\p)\>
  \Bigl\{ \bar\psi_{p\up}\psi_{p\up} 
   \bar\psi_{k\up}\psi_{k\up}+  
   \bar\psi_{p\down}\psi_{p\down} \bar\psi_{k\down}\psi_{k\down}  +
  2 \bar\psi_{p\up}\psi_{p\down} \bar\psi_{k\down}\psi_{k\up}   \Bigr\}  \cr
 &\phantom{=}+{\ts {1\over \kappa^3}} 
   \sum_{k,p}U(\k-\p)\Bigl\{ 
 \bar\psi_{p\up}\bar\psi_{-p\up}\psi_{k\up}  \psi_{-k\up} +
 \bar\psi_{p\down}\bar\psi_{-p\down}\psi_{k\down}  \psi_{-k\down}   +
  2\bar\psi_{p\up}\bar\psi_{-p\down}\psi_{k\up}\psi_{-k\down}   \Bigr\} \cr }$$
We substitute $ U(\k-\p)=\sum_{l=0}^J\lambda_l y_l(\k') \bar y_l(\p')$ and use 
the identities 
$$ e^{{1\over2}X^2}=\ts {1\over\sqrt{2\pi}} \int_{\Bbb R} e^{Xw}
      e^{-{1\over2} w^2} dw $$
$$ e^{-2XY}=\ts {1\over 2\pi} \int_{\Bbb R^2} e^{iX\phi+iY\bar\phi} e^{-{1\over2} 
     |\phi|^2} dudv  $$
to obtain 
$$\eqalignno{ e^{-\U(\psi,\bar\psi)}&=  \cr
 \int &\exp\biggl\{ {\ts
  \left( {2U_0\over \kappa}\right)^{1\over2} w\,{1\over\kappa} } 
  \sum_k\Bigl[ \bar\psi_{k\up}\psi_{k\up}+
   \bar\psi_{k\down}\psi_{k\down}\Bigr]\biggr\}
   \times  \cr
  & \exp\biggl\{ \sum_l\biggl[  
   {\ts  i\left( {\lambda_l\over2\kappa}\right)^{1\over2} \bar\xi^l_{\up\up}\, 
  {1\over\kappa}}\sum_k \bar y_l(\k')\,\bar\psi_{k\up} \psi_{k\up} +
   {\ts  i\left( {\lambda_l\over2\kappa}\right)^{1\over2} \xi^l_{\up\up}\, 
  {1\over\kappa}}\sum_k  y_l(\k')\,\bar\psi_{k\up} \psi_{k\up} \biggr]\biggr\} 
   \times \cr
 & \exp\biggl\{ \sum_l\biggl[  
   {\ts  i\left( {\lambda_l\over2\kappa}\right)^{1\over2} \bar\xi^l_{\down\down}\, 
  {1\over\kappa}}\sum_k \bar y_l(\k')\,\bar\psi_{k\down} \psi_{k\down} +
   {\ts  i\left( {\lambda_l\over2\kappa}\right)^{1\over2} \xi^l_{\down\down}\, 
  {1\over\kappa}}\sum_k  y_l(\k')\,\bar\psi_{k\down} \psi_{k\down} \biggr]\biggr\} 
   \times \cr
 & \exp\biggl\{ \sum_l\biggl[  
   {\ts  i\left( {\lambda_l\over\kappa}\right)^{1\over2} \bar\xi^l_{\up \down}\, 
  {1\over\kappa}}\sum_k \bar y_l(\k')\,\bar\psi_{k\up} \psi_{k\down} +
   {\ts  i\left( {\lambda_l\over \kappa}\right)^{1\over2} \xi^l_{\up\down}\, 
  {1\over\kappa}}\sum_k  y_l(\k')\,\bar\psi_{k\down} \psi_{k\up} \biggr]\biggr\} 
   \times \cr  
 & \exp\biggl\{ \sum_l\biggl[  
   {\ts  i\left( {\lambda_l\over2\kappa}\right)^{1\over2} \bar\phi^l_{\up\up}\, 
  {1\over\kappa}}\sum_k \bar y_l(\k')\,\bar\psi_{k\up} \bar\psi_{-k\up} +
   {\ts  i\left( {\lambda_l\over2\kappa}\right)^{1\over2} \phi^l_{\up\up}\, 
  {1\over\kappa}}\sum_k  y_l(\k')\,\psi_{k\up} \psi_{-k\up} \biggr]\biggr\} 
   \times \cr
 & \exp\biggl\{ \sum_l\biggl[  
   {\ts  i\left( {\lambda_l\over2\kappa}\right)^{1\over2} \bar\phi^l_{\down\down}\, 
  {1\over\kappa}}\sum_k \bar y_l(\k')\,\bar\psi_{k\down} \bar\psi_{-k\down} +
   {\ts  i\left( {\lambda_l\over2\kappa}\right)^{1\over2} \phi^l_{\down\down}\, 
  {1\over\kappa}}\sum_k  y_l(\k')\,\psi_{k\down} \psi_{-k\down} \biggr]\biggr\} 
   \times \cr
 & \exp\biggl\{ \sum_l\biggl[  
   {\ts  i\left( {\lambda_l\over\kappa}\right)^{1\over2} \bar\phi^l_{\up\down}\, 
  {1\over\kappa}}\sum_k \bar y_l(\k')\,\bar\psi_{k\up} \bar\psi_{-k\down} +
   {\ts  i\left( {\lambda_l\over\kappa}\right)^{1\over2} \phi^l_{\up\down}\, 
  {1\over\kappa}}\sum_k  y_l(\k')\,\psi_{k\up} \psi_{-k\down} \biggr]\biggr\} 
   d\nu(w,\xi,\phi) \cr}$$
where 
$$d\nu(w,\xi,\phi)=   {\ts {1\over \sqrt{2\pi}} }
    \,e^{-{1\over2}w^2}dw\, \pro_l \pro_{(\sigma\tau)\in\atop\{\up\up,\down\down,
   \up\down\} }
  \ts {1\over (2\pi)^2} \, e^{-{1\over2}(|\phi_{\sigma\tau}^l|^2
    +|\xi_{\sigma\tau}^l|^2)}
    du_{\sigma\tau}^ldv_{\sigma\tau}^l da_{\sigma\tau}^l
    db_{\sigma\tau}^l    $$
By a substitution of variables and collecting terms, one obtains 
$$\eqalignno{ &e^{-\U(\psi,\bar\psi)}=  \cr
 &\int \exp\biggl\{ {\ts {1\over\kappa}}\sum_k \biggl[ U_0^{1\over2}w+
  i\sum_l \lambda_l^{1\over2}\Bigl(\bar\xi_{\up\up}^l \,\bar y_l(\k')+
    \xi_{\up\up}^l \, y_l(\k')\Bigr) \biggr] 
     \bar\psi_{k\up}\psi_{k\up} \biggr\}\times\cr 
  &\phantom{\int} \exp\biggl\{ {\ts {1\over\kappa}}\sum_k \biggl[ U_0^{1\over2}w+
  i\sum_l \lambda_l^{1\over2}\Bigl(\bar\xi_{\down\down}^l \,\bar y_l(\k')+
    \xi_{\down\down}^l \, y_l(\k')\Bigr) \biggr] 
     \bar\psi_{k\down}\psi_{k\down} \biggr\}\times\cr 
  &\phantom{\int} \exp\biggl\{ \sum_l\biggl[  
   {\ts  i(2\lambda_l)^{1\over2} \bar\xi^l_{\up \down}\, 
  {1\over\kappa}}\sum_k \bar y_l(\k')\,\bar\psi_{k\up} \psi_{k\down} +
   {\ts  i(2\lambda_l)^{1\over2} \xi^l_{\up\down}\, 
  {1\over\kappa}}\sum_k  y_l(\k')\,\bar\psi_{k\down} \psi_{k\up} \biggr]\biggr\} 
   \times \cr  
 &\phantom{\int} \exp\biggl\{ \sum_l\biggl[  
   {\ts  i\lambda_l^{1\over2} \bar\phi^l_{\up\up}\, 
  {1\over\kappa}}\sum_k {\ts  {\bar y_l(\k')-\bar y_l(-\k')\over2} }
    \,\bar\psi_{k\up} \bar\psi_{-k\up} +
   {\ts  i\lambda_l^{1\over2} \phi^l_{\up\up}\, 
  {1\over\kappa}}\sum_k {\ts  {y_l(\k')-y_l(-\k')\over2}} 
      \,\psi_{k\up} \psi_{-k\up} \biggr]\biggr\} 
   \times \cr
 &\phantom{\int} \exp\biggl\{ \sum_l\biggl[  
   {\ts  i\lambda_l^{1\over2} \bar\phi^l_{\down\down}\, 
  {1\over\kappa}}\sum_k {\ts  {\bar y_l(\k')-\bar y_l(-\k')\over2} }
     \,\bar\psi_{k\down} \bar\psi_{-k\down} +
   {\ts  i\lambda_l^{1\over2} \phi^l_{\down\down}\, 
  {1\over\kappa}}\sum_k {\ts  {y_l(\k')-y_l(-\k')\over 2} }
     \,\psi_{k\down} \psi_{-k\down} \biggr]\biggr\} 
   \times \cr
 &\phantom{\int} \exp\biggl\{ \sum_l\biggl[  
   {\ts  i(2\lambda_l)^{1\over2} \bar\phi^l_{\up\down}\, 
  {1\over\kappa}}\sum_k \bar y_l(\k')\,\bar\psi_{k\up} \bar\psi_{-k\down} +
   {\ts  i(2\lambda_l)^{1\over2} \phi^l_{\up\down}\, 
  {1\over\kappa}}\sum_k  y_l(\k')\,\psi_{k\up} \psi_{-k\down} \biggr]\biggr\} 
   d\nu_\kappa(w,\xi,\phi) \cr}$$
where $ d\nu_\kappa(w,\xi,\phi)$ is defined in the statement of the theorem. 
Using the definition of the fields $\Xi$, $\Gamma$ and $\Phi$, the above
expression reads 
$$\eqalignno{ e^{-\U(\psi,\bar\psi)}&=  \int \exp\biggl\{ {\ts {1\over\kappa}} \sum_k 
  \Bigl[  \Gamma_{\k\up}\, \bar\psi_{k\up}\psi_{k\up}+
    \Gamma_{\k\down}\, \bar\psi_{k\down}\psi_{k\down}
  +i\bar\Xi_{\k\up\down}\bar\psi_{k\up}\psi_{k\down} 
   +i\Xi_{\k\up\down}\bar\psi_{k\down}\psi_{k\up}  \cr
 &\phantom{\int \exp}   
    + i\bar\Phi_{\k\up\down}\bar\psi_{k\up}\bar\psi_{-k\down}+
   i\Phi_{\k\up\down}\psi_{k\up}\psi_{-k\down} 
  +{\ts {i\over2}}
   \sum_\sigma\bigl( \bar\Phi_{\k\sigma\sigma}\bar\psi_{k\sigma}
  \bar\psi_{-k\sigma}+
   \Phi_{\k\sigma\sigma}\psi_{k\sigma}\psi_{-k\sigma}\bigr) \Bigr] \biggr\}
   d\nu_\kappa  \cr}$$
We now rewrite the exponent in order to perform the fermionic functional integral. 
Since, if the set of spatial momenta satisfy $\k\in M_\omega=-M_\omega$, 
$$\eqalignno{ \sum_k\Gamma_{k\up}\,\bar\psi_{k\up}\psi_{k\up}&=\sum_{k_0\in 
   {\pi\over\beta}(2\Bbb Z+1)}
    \sum_{\k\in M_\omega}\Gamma_{k\up}\,\bar\psi_{k\up}\psi_{k\up}  
  =\sum_{k_0>0}\sum_{\k\in M_\omega}\Bigl[ \Gamma_{k\up}\,\bar\psi_{k\up}\psi_{k\up}+
     \Gamma_{-k\up}\,\bar\psi_{-k\up}\psi_{-k\up}\Bigr]  \cr
  &={\ts {1\over2}} \sum_{k\atop k_0>0}
     \Bigl[ \Gamma_{k\up}\bigl(\bar\psi_{k\up}\psi_{k\up}-\psi_{k\up}\bar\psi_{k\up}
    \bigr)+ \Gamma_{-k\up}\bigl( \bar\psi_{-k\up}\psi_{-k\up}-
    \psi_{-k\up}\bar\psi_{-k\up}\bigr)  \Bigr]  \cr}$$
one obtains, using the antisymmetry of the $\Phi_{k\sigma\sigma}$, 
 $\bar\Phi_{k\sigma\sigma}$ 
$$\eqalignno{ \sum_k&\Bigl\{ \su_\sigma  \Gamma_{\k\sigma}\, 
  \bar\psi_{k\sigma}\psi_{k\sigma} 
   +i \bar\Phi_{\k\up\down}\bar\psi_{k\up}\bar\psi_{-k\down}+
   i\Phi_{\k\up\down}\psi_{k\up}\psi_{-k\down}  \cr
 &\phantom{m}  +i\bar\Xi_{\k\up\down}\bar\psi_{k\up}\psi_{k\down} 
   +i\Xi_{\k\up\down}\bar\psi_{k\down}\psi_{k\up} 
  +{\ts {i\over2}} 
   \su_\sigma\bigl( \bar\Phi_{\k\sigma\sigma}\bar\psi_{k\sigma}\bar\psi_{-k\sigma}+
   \Phi_{\k\sigma\sigma}\psi_{k\sigma}\psi_{-k\sigma}\bigr) \Bigr\}   \cr  
 ={\ts {1\over2}}  \sum_{k\atop k_0>0}&
       \biggl\{ \su_\sigma \Bigl[  \Gamma_{\k\sigma} 
  \bigl(\bar\psi_{k\sigma}\psi_{k\sigma}  -\psi_{k\sigma}\bar\psi_{k\sigma} \bigr)+ 
  \Gamma_{-\k\sigma} 
  \bigl(\bar\psi_{-k\sigma}\psi_{-k\sigma} -\psi_{-k\sigma}\bar\psi_{-k\sigma}
   \bigr) \Bigr] \cr
 &\phantom{m}   + i\bar\Phi_{\k\up\down}\bigl(\bar\psi_{k\up}\bar\psi_{-k\down}-
     \bar\psi_{-k\down}\bar\psi_{k\up} \bigr)+
  i \bar\Phi_{-\k\up\down}\bigl(\bar\psi_{-k\up}\bar\psi_{k\down}-
     \bar\psi_{k\down}\bar\psi_{-k\up} \bigr)  \cr
 &\phantom{m} + i \Phi_{\k\up\down}\bigl( \psi_{k\up}
   \psi_{-k\down}-\psi_{-k\down}
     \psi_{k\up}\bigr) +
   i\Phi_{-\k\up\down}\bigl( \psi_{-k\up}\psi_{k\down}-\psi_{k\down}
     \psi_{-k\up}\bigr)  \cr
 &\phantom{m}  +i\bar\Xi_{\k\up\down}
    \bigl( \bar\psi_{k\up}\psi_{k\down}- \psi_{k\down}
    \bar\psi_{k\up} \bigr)+
   i\bar\Xi_{-\k\up\down}\bigl( \bar\psi_{-k\up}\psi_{-k\down}- \psi_{-k\down}
    \bar\psi_{-k\up} \bigr)  \cr
 &\phantom{m}  +i\Xi_{\k\up\down}\bigl( \bar\psi_{k\down}\psi_{k\up}- \psi_{k\up}
    \bar\psi_{k\down}  \bigr)+
   i\Xi_{-\k\up\down}\bigl( \bar\psi_{-k\down}\psi_{-k\up}- \psi_{-k\up}
    \bar\psi_{-k\down}  \bigr)  \cr
  &\phantom{m} +i \su_\sigma\Bigl[ \bar\Phi_{\k\sigma\sigma}\bigl( 
    \bar\psi_{k\sigma}\bar\psi_{-k\sigma}
    -\bar\psi_{-k\sigma}\bar\psi_{k\sigma}\bigr) 
   +\Phi_{\k\sigma\sigma}\bigl( \psi_{k\sigma}\psi_{-k\sigma}-
   \psi_{-k\sigma}\psi_{k\sigma} \bigr)\Bigr] \biggr\}   \cr }$$
Since, if $a_k=ik_0-e_\k$ 
$$d\mu_C(\psi,\bar\psi)=\pro_{k,\sigma}{\ts  {\kappa\over a_k}}\; 
   e^{-{1\over \kappa} \sum_{k,\sigma}a_k \bar\psi_{k,\sigma} 
   \psi_{k,\sigma} } \pro_{k,\sigma} d\psi_{k,\sigma} d\bar\psi_{k,\sigma} $$ one 
obtains 
$$\eqalignno{ Z(\beta,L,\{s_{k,\sigma}\})&=\int e^{-\U(\psi,\bar\psi)+{1\over\kappa}
    \sum_{k,\sigma} s_{k,\sigma}\bar\psi_{k,\sigma}\psi_{k,\sigma} }
    d\mu_C(\psi,\bar\psi)  \cr
 &=\pro_{k,\sigma}{\ts  {\kappa\over a_k}}\int
   \int e^{-{1\over2}{1\over\kappa}\sum_{k_0>0}\sum_{\k} 
     \la \Psi_k,S_k \Psi_k\ra} \pro_{k,\sigma} d\psi_{k\sigma}d\bar\psi_{k\sigma}
    \>d\nu_\kappa(w,\xi,\phi)  \cr
 &= \int \prod_{k_0>0\atop\k\in M_\omega}
      \biggl\{ {\ts \left({\kappa^2\over a_k a_{-k}}\right)^2 }
    \int e^{-{1\over2}{1\over\kappa}\la \Psi_k,S_k \Psi_k\ra}
   d\bar\psi_{k\up}d\psi_{k\up} 
      d\bar\psi_{-k\down}d\psi_{-k\down} \times  \cr
 &\phantom{\int \prod_{k_0>0\atop \k\in M } \biggl\{ {\ts
       \left({\kappa^2\over a_k a_{-k}}\right)^2 }
    \int e^{-{1\over2}{1\over\kappa}\la \Psi_k,S_k \Psi_k\ra} }
    d\psi_{k\down} d\bar\psi_{k\down}d\psi_{-k\up}d\bar\psi_{-k\up} \biggr\} 
  d\nu_\kappa(w,\xi,\phi)  \cr
 &=\int \prod_{k_0>0\atop\k\in M_\omega}
    \biggl\{ {\ts \left({1\over a_k a_{-k}}\right)^2 }
    {\rm Pf} S_k \biggr\} d\nu_\kappa(w,\xi,\phi) \cr}$$  
where 
$$\Psi_k=\bigl(\bar\psi_{k\up},\psi_{k\up},\bar\psi_{-k\down},\psi_{-k\down}, 
    \psi_{k\down}, \bar\psi_{k\down},\psi_{-k\up},\bar\psi_{-k\up}\bigr)$$ 
and ${\rm Pf}S_k$ is the Pfaffian of the $8\times 8$ skew symmetric 
 matrix $S_k$ defined in the statement of the theorem.
 We used that
$$\pro_{k,\sigma} d\psi_{k\sigma}d\bar\psi_{k\sigma}
 =\pro_{k_0>0}\pro_{\k}\Bigr\{+d\bar\psi_{k\up}d\psi_{k\up} 
      d\bar\psi_{-k\down}d\psi_{-k\down} 
    d\psi_{k\down} d\bar\psi_{k\down}d\psi_{-k\up}d\bar\psi_{-k\up}\Bigr\}  $$
Part b) of the theorem follows from  
$$ \ts {1\over\kappa}\la \bar\psi_{p\sigma}\psi_{p\sigma}\ra=\ts 
  {\partial\over \partial s_{p\sigma}}_{|s=0} \log Z(\beta,L,\{s_{k\sigma}\}) 
  \eqno \blacksquare$$
\bigskip
\noindent{\bf Lemma II.2:} {\it Let $S_k$ be the skew symmetric $8\times 8$ 
 matrix of Theorem II.1. Then the Pfaffian of $S_k$ is given by  
$$\eqalignno{ {\rm Pf} S_k=&\bigl( A_{k\up}A_{-k\down} +\bar\Phi_{\k\up\down} 
  \Phi_{\k\up\down}\bigr)\bigl(A_{-k\up}A_{k\down} +\bar\Phi_{-\k\up\down} 
  \Phi_{-\k\up\down}\bigr)  \cr
 & +A_{k\up}A_{-k\up}\Phi_{\k\down\down}\bar\Phi_{\k\down\down}
  +A_{k\down}A_{-k\down}\Phi_{\k\up\up}\bar\Phi_{\k\up\up} \cr
 &+\Phi_{\k\up\up}\Phi_{\k\down\down}
    \bar\Phi_{\k\up\down}\bar\Phi_{-\k\up\down}+
  \bar\Phi_{\k\up\up}\bar\Phi_{\k\down\down}
    \Phi_{\k\up\down}\Phi_{-\k\up\down}  \cr
  &+\Phi_{\k\up\up}\Phi_{\k\down\down}
    \bar\Phi_{\k\up\up}\bar\Phi_{\k\down\down}  \cr
 &+i\Xi_{\k\up\down}A_{-k\up}\bar\Phi_{\k\up\down}\Phi_{\k\down\down}
  +i\bar\Xi_{\k\up\down}A_{-k\up}\Phi_{\k\up\down}\bar\Phi_{\k\down\down} \cr
  &-i\Xi_{\k\up\down}A_{-k\down}\Phi_{-\k\up\down}\bar\Phi_{\k\up\up} 
  -i\bar\Xi_{\k\up\down}A_{-k\down}\bar\Phi_{-\k\up\down}\Phi_{\k\up\up} \cr
 &+i\Xi_{-\k\up\down}A_{k\down}\Phi_{\k\up\down}\bar\Phi_{\k\up\up} 
  +i\bar\Xi_{-\k\up\down}A_{k\down}\bar\Phi_{\k\up\down}\Phi_{\k\up\up} \cr
 &-i\Xi_{-\k\up\down}A_{k\up}\bar\Phi_{-\k\up\down}\Phi_{\k\down\down} 
  -i\bar\Xi_{-\k\up\down}A_{k\up}\Phi_{-\k\up\down}\bar\Phi_{\k\down\down} \cr
 &+\Xi_{\k\up\down}\bar\Xi_{\k\up\down}A_{-k\up}A_{-k\down}
   +\Xi_{-\k\up\down}\bar\Xi_{-\k\up\down}A_{k\up}A_{k\down} \cr
 &+\Xi_{\k\up\down}\bar\Xi_{-\k\up\down}
     \bar\Phi_{\k\up\down}\Phi_{-\k\up\down}
  +\Xi_{-\k\up\down}\bar\Xi_{\k\up\down}
     \bar\Phi_{-\k\up\down}\Phi_{\k\up\down}  \cr
  &-\Xi_{\k\up\down}\Xi_{-\k\up\down}\bar\Phi_{\k\up\up}\Phi_{\k\down\down}
   -\bar\Xi_{\k\up\down}\bar\Xi_{-\k\up\down}
     \Phi_{\k\up\up}\bar\Phi_{\k\down\down}  \cr
  &+\Xi_{\k\up\down}\bar\Xi_{\k\up\down}
     \Xi_{-\k\up\down}\bar\Xi_{-\k\up\down}  \cr}$$   }
\bigskip
\noindent{\bf Proof:} The Pfaffian of $S_k$ is given by the sum of all 
 contractions $\sum\pro\la \psi\psi\ra$ of the fields
$$\bar\psi_{k\up},\psi_{k\up},\bar\psi_{-k\down},\psi_{-k\down}, 
    \psi_{k\down}, \bar\psi_{k\down},\psi_{-k\up},\bar\psi_{-k\up}$$
where the value $\la \psi\psi\ra$ is given by the corresponding matrix element. 
That is, the Pfaffian can be evaluated by using Wick's Theorem or integration 
by parts. Since $S_k$ is an $8\times8$ matrix, one has ${\rm Pf}[-S_k]=
  {\rm Pf}S_k$ and 
$$\eqalignno{ {\rm Pf}S_k=&\la \bar\psi_{k\up}\psi_{k\up}\bar\psi_{-k\down}
   \psi_{-k\down} \psi_{k\down}\bar\psi_{k\down}\psi_{-k\up}
     \bar\psi_{-k\up} \ra_{S_k} \cr
 =&-A_{k\up}\la  \bar\psi_{-k\down}
   \psi_{-k\down} \psi_{k\down}\bar\psi_{k\down}\psi_{-k\up}
     \bar\psi_{-k\up} \ra_{S_k} 
  -i\bar\Phi_{\k\up\down} \la
   \psi_{k\up} \psi_{-k\down} \psi_{k\down}\bar\psi_{k\down}\psi_{-k\up}
     \bar\psi_{-k\up} \ra_{S_k}  \cr
  &-i\bar\Xi_{\k\up\down}\la \psi_{k\up}\bar\psi_{-k\down}
   \psi_{-k\down} \bar\psi_{k\down}\psi_{-k\up}
     \bar\psi_{-k\up} \ra_{S_k} 
       +i\bar\Phi_{\k\up\up} 
  \la  \psi_{k\up}\bar\psi_{-k\down}
   \psi_{-k\down} \psi_{k\down}\bar\psi_{k\down}\psi_{-k\up} \ra_{S_k}  \cr
 =&+A_{k\up}A_{-k\down}\la \psi_{k\down}\bar\psi_{k\down}\psi_{-k\up}
     \bar\psi_{-k\up} \ra_{S_k}+iA_{k\up}\bar\Phi_{\k\down\down} \la 
    \psi_{-k\down} \psi_{k\down}\psi_{-k\up}
     \bar\psi_{-k\up} \ra_{S_k} \cr
  &+iA_{k\up}\Xi_{-\k\up\down}\la 
    \psi_{-k\down} \psi_{k\down}\bar\psi_{k\down}
     \bar\psi_{-k\up} \ra_{S_k}      \cr
 &+\bar\Phi_{\k\up\down} \Phi_{\k\up\down} \la 
       \psi_{k\down}\bar\psi_{k\down}\psi_{-k\up}
     \bar\psi_{-k\up} \ra_{S_k} -\bar\Phi_{\k\up\down}\Xi_{\k\up\down} \la 
  \psi_{-k\down} \psi_{k\down}\psi_{-k\up}
     \bar\psi_{-k\up} \ra_{S_k} \cr
  &-\bar\Phi_{\k\up\down}\Phi_{\k\up\up}
     \la\psi_{-k\down} \psi_{k\down}\bar\psi_{k\down}
     \bar\psi_{-k\up} \ra_{S_k} \cr    
 &- \bar\Xi_{\k\up\down}
    \Phi_{\k\up\down}\la \bar\psi_{-k\down}
     \bar\psi_{k\down}\psi_{-k\up} \bar\psi_{-k\up} \ra_{S_k}
  - \bar\Xi_{\k\up\down}\Xi_{\k\up\down}\la \bar\psi_{-k\down}
   \psi_{-k\down} \psi_{-k\up} \bar\psi_{-k\up} \ra_{S_k} \cr
 &-\bar\Xi_{\k\up\down}\Phi_{\k\up\up}\la  \bar\psi_{-k\down}
   \psi_{-k\down} \bar\psi_{k\down} \bar\psi_{-k\up} \ra_{S_k}  \cr
 & +\bar\Phi_{\k\up\up}\Phi_{\k\up\down}\la \bar\psi_{-k\down}
      \psi_{k\down}\bar\psi_{k\down}\psi_{-k\up} \ra_{S_k} 
  -\bar\Phi_{\k\up\up}\Xi_{\k\up\down}\la \bar\psi_{-k\down}
   \psi_{-k\down} \psi_{k\down}\psi_{-k\up} \ra_{S_k} \cr
 &-\bar\Phi_{\k\up\up}\Phi_{\k\up\up} \la \bar\psi_{-k\down}
   \psi_{-k\down} \psi_{k\down}\bar\psi_{k\down} \ra_{S_k} \cr  
 =&+\bigl( A_{k\up}A_{-k\down}+\Phi_{\k\up\down}\bar\Phi_{\k\up\down}\bigr) 
      \la \psi_{k\down}\bar\psi_{k\down}\psi_{-k\up}
     \bar\psi_{-k\up} \ra_{S_k}  \cr
  &+\bigl(A_{k\up}i\Xi_{-\k\up\down}-\bar\Phi_{\k\up\down}\Phi_{\k\up\up}\bigr) 
   \la \psi_{-k\down} \psi_{k\down}\bar\psi_{k\down}
     \bar\psi_{-k\up} \ra_{S_k}  \cr
 &+\bigl( A_{k\up}i\bar\Phi_{\k\down\down}
      -\bar\Phi_{\k\up\down}\Xi_{\k\up\down}\bigr) \la 
    \psi_{-k\down} \psi_{k\down}\psi_{-k\up}
     \bar\psi_{-k\up} \ra_{S_k} \cr 
  &- \bar\Xi_{\k\up\down}
    \Phi_{\k\up\down}\la \bar\psi_{-k\down}
     \bar\psi_{k\down}\psi_{-k\up} \bar\psi_{-k\up} \ra_{S_k}
  - \bar\Xi_{\k\up\down}\Xi_{\k\up\down}               \la \bar\psi_{-k\down}
   \psi_{-k\down} \psi_{-k\up} \bar\psi_{-k\up} \ra_{S_k} \cr
 &-\bar\Xi_{\k\up\down}\Phi_{\k\up\up}\la  \bar\psi_{-k\down}
   \psi_{-k\down} \bar\psi_{k\down} \bar\psi_{-k\up} \ra_{S_k} 
  +\bar\Phi_{\k\up\up}\Phi_{\k\up\down}\la \bar\psi_{-k\down}
      \psi_{k\down}\bar\psi_{k\down}\psi_{-k\up} \ra_{S_k} \cr
 & -\bar\Phi_{\k\up\up}\Xi_{\k\up\down}\la \bar\psi_{-k\down}
   \psi_{-k\down} \psi_{k\down}\psi_{-k\up} \ra_{S_k}
  -\bar\Phi_{\k\up\up}\Phi_{\k\up\up} \la \bar\psi_{-k\down}
   \psi_{-k\down} \psi_{k\down}\bar\psi_{k\down} \ra_{S_k} \cr  
 =&+\bigl( A_{k\up}A_{-k\down}+\Phi_{\k\up\down}\bar\Phi_{\k\up\down}\bigr) 
   \bigl(A_{k\down}A_{-k\up}+\Phi_{-\k\up\down}\bar\Phi_{-\k\up\down}\bigr) \cr 
 &+\bigl(A_{k\up}i\Xi_{-\k\up\down}-\bar\Phi_{\k\up\down}\Phi_{\k\up\up}\bigr) 
  \bigl(-\Phi_{\k\down\down}\bar\Phi_{-\k\up\down}-i\bar\Xi_{-k\up\down} 
   A_{k\down}\bigr)  \cr
 &+\bigl( A_{k\up}i\bar\Phi_{\k\down\down}
      -\bar\Phi_{\k\up\down}\Xi_{\k\up\down}\bigr)\bigl( -i\Phi_{\k\down\down}
   A_{-k\up}-\bar\Xi_{-k\up\down}\Phi_{-\k\up\down}\Bigr)  \cr
 &- \bar\Xi_{\k\up\down}\Phi_{\k\up\down}\bigl(-i\bar\Phi_{\k\down\down}
  A_{-k\up}-\Xi_{-\k\up\down}\bar\Phi_{-\k\up\down}\bigr) \cr
 &- \bar\Xi_{\k\up\down}\Xi_{\k\up\down}\bigl( -A_{-k\down}A_{-k\up}
    -\Xi_{-\k\up\down}\bar\Xi_{-\k\up\down}\bigr) \cr
 &-\bar\Xi_{\k\up\down}\Phi_{\k\up\up}\bigl(A_{-k\down}i\bar\Phi_{-\k\up\down}
   +\bar\Phi_{\k\down\down}\bar\Xi_{-k\up\down}\bigr) \cr
  &+\bar\Phi_{\k\up\up}\Phi_{\k\up\down}\bigl( \bar\Phi_{\k\down\down} 
   \Phi_{-\k\up\down}+i\Xi_{-\k\up\down}A_{k\down} \bigr) \cr
 &-\bar\Phi_{\k\up\up}\Xi_{\k\up\down}\bigl( A_{-k\down}i\Phi_{-\k\up\down} 
   +\Xi_{-\k\up\down}\Phi_{\k\down\down} \bigr) \cr
 &-\bar\Phi_{\k\up\up}\Phi_{\k\up\up} \bigr( -A_{-k\down}A_{k\down} 
   -\bar\Phi_{\k\down\down}\Phi_{\k\down\down}\bigr) \cr}$$
By multiplying out the brackets one obtains the stated formula $\blacksquare$
\bigskip
Before we  specialize Theorem II.1  in section III where the effective potential 
 and the two point functions are computed more explicitly, in the following 
 theorem  we write down the
integral representation for the generating functional of the connected amputated 
 Greens functions. 
\bigskip
\noindent{\bf Theorem II.3:} {\it Let $\U(\psi,\bar\psi)$ be given by (II.1) 
   and let $V$ 
 be the effective potential (II.14). Let 
$$G(\eta)=\log \int e^{-\U(\psi+\eta,\bar\psi+\bar\eta)} d\mu_C(\psi,\bar\psi)
   \eqno (\II.16) $$
be the generating functional for the connected amputated Greens functions. 
Then, if $\Eta k$ denotes the eight component vector 
$$\Eta k=(a_{k}\eta_{k\up},-a_{k}\bar\eta_{k\up},
  a_{-k}\eta_{-k\down},-a_{-k}\bar\eta_{-k\down},
  -a_{k}\bar\eta_{k\down},a_{k}\eta_{k\down},-a_{-k}\bar\eta_{-k\up},
    a_{-k}\eta_{-k\up})$$
and $S_k=S_k(w,\xi,\phi)$ is the $8\times 8$ matrix of Theorem II.1, 
one has the following integral representation
$$G(\eta)-G(0)=-{\ts{1\over\kappa}}\sum_{k\sigma} a_k\bar\eta_{k\sigma}
    \eta_{k\sigma}\>+\>  \log{ \int e^{-{1\over2}{1\over\kappa}
     \sum_k\la \Eta k,S_k^{-1} \Eta k\ra}  e^{-\kappa V(w,\xi,\phi)} dw\pro d\xi
    d\phi  \over  \int^{\phantom{I}} e^{-\kappa V(w,\xi,\phi)} dw\pro d\xi d\phi } 
 \eqno (\II.17) $$  }
\bigskip
\noindent{\bf Proof:}  By a substitution of Grassmann variables, 
$$\int e^{-\U(\psi+\eta,\bar\psi+\bar\eta)}d\mu_C=$$
$$   e^{-{1\over\kappa}\sum_{k\sigma}a_k\bar\eta_{k\sigma}\eta_{k\sigma}} 
  \int e^{-\U(\psi,\bar\psi)}\pro_{k\sigma}{\ts{\kappa\over a_k}}\> 
  e^{{1\over\kappa}\sum_{k\sigma} a_k[\bar\psi_{k\sigma}\eta_{k\sigma} + 
   \bar\eta_{k\sigma}\psi_{k\sigma}-
    \bar\psi_{k\sigma}\psi_{k\sigma}]} \pro_{k\sigma}d\psi_{k\sigma} d\bar 
   \psi_{k\sigma} $$
As in the proof of Theorem II.1, one has 
$$e^{-\U(\psi,\bar\psi)-{1\over\kappa}\sum_{k\sigma}a_k\bar\psi_{k\sigma} 
    \psi_{k\sigma}}= \int e^{-{1\over2}{1\over\kappa} \sum_{k_0>0}\sum_{\k} 
   \la \Psi_k,S_k \Psi_k\ra } d\nu_\kappa(w,\xi,\phi) $$
such that 
$$\eqalignno{ \int e^{-\U(\psi,\bar\psi)}&\pro_{k\sigma}{\ts{\kappa\over a_k}}\> 
  e^{{1\over\kappa}\sum_{k\sigma} a_k[\bar\psi_{k\sigma}\eta_{k\sigma} + 
   \bar\eta_{k\sigma}\psi_{k\sigma}-
    \bar\psi_{k\sigma}\psi_{k\sigma}]} \pro_{k\sigma}d\psi_{k\sigma} d\bar 
   \psi_{k\sigma}   \cr
 &=  \int\int  \pro_{k_0>0,\k}{\ts \left( { {\rm Pf}S_k\over a_k^2a_{-k}^2}\right)}\,
     e^{{1\over\kappa}\sum_{k\sigma} a_k[\bar\psi_{k\sigma}\eta_{k\sigma} + 
   \bar\eta_{k\sigma}\psi_{k\sigma}]} \times  \cr
 &\phantom{+ \int\int } 
        \pro_{k_0>0,\k} {\ts {\kappa^4\over 
    {\rm Pf}S_k }} \>
    e^{-{1\over2}{1\over\kappa} \sum_{k_0>0}\sum_{\k} 
   \la \Psi_k,S_k \Psi_k\ra } d\nu_\kappa(w,\xi,\phi) \pro_{k\sigma} 
   d\psi_{k\sigma} d\bar  \psi_{k\sigma}   \cr
 &=  \int\int  \pro_{k_0>0,\k}{\ts \left( { {\rm Pf}S_k\over a_k^2a_{-k}^2}\right)}\,
     e^{{1\over\kappa}\sum_{k_0>0,\k} \la \Psi_k,\Eta k\ra } \times  \cr
 &\phantom{+ \int\int } 
   \pro_{k_0>0,\k} {\ts {\kappa^4\over 
    {\rm Pf}S_k }} \>
    e^{-{1\over2}{1\over\kappa} \sum_{k_0>0}\sum_{\k} 
   \la \Psi_k,S_k \Psi_k\ra }  \pro_{k_0>0,\k}d\Psi_k\;d\nu_\kappa(w,\xi,\phi)
     \cr
 &=\int  \pro_{k_0>0,\k}{\ts \left( { {\rm Pf}S_k\over a_k^2a_{-k}^2}\right)}\,
   e^{-{1\over2}{1\over\kappa} \sum_{k_0>0,\k} \la \Eta k,S_k^{-1} \Eta k\ra} 
   d\nu_\kappa(w,\xi,\phi)  \cr}$$
where 
$$d\Psi_k=d\bar\psi_{k\up}d\psi_{k\up} 
      d\bar\psi_{-k\down}d\psi_{-k\down} 
    d\psi_{k\down} d\bar\psi_{k\down}d\psi_{-k\up}d\bar\psi_{-k\up} $$
By definition of $V$ 
$$\pro_{k_0>0,\k}{\ts \left( { {\rm Pf}S_k\over a_k^2a_{-k}^2}\right)}\,
    d\nu_\kappa(w,\xi,\phi) =e^{-\kappa V(w,\xi,\phi)} dw\pro d\xi d\phi$$
which proves the theorem $\blacksquare$ 

\bigskip
\bigskip
\bigskip
\goodbreak
\magnification=\magstep1
\font\gross=cmbx12 scaled \magstep0
\font\mittel=cmbx10 scaled \magstep0
\font\Gross=cmr12 scaled \magstep2
\font\Mittel=cmr12 scaled \magstep0
\overfullrule=0pt
\def\Bbb#1{{\bf #1}}
\def\blacksquare{\bullet}
%\baselineskip=12pt
%\nopagenumbers
\def\up{\uparrow}
\def\down{\downarrow}
\def\k{{\bf k}}

\def\p{{\bf p}}
\def\q{{\bf q}}
\def\x{{\bf x}}
\def\y{{\bf y}}
\def\I{{\rm I}}
\def\II{{\rm II}}
\def\III{{\rm III}}
\def\IV{{\rm IV}}
\def\ts{\textstyle}
\def\ds{\displaystyle}
\def\tr{\Delta}
\def\la{\langle}
\def\ra{\rangle}
\def\pro{\mathop\Pi}
\def\1cm{\hskip 1cm}
\def\1k{{\textstyle{1\over\kappa}}}
\def\db#1{{\ts{d^d\k\over (2\pi)^d}}}
\def\vp{\varphi}
\def\sl{g}

\def\U{{\cal U}}
\def\V{{\cal V}}
\def\O{{ O}}
\def\su{\mathop{\Sigma}}
\def\u{\underline}
\def\Eta#1{\u\zeta_{\phantom{.}\!#1}}
\def\Pf{{\rm Pf}}
\def\ep{\epsilon}

%\pageno=22

\noindent {\gross III. Solution for Pure BCS} 
\bigskip
\bigskip
We consider the model 
$$Z(\beta,L,\{s_{k,\sigma}\})=\int e^{-\U(\psi,\bar\psi)+{1\over\kappa}
    \sum_{k,\sigma} s_{k,\sigma}\bar\psi_{k,\sigma}\psi_{k,\sigma} }
    d\mu_C(\psi,\bar\psi)\eqno (\III.1)$$
with $e_\k={\k^2/2m}-\mu$, $C(k)={1/( ik_0-e_\k)}$ and 
$$  \U(\psi,\bar\psi)={\ts {1\over \kappa^3}}\!\!\!
   \sum_{\sigma,\tau\in\{\up,\down\}}
    \sum_{k,p}  U(\k-\p) \>\bar\psi_{k,\sigma}
   \bar\psi_{-k,\tau}\psi_{p,\sigma} 
  \psi_{-p,\tau} \eqno(\III.2)$$
The electron-electron interaction $U$ is given by (I.37). 
\par
To write down the effective potential and the two point functions in this case, 
one first has to compute the Pfaffian of the matrix $S_k$ of Theorem II.1. 
Since we consider only a BCS interaction, the $\Xi$ and $\Gamma$ fields 
 are zero. With Lemma II.2 one obtains 
$$\eqalignno{ {\rm Pf} S_k=&\bigl( a_{k\up}a_{-k\down} +\bar\Phi_{\k\up\down} 
  \Phi_{\k\up\down}\bigr)\bigl(a_{-k\up}a_{k\down} +\bar\Phi_{-\k\up\down} 
  \Phi_{-\k\up\down}\bigr)  \cr
 & +a_{k\up}a_{-k\up}\Phi_{\k\down\down}\bar\Phi_{\k\down\down}
  +a_{k\down}a_{-k\down}\Phi_{\k\up\up}\bar\Phi_{\k\up\up} \cr
 &+\Phi_{\k\up\up}\Phi_{\k\down\down}
    \bar\Phi_{\k\up\down}\bar\Phi_{-\k\up\down}+
  \bar\Phi_{\k\up\up}\bar\Phi_{\k\down\down}
    \Phi_{\k\up\down}\Phi_{-\k\up\down}
   +\Phi_{\k\up\up}\Phi_{\k\down\down}
    \bar\Phi_{\k\up\up}\bar\Phi_{\k\down\down}&(\III.3)  \cr}$$
where  $a_{k\sigma}=a_k-s_{k\sigma}$. In particular, for $s_{k\sigma}=0$ 
$$ {\rm Pf} S_k=(a_ka_{-k}+\Omega_\k^+)(a_ka_{-k}+\Omega_\k^-)
  \eqno (\III.4)  $$
where $\Omega_\k^\pm$ are the solutions of the quadratic equation 
$$\eqalignno{ \Omega^2&-\Bigl(\bar\Phi_{\k\up\down} \Phi_{\k\up\down}+
    \bar\Phi_{-\k\up\down}  \Phi_{-\k\up\down}+
   \Phi_{\k\up\up}\bar\Phi_{\k\up\up}+
   \Phi_{\k\down\down}\bar\Phi_{\k\down\down}\Bigr)\Omega+ 
  \Phi_{\k\up\down}\bar\Phi_{\k\up\down}
   \Phi_{-\k\up\down}\bar\Phi_{-\k\up\down} \cr
  &+\Phi_{\k\up\up}\Phi_{\k\down\down}
    \bar\Phi_{\k\up\down}\bar\Phi_{-\k\up\down}+
  \bar\Phi_{\k\up\up}\bar\Phi_{\k\down\down}
    \Phi_{\k\up\down}\Phi_{-\k\up\down} 
   +\Phi_{\k\up\up}\Phi_{\k\down\down}
    \bar\Phi_{\k\up\up}\bar\Phi_{\k\down\down}=0 &(\III.5) \cr}$$
The effective potential becomes 
$$\eqalignno{ V(\phi)&=\sum_{\sigma\tau\in 
  \{\up\up,\down\down,\up\down\}} \sum_{l=0}^N 
    |\phi_{\sigma\tau}^l|^2 -{\ts {1\over\beta L^d} }
     \sum_{\k\in M}\log\pro_{k_0>0}\ts \left\{ {a_k a_{-k}+\Omega_\k^+\over 
    a_ka_{-k} } {a_k a_{-k}+\Omega_\k^-\over 
    a_ka_{-k} } \right\}  &(\III.6)  \cr} $$
\par
We start in section III.1 with the case of a delta 
  function interaction, that is, $J=0$ in (I.37). 
This reproduces the usual mean field results. In section 
 III.2, we consider a more general electron-electron interaction of the form (I.37) 
 with arbitrary $J$. One finds that the standard approach  based on 
 approximating the Hamiltonian by quadratic terms and imposing a self 
 consistency equation may be misleading. 
\bigskip
\bigskip
\goodbreak
%
%
% CHAPTER III.1
%
%\goodbreak
\noindent{\mittel{III.1 BCS with Delta Function Interaction} }
\bigskip
\noindent{\bf Corollary III.1:} {\it Let $s_{k,\up},s_{k,\down},r_k,\bar r_k$ be some 
real or complex numbers, let 
$$\eqalignno{
    Z(\beta,L,\{s_k\},\{r_k\})&=\int \exp\biggl\{-{\ts  {\lambda\over\kappa^3}} 
    \su_{k,p} \psi_{k\up}\psi_{-k\down}
    \bar\psi_{p\up}\bar\psi_{-p\down} 
     + {\ts {1\over\kappa}}\su_k[s_{k\up}\bar\psi_{k\up}\psi_{k\up}  \cr
  &+
    s_{k\down}\bar\psi_{k\down}\psi_{k\down}   
   +r_k\psi_{k\up}\psi_{-k\down}
   -\bar r_k\bar\psi_{k\up}\bar\psi_{-k\down} ]  \biggr\} 
   d\mu_C(\psi,\bar\psi) &(\III.7)  \cr}$$ 
and let 
$$\eqalignno{\1k \la\bar\psi_{p\sigma}\psi_{p\sigma}\ra_{\beta,L,r}  
 &={\ts {\partial \over \partial s_{p\sigma}}} \log Z(\beta,L,\{s_k\},\{r_k\})\bigr|_{
   s_k=0,r_k=r} &(\III.8) \cr 
  \1k \la \psi_{p\up}\psi_{-p\down}\ra_{\beta,L,r}
 &={\ts {\partial \over \partial r_{p}}} \log Z(\beta,L,\{s_k\},\{r_k\})\bigr|_{
   s_k=0,r_k=r} &(\III.9) \cr  }$$
Define the effective potential 
$$V_{\beta,r}(u,v)=\ts  u^2+(v+{|r|\over\sqrt\lambda})^2-\int_M {d^d\k\over (2\pi)^d} 
 \> W_{\beta,\k}\bigl( \lambda(u^2+v^2)\bigr)  \eqno (\III.10)$$
where
$$W_{\beta,\k}(y)=
   \ts {1\over\beta}\log\left[ {\cosh({\beta\over2}\sqrt{ e_\k^2+y})\over 
    \cosh {\beta\over2}e_\k } \right]^2 \eqno (\III.11)$$
\item{\bf a)} Let $a_p=ip_0-e_\p$. 
  There are the two dimensional integral representations 
$$Z(\beta,L,r)={\ts{\kappa\over\pi}} 
    \ts \int e^{-\kappa V_{\beta,r}(u,v)} dudv  \eqno(\III.12)$$
$$\1k \la\bar\psi_{p,\sigma}\psi_{p,\sigma}\ra_{\beta,L,r}=
   {\int  {-a_{-p}\over a_pa_{-p}+\lambda(u^2+v^2)}
     \>e^{-\kappa V_{\beta,r}(u,v)} dudv 
  \over  \int e^{-\kappa V_{\beta,r}(u,v)} dudv}  \eqno(\III.13)$$
$$\1k \la \psi_{p,\up}\psi_{-p,\down}\ra_{\beta,L,r}={
   \int {\ts { -i \sqrt\lambda\,e^{-i\alpha}(u-iv)
             \over a_pa_{-p}+ \lambda(u^2+v^2)}}\>
      e^{-\kappa V_{\beta,r}(u,v)} dudv
  \over \int e^{-\kappa V_{\beta,r}(u,v)} dudv}    $$
$$\1k \la \bar\psi_{p,\up}\bar\psi_{-p,\down}\ra_{\beta,L,r}={
   \int {\ts {- i \sqrt\lambda\, e^{i\alpha}(u+iv)
             \over a_pa_{-p}+\lambda(u^2+v^2)}}\>
    e^{-\kappa V_{\beta,r}(u,v)} dudv
  \over \int e^{-\kappa V_{\beta,r}(u,v)} dudv}\eqno (\III.14)  $$
\item{\bf b)} Let $V_\beta(\rho):=\rho^2-\int
    {d^d\k\over(2\pi)^d} \>W_{\beta,\k}(\lambda\rho^2)$ 
 and $|\tr|^2=\lambda \rho_0^2$ be given by 
  $V_\beta(\rho_0)=\min_{\rho\ge 0} V_\beta(\rho)$  
Then, if $\tr=\tr(\lambda,\beta)=|\tr|\>e^{-i\alpha}$,  one has 
$$\lim_{|r|\to 0}\lim_{L\to\infty} {\ts{1\over\beta L^d}}\log Z(\beta,L,r)=
      -V_\beta(|\tr|)  \eqno(\III.15) $$
$$\lim_{|r|\to 0}\lim_{L\to\infty}
      {\ts{1\over\beta L^d}}  \la\bar\psi_{p,\sigma}\psi_{p,\sigma}\ra_{\beta,L,r}= 
   \ts {-a_{-p}\over a_pa_{-p}+|\tr|^2}  \eqno(\III.16)$$
$$\lim_{|r|\to 0}\lim_{L\to\infty}
    {\ts{1\over\beta L^d}} \la \psi_{p,\up}\psi_{-p,\down}\ra_{\beta,L,r}=\ts 
    {\tr \over a_pa_{-p}+|\tr|^2} \eqno(\III.17) $$
$$\lim_{|r|\to 0}\lim_{L\to\infty}
    {\ts{1\over\beta L^d}} \la \bar\psi_{p,\up}\bar\psi_{-p,\down}\ra_{\beta,L,r}=\ts 
    {-\bar\tr \over a_pa_{-p}+|\tr|^2}  \eqno(\III.18)  $$
but 
$$\lim_{L\to\infty}\lim_{|r|\to 0}
   \1k \la \psi_{p,\up}\psi_{-p,\down}\ra_{\beta,L,r}=
    \lim_{L\to\infty}\lim_{|r|\to 0}
    \1k \la\bar\psi_{p,\up}\psi_{-p,\down}\ra_{\beta,L,r}=
   0   \eqno (\III.19) $$      }
\bigskip
\noindent{\bf Proof:} {\bf a)} For a $\delta$ function interaction one has $J=0$ in 
 Theorem II.1 and the fields (II.10) are zero,  $\Phi_{\k\sigma\sigma}=0$. 
 The fields (II.9) 
become $\Phi_{\k\up\down}=(2\lambda)^{1\over2}\phi^0_{\up\down}\equiv 
  (2\lambda)^{1\over2}\phi$. Since we added the $r\psi\psi$ and 
 $\bar r\bar\psi\bar\psi$ terms to the exponent in (III.7) which were not present in 
 Theorem II.1, these fields have to be substituted by ($g=\sqrt{2\lambda}$)
$$g\phi\to g\phi-ir_k\>,\;\;\;\; g\bar\phi\to g\bar\phi+i\bar r_k$$ 
The Pfaffian (III.3) becomes 
$$ {\rm Pf} S_k=\Bigl( a_{k\up}a_{-k\down} +[g\bar\phi+i\bar r_k]
  [g\phi-ir_k]\Bigr)\Bigl(a_{-k\up}a_{k\down} +[g\bar\phi+i\bar r_{-k}]
  [g\phi-ir_{-k}] \Bigr) \eqno (\III.20) $$
such that 
$${\ts{{\partial\over \partial s_{p\sigma}}{\rm Pf} S_p\over{\rm Pf} S_p}}_{|_{s_k=0,
  r_k=r}}  =
  {\ts {-a_{-p}\over a_pa_{-p}+[ g \phi-ir]
   [ g \bar\phi+i\bar r]}} \eqno (\III.21)$$
and 
$${\ts{{\partial\over \partial r_{p}}{\rm Pf} S_p\over{\rm Pf} S_p}}_{|_{s_k=0,
  r_k=r}}  =
  {\ts {-i[g\bar\phi+i\bar r]\over a_pa_{-p}+[ g \phi-ir]
   [ g \bar\phi+i\bar r]}} \eqno (\III.22)$$
  The effective potential (III.6) is given by (the $|\phi|^2$ term is not shifted by 
 $r$) 
$$\eqalignno{ V(\phi)&=
    |\phi|^2 -{\ts {1\over\beta L^d} }
     \sum_{\k\in M}\log\pro_{k_0>0}\ts \left\{ {a_k a_{-k}+
  [g\phi-ir] [g\bar\phi+i\bar r] \over 
    a_ka_{-k} }  \right\}^2  &(\III.23)  \cr} $$
The product over $k_0\in {\pi\over\beta}(2\Bbb N+1)$ in (III.23) can be computed  
explicitly using the formula [Hn]
$$\pro_{n=0}^\infty\ts \left( 1+{\xi\over(2n+1)^2+a^2}\right)=
   {\cosh\left({\pi\over2}\sqrt{a^2+\xi}\right)\over \cosh{\pi\over2}a}\>
  \eqno (\III.24)$$
One obtains 
$$\pro_{k_0>0}{\ts \left\{ {a_k a_{-k}+\xi \over 
    a_ka_{-k} }  \right\}^2}
 =\pro_{k_0\in{\pi\over\beta}(2\Bbb N+1)}\ts\left(1+{\xi\over k_0^2+e_\k^2}\right)^2
   =\biggl\{ {\cosh\left( {\beta\over2}\sqrt{e_\k^2+\xi}\right)\over \cosh
    \left({\beta\over2}e_\k\right)} \biggr\}^2 \eqno (\III.25)$$
which gives, if one approximates the Riemannian sum ${1\over L^d}\sum_{\k\in
  M_\omega}$ 
 by an integral 
$$V(\phi)=|\phi|^2-\int_{M_\omega} {\ts {d^d\k\over (2\pi)^d}}\> W_{\beta,\k}\bigl( 
  [g\phi-ir] [g\bar\phi+i\bar r]\bigr)  \eqno (\III.26)$$
Thus one arrives at the integral representations 
$$Z(\beta,L,r)={\ts{\kappa\over\pi}} 
    \ts \int e^{-\kappa V(\phi)} dudv  \eqno(\III.27)$$
$$\1k \la\bar\psi_{p,\sigma}\psi_{p,\sigma}\ra_{\beta,L,r}=
   {\int  {-a_{-p}\over a_pa_{-p}+[g\phi-ir] [g\bar\phi+i\bar r]}
     \>e^{-\kappa V(\phi)} dudv 
  \over  \int e^{-\kappa V(\phi)} dudv}  \eqno(\III.28)$$
$$\1k \la \psi_{p,\up}\psi_{-p,\down}\ra_{\beta,L,r}={
   \int {\ts { -i  [g\bar\phi+i\bar r]
             \over a_pa_{-p}+ [g\phi-ir] [g\bar\phi+i\bar r] }}\>
      e^{-\kappa V(\phi)} dudv
  \over \int e^{-\kappa V(\phi)} dudv}  \eqno(\III.29)  $$
where $V$ is given by (III.26). Part (a) then follows from the substitution of 
 variables 
$$ \int F\left( \sl \phi-i r,\sl \bar\phi+i\bar r\right)
     \> e^{-\kappa |\phi|^2} dudv =   \int F\left(e^{i\alpha}\sl \phi, 
    e^{-i\alpha}\sl\bar\phi \right) 
   e^{-{\kappa} \left(u^2+(v+{|r|\over\sl})^2\right)}
    dudv \eqno (\III.30)$$
which holds for both signs of $\lambda$. 
\par
  To obtain part {\bf (b)}, 
 one has to compute the limit of 
$$\delta_{\kappa,r} (u,v)= { e^{-\kappa V_{\beta,r}(u,v)}\over \int
    e^{-\kappa V_{\beta,r}(u,v)} dudv}  \eqno (\III.31)$$ 
where 
$$\ts V_{\beta,r}(u,v)= u^2+(v+{|r|\over\sl})^2
   -W_\beta\bigl(\lambda(u^2+v^2)\bigr)\eqno (\III.32)$$
if we define 
$$W_\beta(\xi)=\int_M {\ts {d^d\k\over (2\pi)^d}} \> W_{\beta,\k}(\xi)=
   {\ts {1\over \beta}} \int_M \db k\ts \>\log\left[ 
   {\cosh\left( {\beta\over2}\sqrt{e_\k^2+\xi}\right)\over \cosh
    \left({\beta\over2}e_\k\right)}\right]^2\eqno (\III.33)$$
In particular, 
$${\rm sign}W_\beta\bigl(\lambda(u^2+v^2)\bigr)={\rm sign}\lambda$$
For positive $\lambda$ and nonzero $r$, 
   $V_{\beta,r}(u,v)$ is real and has a unique global minimum 
 determined by 
$$\eqalignno{ 2u-2u\lambda W_\beta'\bigl(\lambda(u^2+v^2)\bigr)&=0  \cr
  2(v+|r|)-2v\lambda W_\beta'\bigl(\lambda(u^2+v^2)\bigr)&=0  \cr}$$
Since $\lambda W_\beta'=1$ does not solve the second equation, the only 
solution of the first equation is $u=0$ and one is left with 
$$ v\left[ \lambda W_\beta'(\lambda v^2)-1\right] =|r| \eqno $$
which  has a solution $v=O(|r|)$ which 
 is a local maximum and two  
 nontrivial solutions $\lambda v^2=|\tr|^2+O(|r|)$ where  the positive one is only a 
 local minimum and the negative one, $v_0$,  is the global minimum. 
Therefore 
$$\lim_{\kappa\to\infty} \delta_{\kappa,r}(u,v)=\lim_{\kappa\to\infty} 
   { e^{-\kappa\left[ V_{\beta,r}(u,v)-V_{\beta,r}(0,v_0)\right]}\over \int
    e^{-\kappa\left[ V_{\beta,r}(u,v)-V_{\beta,r}(0,v_0)\right]} dudv}
   =\delta(u)\,\delta(v-v_0)  \eqno $$
and 
$$\lim_{|r|\to 0}\lim_{\kappa\to\infty} \delta_{\kappa,r}(u,v)=\ts 
    \delta(u)\,\delta\left(v+{|\tr|\over g}\right) \eqno (\III.34)$$
This proves the formulae under (b)  for attractive $\lambda$. Since  
$$\lim_{|r|\to 0} V_{r,\beta}(u,v)=u^2+v^2-W_\beta\left(\lambda(u^2+v^2)\right)$$
is an even function in $u$ and $v$, $\lim_{r\to 0}
  \la \psi_{p,\up}\psi_{-p,\down}\ra_{\beta,L,r}$=$\lim_{r\to 0}
   \la \bar\psi_{p,\up}\bar\psi_{-p,\down}\ra_{\beta,L,r}=0$. 
The limit of the logarithm of the partition function becomes 
$$\eqalignno{ {\ts {1\over \kappa}}\log Z&={\ts {1\over \kappa}}\log 
  {\ts{\kappa\over\pi}} \int e^{-\kappa V_{\beta,r}(u,v) }dudv   \cr
 &= {\ts {1\over \kappa}}\log 
  {\ts{\kappa\over\pi}} \int e^{-\kappa \left[V_{\beta,r}(u,v)
     -V_{\beta,r}(0,v_0)\right] }dudv +{\ts {1\over \kappa}}\log e^{-\kappa
   V_{\beta,r}(0,v_0)}  \cr}$$
The first term on the right hand side may be approximated by ($V_{uv}=
  {\partial^2V_{\beta,r}\over \partial u\partial v}(0,v_0)=0$)
$$ {\ts {1\over \kappa}}\log 
  {\ts{\kappa\over\pi}} \int
     e^{-{\kappa\over2}  (V_{uu}u^2+V_{vv}(v-v_0)^2)}dudv=
  {\ts {1\over \kappa}}\log 
  {\ts{1\over\pi }}
    \int e^{- {1\over2}(V_{uu}u^2+V_{vv}v^2)}dudv\buildrel
   \kappa\to\infty\over \to 0$$
which results in 
$$\lim_{\kappa\to\infty} {\ts{1\over \kappa}}\log Z=-V_{\beta,r}(0,v_0)$$
\medskip
Now let $\lambda$ be negative. In that case the effective potential (III.32) 
 is complex:  
$$\eqalignno{ V_{\beta,r}(u,v)&=u^2+v^2-\ts {|r|^2\over |\lambda|}
   -2iv{|r|\over \sqrt{|\lambda|} }-W_\beta\left(\lambda(u^2+v^2)\right) \cr
 &\ts =\rho^2  +|W_\beta\left(\lambda\rho^2\right)|
  -{|r|^2\over |\lambda|}-2i\rho\cos\vp{|r|\over \sqrt{|\lambda|} }  \cr}$$
Since the real part $U_{\beta,r}={\rm Re}V_{\beta,r}$ has 
 a global minimum at $u=v=0$ one has, since $U''={\partial^2U_{\beta,r}
   \over \partial u^2 }(0,0)={\partial^2U_{\beta,r}\over \partial v^2 }(0,0)>0$, 
$$\eqalignno{ \delta_{\kappa,r}(u,v)&={ e^{-\kappa\bigl[U_{\beta,r}(u,v)
     -2iv{|r|\over \sqrt{|\lambda|} } \bigr]} \over \int 
    e^{-\kappa\bigl[U_{\beta,r}(u,v)
     -2iv{|r|\over \sqrt{|\lambda|} } \bigr]} dudv } 
  \approx { e^{-\kappa\bigl[ {U''\over2}(u^2+v^2)
     -2iv{|r|\over \sqrt{|\lambda|} } \bigr]} \over \int 
    e^{-\kappa\bigl[{U''\over 2}(u^2+v^2)
     -2iv{|r|\over \sqrt{|\lambda|} } \bigr]} dudv }  \cr
 &= { e^{-\kappa {U''\over2}u^2  } \over \int 
    e^{-\kappa {U''\over 2}u^2} du } \;
     { e^{-\kappa\bigl[ {U''\over2}v^2
     -2iv{|r|\over \sqrt{|\lambda|} } \bigr]} \over \int 
    e^{-\kappa\bigl[{U''\over 2}v^2
     -2iv{|r|\over \sqrt{|\lambda|} } \bigr]} dv } \cr
 & = { e^{-\kappa {U''\over2}u^2  } \over \int 
    e^{-\kappa {U''\over 2}u^2} du } \;
     { e^{-\kappa  {U''\over2}\bigl[v-i{2|r|\over \sqrt{|\lambda|}U'' }
       \bigr]^2} \over \int 
    e^{-\kappa {U''\over 2}\bigl[v-i{2|r|\over \sqrt{|\lambda|}U'' }
       \bigr]^2} dv } 
   \buildrel \kappa\to\infty \over \to \ts \delta(u)\,\delta\Bigl(
     v-i{2|r|\over \sqrt{|\lambda|}U'' } \Bigr)   \cr  }$$
which results in 
 $$\lim_{|r|\to 0}\lim_{\kappa\to\infty}\delta_{\kappa,r}(u,v)=\delta(u)\,\delta(v)
    \eqno (\III.35)$$
and 
$$\lim_{|r|\to 0} \lim_{\kappa\to\infty}\ts {1\over \kappa}
   \la \bar\psi_{p,\up}\psi_{-p,\down}\ra_{\beta,L,r}=\int\ts   {-a_{-p}\over 
   a_pa_{-p}+\lambda(u^2+v^2)}\>\delta(u)\delta(v)\>dudv 
    ={-a_{-p}\over a_pa_{-p}}= 
   -{1\over a_p} $$ 
$$\lim_{|r|\to 0} \lim_{\kappa\to\infty}{\ts {1\over \kappa}}
   \la \psi_{p,\up}\psi_{-p,\down}\ra_{\beta,L,r}=\lim_{|r|\to 0}
    \lim_{\kappa\to\infty}\ts {1\over \kappa}
   \la \bar\psi_{p,\up}\bar \psi_{-p,\down}\ra_{\beta,L,r}=0\eqno \blacksquare$$
\bigskip\goodbreak
Using (II.17) and (III.34,35) the infinite volume limit for the generating functional 
 for the connected Greens functions can be computed in a similar way. One finds 
\bigskip
\noindent{\bf Corollary III.2:} {\it  Let $r=|r|e^{i\alpha}$, 
 $\gamma=ge^{i\alpha}(u+iv)$,
  $\bar\gamma=g e^{-i\alpha}(u-iv)$ and let 
$$\eqalignno{ G&(\eta)= G_r(\beta,L,\eta)= \log 
  \int e^{ \U_r(\psi+\eta,\bar\psi+\bar\eta)} 
   d\mu_C(\psi,\bar\psi) &(\III.36) \cr} $$
 be the generating functional for the connected amputated Greens functions 
where
$$\U_r(\psi,\bar\psi)={\ts -  {\lambda\over\kappa^3}} 
    \sum_{k,p}\psi_{k\up}\psi_{-k\down}\bar\psi_{p\up} 
  \bar\psi_{-p\down}+{\ts {1\over\kappa}}\sum_k[r\psi_{k\up}\psi_{-k\down} 
   -\bar r\bar\psi_{k\up}\bar\psi_{-k\down}] \eqno (\III.37) $$
Let $V_{\beta,r}$ be the effective potential (III.10) and let $\Delta$ be given as in 
 Corollary III.1. Then there is the integral representation 
$$\eqalignno{ G&(\eta)-G(0)= 
     -\1k\sum_k\Bigl[ a_k\bar\eta_{k\up}\eta_{k\up}
    +a_{-k}\bar\eta_{-k\down}\eta_{-k\down}\Bigr] \;+\;   
  &(\III.38)  \cr
 & + \log{ \int dudv\> 
    e^{-\kappa V_{\beta,r}(u,v) }
      \exp\biggl\{ \1k\sum_k   (a_k\bar\eta_{k\up},
    -a_{-k}\eta_{-k\down} ) {1\over a_ka_{-k}+\gamma\bar\gamma} 
   \left( {a_{-k}\atop i\gamma}\; {-i\bar\gamma \atop  -a_{k}} \right)
    \left( a_k\eta_{k\up} \atop 
    -a_{-k}\bar\eta_{-k\down}\right) \biggr\} 
   \over \int dudv\> 
    e^{-\kappa V_{\beta,r}(u,v) }  }  \cr}$$
  For attractive $\lambda>0$ the infinite volume limit of the generating
functional is given by 
$$\eqalignno{ \lim_{|r|\to 0}\lim_{L\to\infty}
    \{G(\beta,&L,r,\eta)-G(\beta,L,r,0)\}= &(\III.39)  \cr
  &-\1k\sum_k\Bigl[\ts  a_k\,{|\tr|^2\over a_ka_{-k}+|\tr|^2} \>
      \bar\eta_{k\up}\eta_{k\up}
    +a_{-k}\,{|\tr|^2\over a_ka_{-k}+|\tr|^2} \>
    \bar\eta_{-k\down}\eta_{-k\down}  \cr
 &\phantom{ -\1k\sum_k\Bigl[} 
      \ts  +\tr\, {a_ka_{-k} \over a_ka_{-k}+|\tr|^2}\>\eta_{-k\down}\eta_{k\up} 
    +\bar\tr\,   {a_ka_{-k} \over a_ka_{-k}+|\tr|^2} \> \bar\eta_{k\up}
    \bar\eta_{-k\down} \Bigr] \cr}   $$
For repulsive $\lambda<0$ one obtains 
$$\ds\lim_{|r|\to 0}\lim_{L\to\infty} \{G(\beta,L,r,\eta)-G(\beta,L,r,0)\}=0\>. 
   \eqno (\III.40)$$  }
\bigskip
\bigskip
%
%
% CHAPTER III.2
%
%
\noindent{\mittel{III.2 BCS with Higher Angular Momentum Terms}}
\bigskip
We now consider the case where the electron-electron interaction contains higher
angular momentum terms. In this case one finds that the usual mean field approach, 
based on approximating the Hamiltonian by quadratic terms, may be misleading. 
 To simplify the algebra we assume that only even $\ell$ 
terms contribute in (I.37). In that case the fields $\Phi_{\k\sigma\sigma}$ with equal 
spin (II.10) are still zero. So let 
$$U(\k'-\p')=\cases{ \ds {1\over2} \sum_{\ell=-j\atop \ell\;{\rm even}}^j 
  \lambda_{|\ell|}
     e^{i\ell\varphi_\k}e^{-i\ell\varphi_\p}
    +{\lambda_0\over2} & if $d=2$ \cr
 \ds \sum_{\ell=0\atop \ell\;{\rm even}}^j \sum_{m=-\ell}^\ell \lambda_\ell
   \bar Y_{\ell m}\left({\ts{\k'}}\right)
     Y_{\ell m}\left({\ts{\p'}}\right) & if $d=3$ \cr} \;\;=:\;
   \sum_{l=0\atop l\;{\rm even}}^J
  \lambda_l\> y_l(\k')\>\bar y_l(\p')\eqno(\III.41) $$
where $y_l(-\k)=(-1)^l y_l(\k)$. 
We only consider the $\bar\psi\psi$ expectations, so we let $r=0$. The Pfaffian 
 (III.3) becomes 
$$ {\rm Pf} S_k=\bigl( a_{k\up}a_{-k\down} +\bar\Phi_{\k\up\down} 
  \Phi_{\k\up\down}\bigr)\bigl(a_{-k\up}a_{k\down} +\bar\Phi_{-\k\up\down} 
  \Phi_{-\k\up\down}\bigr)  \eqno (\III.42)$$
where
$$\Phi_{\k\up\down}=\sum_{l=0\atop l\;{\rm even}}^J (2\lambda_l)^{1\over2} 
    \phi_{\up\down}^l\,y_l(\k')\>,\;\;\;
  \bar\Phi_{\k\up\down}=\sum_{l=0\atop l\;{\rm even}}^J (2\lambda_l)^{1\over2} 
     \bar\phi_{\up\down}^l\, \bar y_l(\k')  \eqno (\III.43)$$
and the effective potential is given by, using (III.25) again to compute the 
 $k_0$ product
$$V(\phi_{\up\down})=\sum_{l=0}^J |\phi_{\up\down}^l|^2- \int_{M_\omega} \ts 
   {d^d\k\over (2\pi)^d}\>W_{\beta,\k}\bigl(
     \bar\Phi_{\k\up\down} \Phi_{\k\up\down} )  \eqno (\III.44)$$
where $W_{\beta,\k}$ is given by (III.11). By Theorem II.1, the two point function 
is given by 
$$\eqalignno{ {\ts{1\over\kappa}} \la \bar\psi_{p\sigma}\psi_{p\sigma}\ra&
   = -{\int 
    {ip_0+e_\p\over p_0^2+e_\p^2+\bar\Phi_{\k\up\down} \Phi_{\k\up\down} }\, 
    e^{-\kappa V(\phi_{\up\down})} \ds \pro_{l=0}^J du_{\up\down}^l
       dv_{\up\down}^l \over 
   \int e^{-\kappa V(\phi_{\up\down})} \ds \pro_{l=0}^J du_{\up\down}^l
       dv_{\up\down}^l } &(\III.45)  \cr}$$
\par
To compute the infinite volume limit, one has to find the global minimum of the 
 real part of $V(\phi_{\up\down})$. This is easier for $d=2$. However, using 
symmetry arguments, it is possible to give a rather explicit expression also for 
$d=3$ without knowing the exact location of the global minimum.
\par
Consider first the two dimensional case. An analysis done by Albrecht 
 Schuette in his Diploma thesis [Sch] shows the following result:
\par
 Suppose that $\lambda_m>0$ is attractive and $\lambda_m>\lambda_\ell$ for 
 all $\ell\ne m$. Then 
$$\lim_{\kappa\to\infty} {\ts{1\over\kappa}}\la \bar\psi_{p\sigma} \psi_{p\sigma} 
   \ra_{\beta,L}=\ts {ip_0+e_\p\over p_0^2+e_\p^2+\lambda_m\rho_m^2}
   \eqno (\III.46)$$
where $\rho_m$ is determined by the BCS equation 
$$1-{\ts {\lambda_m\over 4\pi}} \int d|\k| |\k| \ts {\tanh( {\beta\over2}\sqrt{ e_\k^2+ 
     \lambda_m \rho_m^2}) \over \sqrt{e_\k^2+\lambda\rho_m^2} }=0\eqno (\III.47)$$
\par
The form of the two point function (III.46) can still be obtained by applying the 
 standard  mean field 
formalism [AB,BW]. However, the situation is different in 3 dimensions. 
Before we state the corresponding theorem, we shortly recall the mean field 
 equations (I.57,58)
\par
The $\la a^+ a\ra$ expectations are given by 
$$\la a_{\k\sigma}^+ a_{\k\sigma} \ra=\ts {1\over2}\left( 1-e_\k 
   \Bigl[ {\tanh({\beta\over2}\sqrt{ e_\k^2+\Delta_\k^*\Delta_\k})\over 
      \sqrt{ e_\k^2+\Delta_\k^*\Delta_\k}} \Bigr]_{\sigma\sigma}\right)
   \eqno (\I.57 )$$
 where the $2\times 2$ matrix $\Delta_\k$, $\Delta_\k^T=-\Delta_{-\k}$, is 
 a solution of the gap equation 
$$\Delta_\p=\int_M \ts{d^d\k\over (2\pi)^d}\, U(\p'-\k')\, \Delta_\k 
   {\tanh({\beta\over2}\sqrt{ e_\k^2+\Delta_\k^*\Delta_\k})\over 
     2 \sqrt{ e_\k^2+\Delta_\k^*\Delta_\k}}  \eqno (\I.58)$$
\par
In 2 dimensions, if one substitutes $U(\p-\k)$ by a single attractive term 
 $\lambda_\ell e^{i\ell(\vp_\p-\vp_\k)}$, then (I.58) has the unitary isotropic 
 solution $\Delta_\k=\Delta\left( {0\atop -e^{i\ell\vp_\k}}\; 
  {e^{i\ell\vp_\k}\atop 0} \right)$ for even $\ell$ and 
  $\Delta_\k=\Delta\left( {\cos\ell\vp_\k \atop \sin\ell\vp_\k}\; 
  {\sin\ell\vp_\k\atop -\cos\ell\vp_\k} \right)$ if $\ell$ is odd. In that case 
 $\Delta_\k^*\Delta_\k=|\Delta|^2 Id$ and (I.57) coincides with (III.46) (after 
 integration of the latter over $p_0$). 
\par
In 3 dimensions, it is proven  [FKT2] 
 that for all $\ell \ge 2$ (I.58) does not have unitary isotropic 
  ($\Delta_\k^*\Delta_\k=const\,Id$) solutions. In view of that result, the symmetry 
 considerations  
 below indicate that in 3 dimensions for $\ell\ge 2$ the standard mean field 
 approach is misleading since one would no longer expect $SO(3)$ 
 invariance for the $\la a_{\k\sigma}^+a_{\k\sigma}\ra$ expectations according to 
 (I.57). But this is indeed the case if there is no external SO(3) symmetry breaking 
 term (which is also not present in the quadratic mean field model, 
 see  the discussion following (I.61) at the end of the introduction). 
\bigskip
\bigskip
\noindent{\bf Theorem III.3:} {\it  Let $\ell$ be even, $\lambda_\ell>0$ be attractive 
and let 
$$\U(\psi,\bar\psi)={\ts {\lambda_\ell\over (\beta L^{d})^3} }\sum_{k,p} 
  \sum_{m=-\ell}^\ell \bar Y_{\ell m} 
   (\k') Y_{\ell m} (\p')\, \bar\psi_{k\up}\bar\psi_{-k\down}
    \psi_{p\up}\psi_{-p\down} \eqno (\III.48) $$
Then, if $e_{R\k}=e_\k$ for all $R\in SO(3)$, one has 
$$\eqalignno{ \lim_{\kappa\to\infty} {\ts{1\over\kappa}}
    \la \bar\psi_{p\sigma} \psi_{p\sigma}  \ra_{\beta,L}&= 
   \lim_{\kappa\to\infty} \int  {\ts{1\over\kappa}}\bar\psi_{p\sigma} \psi_{p\sigma}
   \, e^{-\U(\psi,\bar\psi)} d\mu_C(\psi,\bar\psi) \cr
  &=  \int_{S^2}\ts {ip_0+e_\p\over p_0^2+e_\p^2+
   \lambda_\ell \rho_0^2 | \su_m  \alpha_m^0 Y_{\ell m}(\x)|^2} 
   \, {d\Omega (\x)\over 4\pi} &(\III.49)  \cr}$$
where $\rho_0\ge 0$ and $\alpha^0\in \Bbb C^{2\ell +1}$, 
 $\su_m |\alpha_m^0|^2=1$, are values at the global minimum (which 
 is degenerated) of 
$$ V(\rho,\alpha)=\rho^2-\int_M\ts  {d^d\k\over (2\pi)^d}\, {1\over\beta} \log \left[
   { \cosh ( {\beta\over2}\sqrt{ e_\k^2+
   \lambda_\ell \rho^2 | \su_m \alpha_m Y_{\ell m}(\k')|^2} ) \over 
   \cosh {\beta\over2} e_\k } \right]^2 $$
In particular, the momentum distribution $n_\p$ is given by 
$$n_\p=\lim_{L\to\infty}{\ts{1\over L^d}} \la a_{\p\sigma}^+ a_{\p\sigma}\ra_{\beta,L}=
   \int_{S^2} \ts {1\over2}\left( 1- {e_\p}{\tanh({\beta\over2}\sqrt{e_\p^2+
    |\Delta(\x)|^2}) \over \sqrt{ e_\p^2+|\Delta(\x)|^2} } \right)
   {d\Omega(\x)\over 4\pi}  \eqno (\III.50)$$
and has $SO(3)$ symmetry. Here $\Delta(\x)=\lambda_\ell^{1\over2} \rho_0 
   \su_m  \alpha_m^0 Y_{\ell m}(\x)$.  }
\bigskip
\noindent{\bf Proof:}  Substituting $u_{\up\down}^l$ by $u_m$ and 
  $u_{\up\down}^l$ by $-v_m$ in (III.45), 
   one has to compute the infinite volume limit of 
$$ {\ts{1\over\kappa}}\la \bar\psi_{p\sigma}\psi_{p\sigma}\ra= 
 -{\int_{\Bbb R^{4\ell+2}} 
    {a_{-p}\over a_pa_{-p}+ |\Phi_{\p}|^2 }\, 
    e^{-\kappa V(\phi)}  \pro_{m=-\ell}^\ell du_m
       dv_m \over 
   \int_{\Bbb R^{4\ell+2}} e^{-\kappa V(\phi)}  \pro_{m=-\ell}^\ell du_m
       dv_m } \eqno (\III.51) $$
where 
$$V(\phi)=\sum_{m=-\ell}^\ell |\phi_m|^2-
    \int_M\ts  {d^d\k\over (2\pi)^d}\, {2\over\beta} \log \left[
   { \cosh ( {\beta\over2}\sqrt{ e_\k^2+ |\Phi_\k|^2} ) \over 
   \cosh {\beta\over2} e_\k } \right]$$ 
and 
$$\Phi_\k=\lambda_\ell^{1\over2} 
      \sum_{m=-\ell}^\ell \bar \phi_m Y_{\ell m}(\k')$$
Let $U\!(\!R)$ be the unitary representation of $SO(3)$ given by 
$$ Y_{\ell m}(R\k')=\sum_{m'} U\!(\!R)_{mm'} Y_{\ell m'}(\k')$$
and let $\sum_m (\overline{U\phi})_m Y_{\ell m}(\k')=: (U\Phi)_{\k}$. Then 
 for all $R\in SO(3)$ one has $\bigl(U\!(\!R)\Phi\bigr)_{\k}=\Phi_{R^{-1}\k}$ and 
$$\eqalignno{ V\bigl(U\!(\!R)\phi\bigr)&= \sum_m  |[U\!(\!R)\phi]_m|^2-
    \int_M\ts  {d^d\k\over (2\pi)^d}\, {2\over\beta} \log \left[
   { \cosh ( {\beta\over2}\sqrt{ e_\k^2+ |[U\Phi]_\k|^2} ) \over 
   \cosh {\beta\over2} e_\k } \right]  \cr
 &=\sum_m  |\phi_m|^2-
    \int_M\ts  {d^d\k\over (2\pi)^d}\, {2\over\beta} \log \left[
   { \cosh ( {\beta\over2}\sqrt{ e_\k^2+ |\Phi_{R^{-1}\k}|^2} ) \over 
   \cosh {\beta\over2} e_\k } \right]  \cr
 &=\sum_m  |\phi_m|^2-
    \int_M\ts  {d^d\k\over (2\pi)^d}\, {2\over\beta} \log \left[
   { \cosh ( {\beta\over2}\sqrt{ e_\k^2+ |\Phi_{\k}|^2} ) \over 
   \cosh {\beta\over2} e_\k } \right] \>=\> V(\phi) &(\III.52) \cr}$$
Let $S^{4\ell+1}=\{\phi\in \Bbb C^{2\ell +1}\>|\> \sum_m |\phi_m|^2=1\}$. Since 
 $U\!(\!R)$ leaves $S^{4\ell+1}$ invariant, $S^{4\ell+1}$ can be
   written as the union 
 of disjoint orbits, 
$$S^{4\ell+1}=\cup_{[\alpha]\in \O} [\alpha]$$
where $[\alpha]=\{ U\!(\!R)\alpha\>|\> R\in SO(3)\}$ is the orbit of $\alpha\in 
  S^{4\ell+1}$ under the action of $U\!(\!R)$ and $\O$  
 is the set of all orbits. If one chooses a fixed representant $\alpha$ in 
 each orbit $[\alpha]$, that is, if one chooses a fixed section $\sigma:\O\to
S^{4\ell+1}$, $[\alpha]\to\sigma_{[\alpha]}$  with $[\sigma_{[\alpha]}]=[\alpha]$, 
  every $\phi\in \Bbb C^{2\ell +1}$ can be uniquely 
 written as 
$$\phi=\rho\, U\!(\!R)\sigma_{[\alpha]}\>,\;\;\;\; \rho=\|\phi\|\ge 0,\;\;
   \ts \alpha={\phi\over 
   \|\phi\|},\;\;
   \;[\alpha]\in\O \;\hbox{and}\;  R\in SO(3)/I_{[\alpha]}$$
where $I_{[\alpha]}=I_{[\alpha]}^\sigma=\{S\in SO(3)\>|\>U(S)\sigma_{[\alpha]}
  =\sigma_{[\alpha]}\}$ 
 is the isotropy subgroup of 
  $\sigma_{[\alpha]}$. 
Let 
$$\ts \int_{\Bbb R^{4\ell +2}} \pro_m du_m dv_m\>f(\phi)=
 \int_{\Bbb R^+} D\!\rho \int_\O D[\alpha] \int_{[\alpha]} D\!R \>
    f\bigl( \rho\, U\!(\!R)\sigma_{[\alpha]}  \bigr)$$
be the integral in (III.51)  over $\Bbb R^{4\ell +2}$ in the new coordinates. That is, 
 for example, $D\!\rho=\rho^{4\ell +1}d\rho$. In the new coordinates  
$$  |\Phi_\p|^2   
   =  \lambda_\ell
   \rho^2 \Bigl|\sum_m \left(\overline{U\!(\!R)\sigma_{[\alpha]} }\right)_m 
    Y_{\ell m}(\p')\Bigr|^2  
    =\lambda_\ell
   \rho^2 \Bigl|\sum_m \bar\sigma_{[\alpha],m} Y_{\ell m}\left(R^{-1}\p'\right)
    \Bigr|^2   $$
such that 
$$\ts {-a_{-p}\over a_pa_{-p}+|\Phi_\p|^2}={-a_{-p}\over a_pa_{-p}+
 \lambda_\ell
   \rho^2 |\su_m \bar\sigma_{[\alpha],m} Y_{\ell m}\left(R^{-1}\p'\right)|^2   }
   \equiv f(\rho,[\alpha],R^{-1}\p)$$
Since $ V(\phi)=      
    V(\rho,[\alpha])$    is independent of $R$, one obtains 
$$\eqalignno{  {\ts{1\over\kappa}}\la \bar\psi_{p\sigma}\psi_{p\sigma}\ra&=  
 {\int   
    {-a_{-p}\over a_pa_{-p}+ |\Phi_{\p}|^2 }\, 
    e^{-\kappa V(\phi)} \ds \pro_{m=-\ell}^\ell du_m
       dv_m \over  \int e^{-\kappa V(\phi)} \ds \pro_{m=-\ell}^\ell du_m
       dv_m } \cr
  & =  {\int_{\Bbb R^+}D\rho \int_{\O} D[\alpha] \int_{[\alpha] } D\!R \; 
    f(\rho,[\alpha],R^{-1}\p)\; e^{-\kappa V(\rho,[\alpha])}
       \over 
   \int_{\Bbb R^+}D\rho \int_{\O} D[\alpha] \int_{[\alpha] } D\!R \;
     e^{-\kappa V(\rho,[\alpha])}   } \cr
 &=  { \int_{\Bbb R^+}D\rho \int_\O D[\alpha]\;{\rm vol}([\alpha]) \;
    { \int_{[\alpha] }D\!R
     f(\rho,[\alpha],R^{-1}\p)\over \int_{[\alpha] }D\!R  }\;
       e^{-\kappa V(\rho,[\alpha])}    \over 
   \int_{\Bbb R^+}D\rho \int_\O D[\alpha]\;{\rm vol}([\alpha]) \;
      e^{-\kappa V(\rho,[\alpha])} } &(\III.53) \cr}$$
It is plausible to assume that at the global minimum of $V(\rho,[\alpha])$ $\rho$ is 
uniquely determined, say $\rho_0$. Let $\O_{\rm min}\subset\O$ be the set of all 
 orbits at which $V(\rho_0,[\alpha])$ takes its global minimum. Then in the infinite 
 volume limit (III.53) becomes 
$$\eqalignno{ \lim_{\kappa\to\infty} 
   {\ts{1\over\kappa}}\la \bar\psi_{p\sigma}\psi_{p\sigma}\ra&=  
 { \int_{\O_{\rm min}} D[\alpha]\;{\rm vol}([\alpha]) \;
    { \int_{[\alpha] }D\!R\>
     f(\rho,[\alpha],R^{-1}\p)\over \int_{[\alpha] }D\!R  }   \over 
   \int_{\O_{\rm min}} D[\alpha]\;{\rm vol}([\alpha])   } &(\III.54) \cr}$$
Consider the quotient of integrals in the numerator of (III.54). Since 
$$\eqalignno{  f(\rho,[\alpha],R^{-1}\p)&=f\Bigl(
   \rho^2 \bigl|\su_m \bar\sigma_{[\alpha],m} Y_{\ell m}\left(R^{-1}\p'\right)
    \bigr|^2 \Bigr)  \cr
 &=f\Bigl( \rho^2 \bigl|\su_m \bigr(\overline{U(S)\sigma_{[\alpha]}} \bigr)_m
     Y_{\ell m}\left(R^{-1}\p'\right)   \bigr|^2 \Bigr)  \cr
 &=f\Bigl( \rho^2 \bigl|\su_m  \bar\sigma_{[\alpha],m}
     Y_{\ell m}\left((RS)^{-1}\p'\right)   \bigr|^2 \Bigr) =
     f(\rho,[\alpha],(RS)^{-1}\p) \cr}$$
for all $S\in I_{[\alpha]}$, one has, since $[\alpha]\simeq SO(3)/I_{[\alpha]}$ 
$$\eqalignno{  { \int_{[\alpha] }D\!R\>
     f(\rho,[\alpha],R^{-1}\p)\over \int_{[\alpha] }D\!R  } &= 
   { \int_{SO(3)/I_{[\alpha]} }D\!R\>
     f(\rho,[\alpha],R^{-1}\p) \int_{I_{[\alpha]}} DS 
     \over \int_{SO(3)/I_{[\alpha]} }D\!R \int_{I_{[\alpha]}} DS } \cr
 &= { \int_{SO(3)/I_{[\alpha]} }D\!R\int_{I_{[\alpha]}}DS\>
     f(\rho,[\alpha],(RS)^{-1}\p)   
     \over \int_{SO(3)/I_{[\alpha]} }\int_{I_{[\alpha]}} D\!R\> DS } \cr
 &={ \int_{SO(3)} DR\> f(\rho,[\alpha],R^{-1}\p)  \over \int_{SO(3)} D\!R} \cr
 &={ \int_{S^2} d\Omega(\x) \int_{SO(3)_{\x\to\p}} D\!R\>
      f(\rho,[\alpha],R^{-1}\p)  \over
     \int_{S^2} d\Omega(\x)  \int_{SO(3)_{\x\to\p}} D\!R} \cr
  &={ \int_{S^2} d\Omega(\x) \>  f(\rho,[\alpha],\x) \int_{SO(3)_{\x\to\p}} D\!R\>
      \over \int_{S^2} d\Omega(\x)  \int_{SO(3)_{\x\to\p}} D\!R}&(\III.55)  \cr}$$
 where 
   $SO(3)_{\x\to\p}=\{R\in SO(3)\>|\> R\x=\p\}$.  If one assumes 
 that $D\!R$ has the usual
invariance properties of the Haar measure, then $\int_{SO(3)_{\x\to\p}} D\!R$ 
does not depend on $\x$ such that it cancells out in (III.55).  Then (III.54) 
  gives 
$$\eqalignno{ \lim_{\kappa\to\infty} 
   {\ts{1\over\kappa}}\la \bar\psi_{p\sigma}\psi_{p\sigma}\ra&=  
 { \int_{\O_{\rm min}} D[\alpha]\;{\rm vol}([\alpha]) \;
    { \int_{S^2 }d\Omega(\x)\>
     f(\rho,[\alpha],\x)\over \int_{S^2 }d\Omega(\x)  }   \over 
   \int_{\O_{\rm min}} D[\alpha]\;{\rm vol}([\alpha])   } &(\III.56) \cr}$$
Now, since the effective potential, which is constant on $O_{\rm min}$, 
   may be written as 
$$V(\rho,[\alpha])=\int_{S^2}\ts {d\Omega(\x)\over4\pi} \> G\Bigl( 
   \rho^2 \bigl|\su_m \bar\sigma_{[\alpha],m} Y_{\ell m}\left(\x\right)\bigr|^2
   \Bigr)$$
with $G(X)=\rho^2-\int\ts{dk\, k^2\over 2\pi^2}\> 
   \log\left[ {\cosh({\beta\over2}\sqrt{e_k^2+\lambda_\ell X}) \over 
   \cosh{\beta\over2}e_k}\right] $, it is plausible to assume that also 
$$ { \int_{S^2 }d\Omega(\x)\>
     f(\rho,[\alpha],\x)\over \int_{S^2 }d\Omega(\x)  } = 
  \int_{S^2}\ts {d\Omega(\x)\over4\pi} \ts {ip_0+e_\p\over p_0^2+e_\p^2+
   \lambda_\ell
   \rho^2 |\su_m \bar\sigma_{[\alpha],m} 
      Y_{\ell m}(x)|^2}$$
is constant on $O_{\rm min}$. In that case also the integrals over $O_{\rm min}$ 
 in (III.56) cancel out and the theorem is proven $\blacksquare$  
\bigskip
\bigskip
\bigskip
\goodbreak
\magnification=\magstep1
\font\gross=cmbx12 scaled \magstep0
\font\mittel=cmbx10 scaled \magstep0
\font\Gross=cmr12 scaled \magstep2
\font\Mittel=cmr12 scaled \magstep0
\overfullrule=0pt
\def\Bbb#1{{\bf #1}}
%\def\blacksquare{\bullet}
%\baselineskip=12pt
%\nopagenumbers
\def\up{\uparrow}
\def\down{\downarrow}
\def\k{{\bf k}}

\def\p{{\bf p}}
\def\q{{\bf q}}
\def\x{{\bf x}}
\def\y{{\bf y}}
\def\I{{\rm I}}
\def\II{{\rm II}}
\def\III{{\rm III}}
\def\IV{{\rm IV}}
\def\ts{\textstyle}
\def\ds{\displaystyle}
\def\tr{\Delta}
\def\la{\langle}
\def\ra{\rangle}
\def\pro{\mathop\Pi}
\def\1cm{\hskip 1cm}
\def\1k{{\textstyle{1\over\kappa}}}
\def\db#1{{\ts{d^d\k\over (2\pi)^d}}}
\def\vp{\varphi}
\def\sl{g}

\def\U{{\cal U}}
\def\V{{\cal V}}
\def\O{{ O}}
\def\su{\mathop{\Sigma}}
\def\u{\underline}
\def\Eta#1{\u\zeta_{\phantom{.}\!#1}}
\def\Pf{{\rm Pf}}
\def\ep{\epsilon}
\def\A{{\rm A}}

%\pageno=33

\noindent {\gross IV. Solution with Delta Function Interaction and }
\par\noindent{\gross $\phantom{IV.I}$   Its Perturbation Theory} 
\bigskip
\bigskip
\noindent{\mittel IV.1 Solution with Delta Function}
\bigskip
We consider the model  
$$Z(\beta,L,\{s_{k,\sigma}\})=\int e^{-\U(\psi,\bar\psi)+{1\over\kappa}
    \sum_{k,\sigma} s_{k,\sigma}\bar\psi_{k,\sigma}\psi_{k,\sigma} }
    d\mu_C(\psi,\bar\psi)  \eqno (\IV.1)$$
where 
$$\eqalignno{   \U(\psi,\bar\psi)&={\ts {\lambda\over \kappa^3}}
   \sum_{\sigma,\tau\in\{\up,\down\}}
    \sum_{k,p,q}  \bigl[\delta_{k,p}+\delta_{k,q-p}
  +\delta_{q,0} \bigr] \>\bar\psi_{k\sigma}
   \bar\psi_{{q}-k\tau}\psi_{p\sigma} 
  \psi_{{q}-p\tau} \cr
 &={\ts {\lambda\over \kappa^3}}
    \sum_{k,p,q}  \bigl[\delta_{k,p}+\delta_{k,q-p}
  +\delta_{q,0} \bigr] \>\bar\psi_{k\up}
   \bar\psi_{q-k\down}\psi_{p\up} 
  \psi_{q-p,\down} &(\IV.2)  \cr}$$
It is useful to explicitly cancel the $\bar\psi_\up\bar\psi_\up\psi_\up\psi_\up$ and 
 $\bar\psi_\down\bar\psi_\down\psi_\down\psi_\down$ 
terms before directly applying Theorem II.1 because then the integral representation 
becomes 4 dimensional instead of 5 dimensional. Both representations are of course 
 equivalent. One obtains 
$$\la \bar\psi_{p\sigma}\psi_{p'\sigma}\ra=\beta L^d \delta_{p,p'} \, 
    { \int\ts { {\partial\over \partial s_{p\sigma}}{\rm Pf}S_p \over {\rm Pf}S_p} \>
     e^{-\kappa V_\beta(z,x,\rho) } dv\,dw\,
     xdx\, \rho d\rho \over \int  e^{-\kappa V_\beta(z,x,\rho) } dv\,dw\,
     xdx\, \rho d\rho }  \eqno (\IV.3)$$ 
where $z=v+iw$,
$$\eqalignno{ \Pf S_k=\,&(A_{k\up}A_{-k\down}+\lambda\rho^2)(A_{-k\up}A_{k\down}+
  \lambda\rho^2)+(A_{k\up}A_{k\down}+A_{-k\up}A_{-k\down}) \lambda x^2 \cr
  &+2\lambda^2x^2\rho^2+\lambda^2x^4 &(\IV.4) \cr}$$
$$A_{k\up}=a_k-s_{k\up}-\sqrt\lambda\, z,\;\;\;\;\; 
   A_{k\down}=a_k-s_{k\down}-\sqrt\lambda\,\bar z \eqno $$
as before $a_k=ik_0-e_\k$ and the effective potential is given by 
$$V_\beta(z,x,\rho)=|z|^2+x^2+\rho^2-{\ts {1\over \beta L^d}} \sum_{\k\in M_\omega} 
    \log\pro_{k_0>0}\ts {{\rm Pf}S_k\over a_k^2a_{-k}^2} \eqno (\IV.5) $$
The product over $k_0$ is computed in Lemma A1 in the appendix. One finds  
$$\eqalignno{ V_\beta(&v,w,x,\rho)= & (\IV.6)  \cr
  & v^2+w^2+x^2+\rho^2-{\ts {1\over L^d}} \sum_{\k\in M_\omega} 
   \ts {1\over\beta} \log\left[ 
   { \cosh^2\left({\beta\over2}
    \sqrt{(e_\k+\sqrt\lambda v)^2+\lambda\rho^2}\right)-\,
   \sin^2\left({\beta\over2}\sqrt{\lambda(w^2+x^2)}\right) \over 
    \cosh^2{\beta\over2}e_\k  } \,e^{-\beta\sqrt\lambda v}  \right] \cr}$$
The $z$ variable sums up forward contributions, $x=|\xi|$ sums up exchange
contributions and $\rho=|\phi|$ collects the BCS contributions. Pure BCS is given 
by $z=x=0$ and (IV.6) coincides with (I.47) or (III.10) (for $r=0$).
\par\goodbreak
To compute the infinite volume limit of (IV.3) one has to find the global minimum 
 of the effective potential (IV.6). 
\par
Consider first the case of an attractive coupling $\lambda=g^2>0$. 
   To make (IV.6) small, 
 the numerator in the logarithm has to be large. Hence for $\lambda$ positive 
$w^2+x^2$ has to be zero. This gives 
$$\ts { {\partial\over \partial s_{p\up}}{\rm Pf}S_p \over {\rm Pf}S_p}= 
  {a_{-p}-gv\over [a_p-gv][a_{-p}-gv]+g^2\rho^2}  \eqno (\IV.7)$$
As in section III.1 $\rho$ is positive for sufficiently small $T={1\over\beta}$ and 
 $v$ may or may not be nonzero to renormalize $e_\p$. Thus for attractive coupling the 
two point function becomes 
$$\lim_{L\to\infty} 
   \ts {1\over\kappa} \la \bar\psi_{p\up}\psi_{p\up}\ra=-{ip_0+e_{\p g}\over 
  p_0^2+e_{\p g}^2+|\Delta|^2}  \eqno  (\IV.8)$$
if $e_{\p g}=e_\p+gv_0$, $|\Delta|^2=\lambda\rho_0^2$ and $v_0$, $\rho_0$ are the 
 ($\lambda$ dependent) values where $V$ takes its global minimum. 
\par
Now consider a repulsive coupling $\lambda=-g^2<0$. In that case the numerator in 
the logarithm in (IV.6) reads $\cosh^2({\beta\over2}
    \sqrt{(e_\k+igv)^2-g^2\rho^2}\,)+\,
   \sinh^2({\beta\over2}g\sqrt{w^2+x^2}\,)$ and the value  
$\rho=0$ is favourable. Furthermore $v=0$ seems favourable. In that case 
$$\ts { {\partial\over \partial s_{p\up}}{\rm Pf}S_p \over {\rm Pf}S_p}= 
  {a_{p}-gw\over a_p^2-g^2(w^2+x^2)}  \eqno (\IV.9)$$
and the two point function becomes, if $\delta_g^2=g^2(w^2+x^2)\ge 0$
$$\lim_{L\to\infty} 
   \ts {1\over\kappa} \la \bar\psi_{p\up}\psi_{p\up}\ra=
  {a_{p}\over a_p^2-\delta_g^2}={1\over2}\left[ {1\over ip_0-e_\p-\delta_g}+
   {1\over ip_0-e_\p+\delta_g}\right] \eqno (\IV.10)$$
since the effective potential is even in $w$ and therefore the $w$ term in (IV.9) 
 cancels. At zero temperature, this results in a momentum distribution 
$$\eqalignno{ n_\k&=\lim_{L\to\infty\atop \beta\to\infty } 
  {\ts{1\over L^d}} \la a_\k^+a_\k\ra_{\beta,L} =\lim_{\ep\to 0\atop \ep<0} 
    \int_{\Bbb R}\ts {dk_0\over 2\pi} 
   \ts {1\over2}\left[ 
   {1\over e^{ik_0\ep}( ik_0-e_{\k })-\delta_g}
   + {1\over e^{ik_0\ep}( ik_0-e_{\k })+\delta_g}\right] \cr
  &=\ts {1\over2}\Bigl[
    \chi(e_{\k }-\delta_g<0)+\chi(e_{\k }+\delta_g<0)\Bigr]
  =\cases{ 1&if $e_{\k }<-\delta_g$ \cr {1\over2} &if $-\delta_g< e_{\k }< \delta_g$\cr 
   0&if $\delta_g<e_{\k }$ \cr}  &(\IV.11)
   \cr}$$
Thus, if $\delta_g$ is nonzero, the exchange and forward contributions lead to a 
 splitting of the Fermi surface. A nonzero $\delta_g$ can be achieved by 
 making the coupling $g^2$ sufficiently big: Let 
  $K={|S^{d-1}|\over (2\pi)^d} mk_F^{d-2}$ $\sim k_F^d/\mu$ and let 
$$V_\beta(y)=y^2-\int_{M_\omega} \ts {d^d\k\over (2\pi)^d} 
  \> {1\over\beta} \log\left[ 
   { \cosh^2({\beta\over2}e_\k)+\,
   \sinh^2({\beta\over2}gy) \over 
    \cosh^2{\beta\over2}e_\k  }  \right] \eqno(\IV.12)$$ 
Then 
$$\lim_{\beta\to\infty} V_\beta(y)= \ts ( 1-{K}g^2)y^2
  +{K}[gy-\omega]_+^2    \eqno(\IV.13)$$ 
where $[x]_+=x$ for positive $x$ and 0 otherwise. 
 The global minimum of $V_\infty$ is at $y_{\rm min}$ where 
$$y_{\rm min} =\cases{ 0& if $g^2<1/K$ \cr K\omega g& if $g^2>1/K$ \cr}
  \eqno (\IV.14)$$  
\bigskip
\bigskip
%
%
% IV.2 PERTURBATION THEORY
%
%
\noindent{\mittel IV.2 Perturbation Theory}
\bigskip
In this section we consider the perturbation theory of the model 
$$Z_{\rm ex / \rm BCS}(\beta,L)=\int e^{-\V_{\rm ex/ \rm BCS}(\psi,\bar\psi) }
    d\mu_C(\psi,\bar\psi)  \eqno (\IV.15)$$
For BCS, one obtains a power series in $C(k)C(-k)$ and for $\V_{\rm ex}$ one gets 
 a power series in $C(k)^2$. 
\par
We will find that the linked cluster theorem, $\log Z$ is given by the sum of all 
 connected diagrams, cannot be applied to (IV.15). More specifically, whereas 
the series for  $Z(\lambda)$, the sum of all diagrams, converges for sufficiently 
 small $\lambda$ (for finite $\beta$ and $L$) with $Z(0)=1$ which implies that also 
 $\log Z$ is analytic for $\lambda$ sufficiently small, the sum of all connected 
 diagrams has radius of convergence zero (for finite $\beta$ and $L$).  
\par
  We start with BCS. First we show how the 
integral representation (III.12) (for $r=0$)
$$Z_{\rm BCS}= {2\kappa} \int e^{-\kappa\Bigl( \rho^2-{1\over L^d}\sum_\k
    \log\left[{ \cosh({\beta\over2}\sqrt{e_\k^2+\lambda\rho^2})
     \over \cosh{\beta\over2}e_\k} \right] \Bigr) } \rho \>d\rho \eqno(\IV.16)  $$
is obtained 
by direct summation of the diagrams without making a Hubbard Stratonovich 
 transformation. That is, without using the identity 
$$ e^{ -{\ts{\lambda\over\kappa^3}}
    \sum_{k,p} \psi_{k\up}\psi_{-k\down}
    \bar\psi_{p,\up}\psi_{-p\down}  }= 
  \int e^{ 
    {i\phi\>g
     \1k \sum_{k} \psi_{k\up}\psi_{-k\down} +i\bar\phi \>
   g \1k
    \sum_{k}  \bar\psi_{k\up}\bar\psi_{-k\down}} 
      }  {\ts{\kappa\over\pi}} \>e^{-\kappa|\phi|^2}dudv   $$
\par
One has 
$$\eqalignno{Z_{\rm BCS}&= \int e^{-{\lambda\over\kappa^3}
      \sum_{k,p}\psi_{k\up}\psi_{-k\down}
      \bar\psi_{p\up} \bar\psi_{-p\down} } d\mu_C(\psi,\bar\psi)&(\IV.17)   \cr
  &= \sum_{n=0}^\infty { \left( {\lambda\over\kappa^3}\right)^n 
  \over n!} \sum_{k_1,\cdots,k_n\atop p_1,\cdots,p_n} 
  \det\left[\kappa \delta_{k_i,p_j}C(k_i)\right]_{1\le i,j\le n}  
 \det\left[\kappa \delta_{k_i,p_j}C(-k_i)\right]_{1\le i,j\le n} \cr
 &= \sum_{n=0}^\infty  {\ts \left( {\lambda\over\kappa}\right)^n } 
  \sum_{\pi\in S_n}\epsilon_\pi 
     \sum_{k_1,\cdots,k_n} C(k_1)C(-k_1)\cdots C(k_n)C(-k_n)
    \>\delta_{k_1,k_{\pi 1}}\cdots \delta_{k_n,k_{\pi n}}  \cr}$$
This is the expansion into Feynman diagrams. It can be summed up by collecting the 
fermion loops: 
\par
Say that the permutation $\pi$ is of type $t(\pi)=1^{b_1}\cdots n^{b_n}$ if the 
 decomposition into disjoint cycles contains $b_r$ $r$-cycles for  $1\le r\le n$. 
Necessarily one has $1b_1+\cdots +nb_n=n$. The number of permutations 
which have $b_r$ $r$-cycles for  $1\le r\le n$ is 
$${n!\over b_1!\cdots b_n!\>1^{b_1}\cdots n^{b_n} }$$
The sign of such a permutation is given by 
$$\epsilon_\pi= (-1)^{(1-1)b_1+(2-1)b_2+\cdots+(n-1)b_n}=
   (-1)^{n-\sum_{r=1}^n b_r} $$
Therefore one obtains 
$$\eqalignno{ Z  
 &=\sum_{n=0}^\infty {\ts  \left( {\lambda\over\kappa}\right)^n } 
   \sum_{b_1,\cdots,b_n=0\atop 1b_1+\cdots +nb_n=n}^n 
    {n!\over b_1!\cdots b_n!\>1^{b_1}\cdots n^{b_n} }\>
     (-1)^{n-\sum_{r=1}^n b_r} \times \cr
 &\phantom{=\sum_{n=0}^\infty {\ts  \left( -{\lambda\over\kappa}\right)^n } 
   \sum_{b_1,\cdots,b_n=0\atop 1b_1+\cdots +nb_n=n}^n} 
   \biggl\{ \sum_k[C(k)C(-k)]^1\biggr\}^{b_1}
   \cdots \biggl\{ \sum_k[C(k)C(-k)]^n\biggr\}^{b_n}  \cr
 &=\sum_{n=0}^\infty 
   \sum_{b_1,\cdots,b_n=0\atop 1b_1+\cdots +nb_n=n}^n 
    n! \>  \prod_{r=1}^n {1\over b_r!}\biggl\{ -{\ts{1\over r}}
   \sum_k\left[{\ts  -{\lambda\over\kappa} }
     C(k)C(-k)\right]^r\biggr\}^{b_r} &(\IV.18) \cr}$$
The only factor which prevents us from an  explicit summation of 
 the above series is the $n!$. Therefore we substitute 
$$n!=\int_0^\infty e^{-x} x^n dx$$
and obtain 
$$\eqalignno{Z &=\int_0^\infty e^{-x}
     \sum_{n=0}^\infty 
   \sum_{b_1,\cdots,b_n=0\atop 1b_1+\cdots +nb_n=n}^n 
   \prod_{r=1}^n {1\over b_r!} \biggl\{ -{\ts{1\over r}}\sum_k
    \left[ {\ts -{\lambda\over\kappa}x}\> C(k)C(-k)\right]^r\biggr\}^{b_r}dx  \cr
 &=\int_0^\infty e^{-x}\prod_{r=1}^\infty \sum_{b_r=0}^\infty 
   {1\over b_r!} \biggl\{ -{\ts{1\over r}}\sum_k
    \left[ {\ts -{\lambda\over\kappa}x}\> C(k)C(-k)\right]^r\biggr\}^{b_r}dx  \cr
 &=\int_0^\infty e^{-x} e^{ -\sum_k\sum_{r=1}^\infty {\ts{1\over r}}
    \left[ {\ts -{\lambda\over\kappa}x}\> C(k)C(-k)\right]^r } dx \cr
 &=\int_0^\infty e^{-x} e^{  \sum_k \log\left[ 1+
    {\ts {\lambda x\over\kappa}}\> C(k)C(-k)\right] } dx &(\IV.19) \cr
 &=2\kappa \int_0^\infty  e^{ -\kappa\Bigl(\rho^2 -{1\over\kappa} 
     \sum_k \log\left[ 1+ {\lambda \rho^2\over k_0^2+e_\k^2} \right]\Bigr) }
   \rho\,d\rho  \cr
 &=2\kappa \int_0^\infty  e^{ -\kappa\Bigl(\rho^2 -{1\over L^d} 
     \sum_\k{1\over\beta} 
    \log\left[ {\cosh({\beta\over2}\sqrt{e_\k^2+\lambda \rho^2}) \over 
     \cosh{\beta\over2}e_\k }  \right]^2\Bigr) }
   \rho\,d\rho  \cr}$$
which  coincides with (IV.16). In the last line we used (III.25) again. The case of an
exchange interaction is treated in the same way: 
$$\eqalignno{ Z_{\rm ex}&=  \int e^{-{\lambda\over\kappa^3}
      \sum_{k,p}\psi_{k\up}\psi_{p\down}
      \bar\psi_{p\up} \bar\psi_{k\down} } d\mu_C(\psi,\bar\psi)   \cr
 &= \sum_{n=0}^\infty  {\ts \left( {\lambda\over\kappa}\right)^n } 
  \sum_{\pi\in S_n}\epsilon_\pi 
     \sum_{k_1,\cdots,k_n} C(k_1)^2\cdots C(k_n)^2
    \>\delta_{k_1,k_{\pi 1}}\cdots \delta_{k_n,k_{\pi n}}  \cr
&=\sum_{n=0}^\infty 
   \sum_{b_1,\cdots,b_n=0\atop 1b_1+\cdots +nb_n=n}^n 
    n! \>  \prod_{r=1}^n {1\over b_r!}\biggl\{ -{\ts{1\over r}}
   \sum_k\left[{\ts   -{\lambda\over\kappa} }
     C(k)^2\right]^r\biggr\}^{b_r} &(\IV.20) \cr
 &=\int_0^\infty e^{-x} e^{ -\sum_k\sum_{r=1}^\infty {\ts{1\over r}}
    \left[ {\ts -{\lambda\over\kappa}x}\> C(k)^2\right]^r } dx \cr
  &=2\kappa \int_0^\infty  e^{ -\kappa\Bigl(y^2 -{1\over L^d} 
     \sum_\k{1\over\beta} 
    \log\left[ {\cosh^2({\beta\over2}e_\k)-\sin^2({\beta\over2}\sqrt\lambda\,y) \over 
     \cosh^2{\beta\over2}e_\k }  \right]^2\Bigr) }
   y\,d y &(\IV.21) \cr}$$
where in the last line Lemma A1 in the appendix has been used to compute the $k_0$ 
  product. Observe that 
 $\lim_{\beta\to\infty}{\ts {1\over \kappa}}\sum_k C(k)^{2r}=0$ 
for all $r\ge 1$ since $\int{dk_0\over 2\pi} {1\over (ik_0-e)^j}=0$ for all $j\ge 2$, 
but for  large repulsive coupling
 $\lim_{\beta\to\infty} {\ts{1\over\beta L^d}}
   \log Z_{\rm ex}=-V_\infty(y_{\rm min})>0$ 
by (IV.13,14). 
\smallskip
We now consider the linked cluster theorem. It states that, if the partition 
 function 
$$Z(\lambda)=\sum_{n=0}^\infty \lambda^n \sum_{G\in \Gamma_n} {\rm val}(G) 
    \eqno (\IV.22)$$
is given by a sum of  diagrams, $\Gamma_n$ being the set of all $n$'th order diagrams, 
then the logarithm 
$$\log Z(\lambda) =\sum_{n=0}^\infty \lambda^n \sum_{G\in \Gamma_n^c} {\rm val}(G) 
     \eqno (\IV.23)$$
is given by the sum of all connected diagrams. This theorem is easily illustrated 
for a quadratic perturbation. Namely, if 
$$\eqalignno{ Z&=\int
    e^{-{\lambda\over\kappa}\sum_{k\sigma}\bar\psi_{k\sigma}\psi_{k\sigma}}
     d\mu_C(\psi,\bar\psi)  \cr
  &=\sum_{n=0}^\infty { (-\lambda)^n\over n!}\sum_{k_1\cdots k_n\atop 
  \sigma_1\cdots \sigma_n}  \det[\delta_{\sigma_i,\sigma_j}\delta_{k_i,k_j}
    C(k_i)]_{1\le i,j\le n}  \cr
  &=\sum_{n=0}^\infty 
   \sum_{b_1,\cdots,b_n=0\atop 1b_1+\cdots +nb_n=n}^n 
      \prod_{r=1}^n {1\over b_r!}\biggl\{ -{\ts{1\over r}}
   \sum_k\left[{\ts  -{\lambda} }
     C(k)\right]^r\biggr\}^{b_r} &(\IV.24) \cr
  &=e^{\sum_k\log\left[ 1+\lambda C(k) \right]} &(\IV.25) \cr}$$
then the sum of all connected diagrams is obtained by summing all the terms with 
 $b_1=\cdots =b_{n-1}=0$ and $b_n=1$ in (IV.24). That is, one gets 
$$\sum_{n=1}^\infty {1\over 1!}\biggl\{ -{\ts{1\over n}}
   \sum_k\left[{\ts  -{\lambda} }
     C(k)\right]^n\biggr\}^{1}= \sum_k\log\left[ 1+\lambda C(k) \right] 
   \eqno (\IV.26) $$
which coincides with $\log Z$. Now consider $Z_{\rm BCS}$. The sum of all connected 
 diagrams is given by the sum of all the terms with 
 $b_1=\cdots =b_{n-1}=0$ and $b_n=1$ in (IV.18). That is, one obtains 
$$\sum_{n=0}^\infty 
    (-1)^{n+1}\left({\ts  {\lambda\over\kappa} }\right)^n n! \, {\ts{1\over n}}
   \sum_k\left[
     C(k)C(-k)\right]^n \eqno (\IV.27) $$
which has radius of convergence zero, at finite temperature and finite volume. 
However, $Z_{\rm BCS}(\lambda)$ has positive radius of convergence and 
 $Z_{\rm BCS}(0)=1$ which means that also $\log Z_{\rm BCS}(\lambda)$ is analytic 
 for sufficiently small (volume and temperature dependent) $\lambda$. That is, in this 
 case the linked cluster theorem does not apply since the right hand side of 
 (IV.23) is infinite. 
\par
We remark that we think that this is an artefact of the specific model at hand for the 
 following reason. Suppose for the moment that the $k$ sums in (IV.17,18) are finite 
with $N$ different values of $k$. Then $Z_{\rm BCS}(\lambda)$ is a polynomial 
 in $\lambda$ of degree (at most) $N$. In particular, the coefficients 
$$    \sum_{b_1,\cdots,b_n=0\atop 1b_1+\cdots +nb_n=n}^n 
   \prod_{r=1}^n {1\over b_r!}\biggl\{ -{\ts{1\over r}}
   \sum_k\left[ C(k)C(-k)\right]^r\biggr\}^{b_r}\;=\;0\;\;\;\;{\rm if}\;\;n>N\>.
  \eqno (\IV. 28) $$
That is, there are strong cancellations between fermion loops of different orders. 
 However, for $\V_{\rm int}\approx\V_{\rm BCS}$ (or forward or exchange), an $n$'th 
 order connected diagram is necessarily a single $n$'th order fermion loop such 
that there are no cancellations at all. In a more realistic model a connected diagram 
 contains fermion loops of different orders and cancellations are present. In fact, 
 a careful diagrammatic analysis [FT,FKLT,FST] shows that the renormalized sum of 
all connected diagrams of the two dimensional electron system with anisotropic 
 dispersion relation $e_\k$ (such that Cooper pairs are suppressed) 
  has positive radius of convergence (which is, in particular, 
 independent of volume and temperature). 
%
%
%  APPENDIX
%
%
\bigskip
\bigskip
\bigskip
\noindent {\gross Appendix}
\bigskip
\bigskip
\noindent{\bf Lemma A.1:} {\bf a)} {\it  Let $a_k=ik_0-e_\k$, $e_{-\k}=e_\k$ 
   and let $b,c,d$ be some
complex numbers. Then 
$$\eqalignno{ \prod_{k_0\in{\pi\over\beta}(2\Bbb Z+1)} 
  {\ts { a_ka_{-k}+ba_k+ca_{-k}+d\over a_ka_{-k} }}\>&:=\>\lim_{\epsilon\to 0\atop 
  \epsilon<0} \prod_{k_0\in{\pi\over\beta}(2\Bbb Z+1)} 
 {\ts  { a_ka_{-k}+be^{i\ep k_0}a_k+ce^{-i\ep k_0}a_{-k}+d\over a_ka_{-k} }} \cr
 &\>=\;\ts{\cosh{\beta\over2}\sqrt{e_\k^2+\omega_+}\over \cosh{\beta\over2}e_\k}\,
 \, {\cosh{\beta\over2}\sqrt{e_\k^2+\omega_-}\over \cosh{\beta\over2}e_\k}\>
    e^{{\beta\over2}(b+c)} &(\A.1) \cr}$$
where $\omega_{\pm}={p\pm\sqrt{p^2-4q}\over2}$ with $p=(b+c-e_\k)^2-e_k^2 
  -4bc+2d$ and $q=4bce_\k^2-2(b+c)e_\k d+d^2$. 
In particular, for $b=c$ and $d=c^2$ 
$$\prod_{k_0\in{\pi\over\beta}(2\Bbb Z+1)} {\ts 
  \left( 1+{c\over a_k}\right)}\>:=\>\lim_{\epsilon\to 0\atop 
  \epsilon<0} \prod_{k_0\in{\pi\over\beta}(2\Bbb Z+1)} \ts 
  \left( 1+e^{-i\ep k_0}{c\over a_k}\right)\>=\>  {1+e^{-\beta(e_\k-c)}\over 
    1+e^{-\beta e_\k} } \eqno (\A.2) $$ 
{\bf b)} Let $A_{k\up}=a_k-\sqrt\lambda\,z$,  $A_{k\down}=a_k-\sqrt\lambda\,\bar z$, 
and let 
$$\eqalignno{ \Pf S_k&=\Pf S_k(a_k,a_{-k})=(A_{k\up}A_{-k\down}+\lambda\rho^2)
  (A_{k\down}A_{-k\up}+\lambda\rho^2) & (\A.3) \cr
 &\phantom{=\Pf S_k(a_k,a_{-k})=(}  
    +(A_{k\up} A_{k\down}+A_{-k\up}A_{-k\down}+2\lambda\rho^2) \lambda x^2 
    +\lambda^2x^4 \cr} $$
Then 
$$\eqalignno{ \prod_{k_0\in{\pi\over\beta}(2\Bbb Z+1)\atop k_0>0}
  {\ts  { \Pf S_k(a_k,a_{-k})\over 
   a_k^2 a_{-k}^2}}\>&:=\>\lim_{\epsilon\to 0\atop 
  \epsilon<0} \prod_{k_0\in{\pi\over\beta}(2\Bbb Z+1)\atop k_0>0} 
 \ts { \Pf S_k(e^{i\ep k_0}a_k,e^{-i\ep k_0} a_{-k})\over 
   a_k^2 a_{-k}^2}  \cr
 &\>= \ts { \cosh^2\left({\beta\over2}
    \sqrt{(e_\k+\sqrt\lambda\,v)^2+\lambda\rho^2}\,\right)-
  \sin^2\left({\beta\over2}\sqrt{\lambda(w^2+ x^2)}\right)  \over 
    \cosh^2{\beta\over2}e_\k  } \,e^{-\beta \sqrt\lambda v}    &(\A.4)   \cr}$$   }
\bigskip
\noindent{\bf Proof:} We first explain why the definition of the product given in 
 the lemma is the right one. 
\par
Consider the quadratic perturbation 
$$H=H_0+\lambda N={\ts {1\over L^d}} \sum_{\k\sigma} e_\k \, a_{\k\sigma}^+ 
  a_{\k\sigma}+{\ts {\lambda\over L^d}}\sum_{\k\sigma} a_{\k\sigma}^+ 
  a_{\k\sigma}$$
One finds by explicit computation of the trace 
$$Z(\lambda,\beta)={\ts { Tr\, e^{-\beta(H_0+\lambda N)}\over Tr\, e^{-\beta H_0}}}
  =\pro_{\k\sigma}\ts {1+e^{-\beta(e_\k+\lambda)}\over 1+e^{-\beta e_\k}} $$
On the other hand, by explicit summation of the perturbation series or
equivalently, by performing the fermionic functional integral, one obtains 
$$\eqalignno{ Z(\lambda,\beta)&=\int e^{{\lambda\over\beta L^d} \sum_{k\sigma} 
  \bar\psi_{k\sigma}  \psi_{k\sigma} } d\mu_C \cr 
 &=\pro_{k\sigma} {\ts {\beta L^d\over ik_0-e_\k}} \int 
   e^{{\lambda\over\beta L^d} \sum_{k\sigma} 
  \bar\psi_{k\sigma}  \psi_{k\sigma} }
  e^{-{1\over \beta L^d}\sum_{k\sigma}(ik_0-e_\k)\bar\psi_{k\sigma} 
    \psi_{k\sigma}} \pro_{k\sigma} d\psi_{k\sigma} d\bar\psi_{k\sigma}  \cr
 &=\pro_{k\sigma} {\ts {ik_0-e_\k-\lambda\over ik_0-e_\k}}= 
   \pro_{\k\sigma} \pro_{k_0}\ts \left( 1-{\lambda\over a_k}\right)&(\A.5) \cr}$$
In the perturbation expansion the product in the last equation results from 
$$\pro_{k\sigma}{\ts \left( 1-{\lambda\over a_k}\right)}= 
  e^{\sum_{k\sigma} \log\left( 1-{\lambda\over a_k}\right)} =
   e^{-\sum_{r=1}^\infty {\lambda^r\over r} \sum_{k\sigma}{1\over {a_k}^r}}$$
The term for $r=1$ requires special attention since $\sum_{k_0} {1\over 
  ik_0-e_\k}$ is not absolutely convergent. This sum enters the perturbation 
 series as the propagator 
$$C(x_0-x_0',\x-\x')=\ts {Tr\, e^{-\beta H_0} T\psi(x_0\x)\psi^+(x_0'\x')\over 
    Tr\,e^{-\beta H_0} }$$
evaluated at $x_0-x_0'=0$. Here $T\psi(x_0\x)\psi^+(x_0'\x')$ is
defined to be 
 $\psi(x_0\x)\psi^+(x_0'\x')$ if $x_0\ge x_0'$ and $-\psi^+(x_0'\x')\psi(x_0\x)$ 
 if $x_0<x_0'$. By explicit computation one finds that for $x_0\ne x_0'$ 
$$\eqalignno{ C(x_0-x_0',\x-\x')&={\ts {1\over L^d}}\sum_\k 
  e^{i\k(\x-\x')} e^{-e_\k(x_0-x_0')}\left[ \theta_\beta(-e_\k)\theta(x_0'-x_0)-
  \theta_\beta(e_\k)\theta(x_0-x_0')\right]  \cr
 &={\ts {1\over \beta L^d}} \sum_{\k,k_0} e^{i\k(\x-\x')-ik_0(x_0-x_0')}
   \ts {1\over ik_0-e_\k} \cr}$$
and for $x_0=x_0'$
$$\eqalignno{ C(0,\x-\x')&={\ts {1\over L^d}} \sum_\k e^{i\k(\x-\x')}
    \theta_\beta(-e_\k) 
 =\lim_{\ep\to 0\atop \ep<0} {\ts {1\over \beta  L^d}}\sum_{\k,k_0} 
   e^{i\k(\x-\x')-ik_0\ep} \ts {1\over ik_0-e_\k} \cr}$$
Here $\theta$ is a step function being 1 for positive arguments and 
 $\theta_\beta(-v)={e^{-\beta v}\over 1+e^{-\beta v}}$ is an approximate step 
function. Therefore the $k_0$-sums have to be evaluated according to 
$$\sum_{k_0}{\ts {1\over ik_0-e_\k}}={\sum_{k_0}}^\ep{\ts {1\over ik_0-e_\k}}
  :=\lim_{\ep\to0\atop \ep<0} e^{-ik_0\ep}{\ts {1\over ik_0-e_\k}}=
  \beta \theta_\beta(-e_\k)=\beta\> \ts {e^{-\beta e_\k}\over 1+e^{-\beta e_\k}}$$
We emphasize this point because the following computation leads to a wrong 
 result 
$$\sum_{k_0}{\ts {1\over ik_0-e_\k}}={\sum_{k_0}}^{sym}{\ts {1\over ik_0-e_\k}}
  :={\ts{1\over2}} \sum_{k_0} \left[ {\ts {1\over ik_0-e_\k}}+
    {\ts {1\over -ik_0-e_\k}}\right] 
  =\sum_{k_0}{\ts {-e_\k\over k_0^2+e_\k^2}}=\ts -{\beta\over2}\>
     {1-e^{-\beta e_\k}\over 1+e^{-\beta e_\k}}$$
Therefore the product in (A.5) is found to be   
$$\eqalignno{ \pro_{k_0}{\ts\left(1-{\lambda\over ik_0-e_\k}\right)}&=
  {\pro_{k_0}}^\ep {\ts\left(1-{\lambda\over ik_0-e_\k}\right)}=
  e^{-\sum_{r=1}^\infty{\lambda^r\over r}
         \sum_{k_0}^\ep{1\over (ik_0-e_\k)^r} } \cr
  &=e^{-\sum_{r=1}^\infty{\lambda^r\over r}\sum_{k_0}^{sym} 
   {1\over (ik_0-e_\k)^r} }
  \; e^{\sum_{k_0}^{sym}{\lambda\over ik_0-e_\k}- 
   \sum_{k_0}^{\ep}{\lambda\over ik_0-e_\k} }  \cr 
 &={\pro_{k_0}}^{sym} {\ts\left(1-{\lambda\over ik_0-e_\k}\right)}\; 
 e^{-{\beta\over2}\lambda}=\pro_{k_0>0} {\ts {k_0^2+(e_\k+\lambda)^2 \over 
   k_0^2+e_\k^2 }\;e^{-{\beta\over2}\lambda}} \cr
 &\buildrel (III.25)\over = 
 \ts {\cosh{\beta\over2}(e_\k+\lambda)\over \cosh{\beta\over2}e_\k}\;
   e^{-{\beta\over2}\lambda} =\ts {1+e^{-\beta(e_\k+\lambda)}\over 
   1+e^{-\beta e_\k}} &(\A.6) \cr}$$
which proves (A.2).  
\par
In the general case one may proceed similarly 
  to obtain (using $a_k+a_{-k}=-2e_\k$) 
$$\eqalignno{ \pro_{k_0\in{\pi\over\beta}(2\Bbb Z+1)} &
  {\ts { a_ka_{-k}+b a_k+ca_{-k}+d\over a_ka_{-k} }} 
   = \lim_{\ep\to 0\atop \ep<0} 
    \pro_{ k_0} {\ts \left( 1+{be^{i\ep k_0}a_k+c
   e^{-i\ep k_0}a_{-k}+d\over a_ka_{-k} }\right)}=:{\pro_{ k_0}}^\ep \cdots\cr
 &={\pro_{ k_0}}^{ sym}  {\ts \left( 1+{ba_k+c
       a_{-k}+d\over a_ka_{-k} }\right)}\> e^{b\sum_{k_0}^\ep-
   \sum_{k_0}^{ sym}{1\over a_{-k}} +c\sum_{k_0}^\ep-
   \sum_{k_0}^{ sym}{1\over a_{k}} }  \cr
 &=\pro_{k_0>0} {\ts { (a_ka_{-k}+b a_k+ca_{-k}+d)
    (a_ka_{-k}+b a_{-k}+ca_{k}+d)\over (a_ka_{-k})^2 }} \>e^{{\beta\over2}(b+c)} \cr
 &=\pro_{k_0>0} {\ts { (a_ka_{-k})^2 
  +a_ka_{-k}(-2e_\k(b+c)+ 2d+b^2+c^2)+bc(a_k^2+a_{-k}^2)
   -2e_\k d(b+c)+d^2
   \over (a_ka_{-k})^2 }} \>e^{{\beta\over2}(b+c)} \cr}$$
Since $a_k^2+a_{-k}^2=-2k_0^2+2e_\k^2=-2a_ka_{-k}+4e_\k^2$ one obtains 
$$\eqalignno{ \mathop{{\pro}^{\ep}}_{k_0\in{\pi\over\beta}(2\Bbb Z+1)} 
  {\ts { a_ka_{-k}+b a_k+ca_{-k}+d\over a_ka_{-k} }} 
   &= \pro_{k_0>0} {\ts { (a_ka_{-k})^2+pa_ka_{-k}+q\over (a_ka_{-k})^2 }}
        \>e^{{\beta\over2}(b+c)} \cr
 &=\pro_{k_0>0}  \ts \left(1+{\omega_+\over a_ka_{-k}}\right) 
     \left(1+{\omega_-\over a_ka_{-k}}\right) \>e^{{\beta\over2}(b+c)}  \cr    } $$
where $p=-2e_\k(b+c)+ 2d+(b+c)^2-4bc$ and $q=4bce_\k^2-2e_\k d(b+c)+d^2$. 
 Application of (III.25) proves part (a) of the lemma. 
\smallskip\noindent
{\bf Part b)} First, if $b_k=a_k-\sqrt\lambda\, v$, one computes that 
$${\rm Pf}S_k=(b_kb_{-k}+\lambda\rho^2)^2+(b_k^2+b_{-k}^2+2\lambda\rho^2)\lambda 
  (w^2+x^2) +\lambda^2(w^2+x^2)^2 \eqno (\A.7 )$$
Namely, since 
$$A_{k\up}A_{-k\down}=a_ka_{-k}-\sqrt\lambda\,v(a_k+a_{-k})+i\sqrt\lambda\,w 
  (a_k-a_{-k})+\lambda(v^2+w^2) \;\;\;\Rightarrow$$
$$\eqalignno{ (A_{k\up}A_{-k\down}+\lambda\rho^2)
   (A_{k\down}A_{-k\up}+\lambda\rho^2)&= \left[a_ka_{-k}-\sqrt\lambda\,v(a_k+a_{-k})
   +\lambda(v^2+w^2+\rho^2)\right]^2   \cr
 &\phantom{mm} +\lambda w^2(a_k-a_{-k})^2  \cr
 &= \left[b_kb_{-k}+\lambda(w^2+\rho^2)\right]^2+\lambda w^2(b_k-b_{-k})^2  \cr}$$
and 
$$A_{k\up} A_{k\down}=(a_k-\sqrt\lambda\, v)^2 +\lambda w^2= 
  b_k^2+\lambda w^2$$ 
one has 
$$\eqalignno{ {\rm Pf}S_k&=\left[b_kb_{-k}+\lambda(w^2+\rho^2)\right]^2
  +\lambda w^2(b_k-b_{-k})^2 
   +\left[b_k^2+b_{-k}^2+2\lambda (w^2+\rho^2)\right]\lambda x^2 
  +\lambda^2 x^4 \cr
&=\left[b_kb_{-k}+\lambda\rho^2\right]^2 
  +( b_k^2+b_{-k}^2+2\lambda\rho^2)\lambda(w^2+x^2)+\lambda^2 (w^2+x^2)^2  \cr}$$
Now let $y^2=w^2+x^2$ and 
$$ F_k(a_k,a_{-k})=\left[b_kb_{-k}+\lambda\rho^2\right]^2 
  +( b_k^2+b_{-k}^2+2\lambda\rho^2)\lambda y^2+\lambda^2 y^4 $$
Then part (b) follows from the following formula
$$\lim_{\ep\to 0\atop \ep>0} \prod_{k_0\in {\pi\over\beta}(2\Bbb Z+1)\atop k_0>0} 
  \ts {F_k( e^{i\ep k_0}a_k, e^{-i\ep k_0}a_{-k})\over a_k^2 a_{-k}^2} = 
 \ts {  \cosh^2\left({\beta\over2}
    \sqrt{(e_\k+\sqrt\lambda\,v)^2+\lambda\rho^2}\,\right)-
  \sin^2\left({\beta\over2}\sqrt{\lambda y^2}\right)  \over 
    \cosh^2{\beta\over2}e_\k  } \,e^{-\beta \sqrt\lambda v}  \eqno (\A.8 )$$
Proof of (A.8): 
 One has $A_k A_{-k}=a_k a_{-k} 
    -\sqrt\lambda\,v(a_k+a_{-k})+\lambda v^2$ and 
$$A_k^2+A_{-k}^2=a_k^2+a_{-k}^2-2\sqrt\lambda\,v(a_k+a_{-k})+2\lambda v^2= 
   2A_k A_{-k} +(a_k-a_{-k})^2$$
which gives 
$$\eqalignno{ F_k& =(A_kA_{-k}+\lambda\rho^2+\lambda y^2)^2
   +(a_k-a_{-k})^2\lambda y^2   \cr
 &=[A_kA_{-k}+\lambda\rho^2+\lambda y^2+i\lambda^{1\over2}y(a_k-a_{-k})]\> 
  [A_kA_{-k}+\lambda\rho^2+\lambda y^2-i\lambda^{1\over2}y(a_k-a_{-k})]
    \cr
 &=\bigl[ a_ka_{-k}-(\sqrt\lambda\,v-i\lambda^{1\over2}y)a_k
    -(\sqrt\lambda\,v+i\lambda^{1\over2}y)a_{-k}+\lambda v^2 +\lambda\rho^2
   +\lambda y^2 \bigr]\times \cr
 &\phantom{mm}\bigl[ a_ka_{-k}-(\sqrt\lambda\,v+i\lambda^{1\over2}y)a_k
    -(\sqrt\lambda\,v-i\lambda^{1\over2}y)a_{-k}+\lambda v^2 +\lambda\rho^2
   +\lambda y^2 \bigr]   \cr}$$
Therefore one obtains
$$\eqalignno{ \mathop{{\pro}^\ep}_{k_0>0} {\ts  {F_k(a_k,a_{-k})\over 
   a_k^2 a_{-k}^2}}&=\lim_{\ep\to 0\atop \ep<0}  \pro_{k_0}\ts {
  a_ka_{-k}-(\sqrt\lambda\,v-i\lambda^{1\over2}y)e^{ik_0\ep}a_k
    -(\sqrt\lambda\,v+i\lambda^{1\over2}y)e^{-ik_0\ep}a_{-k}+\lambda v^2 
   +\lambda\rho^2
   +\lambda y^2 \over a_ka_{-k} }  \cr 
 &\buildrel (\A.)\over = 
  \ts{\cosh{\beta\over2}\sqrt{e_\k^2+\omega_+}\over \cosh{\beta\over2}e_\k}\,
  {\cosh{\beta\over2}\sqrt{e_\k^2+\omega_-}\over \cosh{\beta\over2}e_\k}\>
    e^{-\beta\sqrt\lambda\,v} \cr}$$
where 
$$\eqalignno{ p&=(e_\k+2\sqrt\lambda\,v)^2-e_\k^2-4(\lambda v^2+\lambda
    y^2)+2\lambda v^2
 +2\lambda\rho^2+2\lambda y^2 \cr
 &=4\sqrt\lambda\,v e_\k+2\lambda v^2+2\lambda\rho^2
 -2\lambda y^2  \cr
  q&  =4(\lambda v^2+\lambda y^2)e_\k^2+4\sqrt\lambda\,v e_\k
 ( \lambda v^2 +\lambda\rho^2 +\lambda y^2)+
  ( \lambda v^2 +\lambda\rho^2 +\lambda y^2)^2  \cr
  &=  (2\sqrt\lambda\,v e_\k+ 
  \lambda v^2 +\lambda\rho^2 +\lambda y^2)^2 +4\lambda y^2e_\k^2  \cr}$$
such that 
$$\ts {p^2\over4}-q= -4(2\sqrt\lambda\,v e_\k+ \lambda v^22 
    +\lambda\rho^2)\lambda y^2 
   -4\lambda y^2 e_\k^2=-4\bigl[ (e_\k+\sqrt\lambda\,v)^2+\lambda\rho^2\bigr] 
   \lambda y^2 $$
and therefore 
$$\eqalignno{ e_\k^2+\omega_{\pm}&=(e_\k+\sqrt\lambda\,v)^2 +\lambda\rho^2
 -\lambda y^2\pm 2i\lambda^{1\over2} y
      \bigl[ (e_\k+\sqrt\lambda\,v)^2+\lambda\rho^2\bigr]^{1\over2} \cr
 &=\left\{ \bigl[ (e_\k+\sqrt\lambda\,v)^2+\lambda\rho^2\bigr]^{1\over2}\pm
       i\lambda^{1\over2} y \right\}^2 \cr}$$
which results in 
$$\eqalignno{ \mathop{{\pro}^\ep}_{k_0>0} {\ts  { F_k(a_k,a_{-k})\over 
   a_k^2 a_{-k}^2}}&= 
   \ts{\cosh{\beta\over2}\left\{[ (e_\k+\sqrt\lambda\,v)^2
     +\lambda\rho^2]^{1\over2}+
       i\lambda^{1\over2} y \right\}    \over \cosh{\beta\over2}e_\k}\,
  {\cosh{\beta\over2}\left\{ [ (e_\k+\sqrt\lambda\,v)^2+\lambda\rho^2]^{1\over2}-
       i\lambda^{1\over2} y \right\}   \over \cosh{\beta\over2}e_\k}\>
    e^{-\beta\sqrt\lambda\,v} \cr
 &= \ts { \sinh^2({\beta\over2}i\sqrt\lambda \,y)\,+\,\cosh^2({\beta\over2}
    \sqrt{(e_\k+\sqrt\lambda\,v)^2+\lambda\rho^2}\,) \over 
    \cosh^2{\beta\over2}e_\k  } \,e^{-\beta \sqrt\lambda\,v} &\blacksquare \cr}$$
\bigskip
\bigskip
\bigskip\goodbreak
\noindent{\gross References}
\bigskip
\item{[AB]} P.W. Anderson, W.F. Brinkman, {\it Theory of Anisotropic 
    Superfluidity} in {\it Basic Notions of Condensed Matter Physics} by P.W. 
    Anderson, Benjamin/Cummings, 1984. 
\item{[B]} N.N. Bogoliubov, {\it On Some Problems of the Theory of 
 Superconductivity},   Physica 26, Supplement,  p.1-16 (1960).
\item{[BZT]} N.N. Bogoliubov, D.B. Zubarev, Iu.A. Tserkovnikov, {\it On the Theory 
 of Phase Transitions}, 
   Sov. Phys. Doklady 2, p.535 (1957).
\item{[BR]} J. Bardeen, G. Rickayzen, {\it Ground State Energy and Green's
Function for Reduced Hamiltonian for Superconductivity}, Phys. Rev. 118, 
  p.936 (1960).
\item{[BW]} R. Balian, N.R. Werthamer, {\it Superconductivity with Pairs in a 
    Relative p Wave}, Phys. Rev. 131, p.1553-1564, 1963.
\item{[FKLT]} J. Feldman, H. Kn\"orrer, D. Lehmann, E. Trubowitz, 
   {\it Fermi Liquids 
    in Two Space Dimensions}, in {\it Constructive Physics}, 
   Springer Lecture Notes in Physics, Bd. 446, 1994. 
\item{[FKT1]} J. Feldman, H. Kn\"orrer, E. Trubowitz, {\it Mathematical Methods of 
  Many Body Quantum Field Theory}, Lecture Notes, ETH Z\"urich.
\item{[FKT2]} J. Feldman, H. Kn\"orrer, E. Trubowitz, {\it A Remark on 
    Anisotropic Superconducting States}, Helv. Phys. Acta 64, p.695-699, (1991). 
\item{[FST]}  J. Feldman, M. Salmhofer, E. Trubowitz, 
  {\it  Perturbation Theory around Non-nested Fermi Surfaces I. 
    Keeping the Fermi Surface Fixed}, J.Stat.Phys. 84, p.1209-1336, (1996).  
\item{[FT]} J. Feldman, E. Trubowitz,  {\it Perturbation Theory for Many Fermion 
   Systems}, Helvetia Physica Acta 63,  p.156-260 (1990); 
     {\it The Flow of an Electron-Phonon 
    System to the Superconducting State}, Helv. Phys. Acta 64, p.214-357, (1991). 
\item{[G]} M.D. Girardeau, {\it Variational Method for the Quantum Statistics 
  of Many Particle Systems}, Phys. Rev. A 42, p.3303, (1990).
\item{[Ha]} R. Haag, {\it The Mathematical Structure of the Bardeen Cooper 
 Schrieffer model}, Nuovo Cimento 25, p.287 (1962). 
\item{[Hn]} E.R. Hanson, {\it A Table of Series and Products}, Prentice-Hall, 1975, 
              Sec. 89.5. 
\item{[Sch]} A. Sch\"utte, {\it The Symmetry of the Gap in the BCS Model for Higher 
  l-Wave Interactions}, Diploma thesis, ETH Z\"urich, 1997.
\item{[W]} G. Wentzel, Phys. Rev. 120, p.1572, (1960).

\end